\documentclass{aastex61}

\usepackage{graphicx}

\newcommand{\kms}{\ensuremath{\,\mathrm{km\,s^{-1}}}}
\newcommand{\mpc}{\ensuremath{\,\mathrm{Mpc}}}

\newcommand{\jy}{\ensuremath{\,\mathrm{Jy}}}
\newcommand{\mjy}{\ensuremath{\,\mathrm{mJy}}}
\newcommand{\cm}{\ensuremath{\,\mathrm{cm}}}
\newcommand{\mum}{\ensuremath{\,\mathrm{\mu m}}}

\newcommand{\wmm}{\ensuremath{\,\mathrm{W\,m^{-2}}}}

\newcommand{\ha}{\ensuremath{\mathrm{H\alpha}}}

\newcommand{\msun}{\ensuremath{\mathrm{M_\odot}}}

\newcommand{\hi}{{\ensuremath{\mathrm{H}\textsc{i}}}}

\begin{document}

\title{Local Volume \hi\ Survey: the far-infrared radio correlation}

\author[0000-0003-2015-777X]{Li Shao}
\affiliation{CSIRO Astronomy \& Space Science, Australia Telescope National Facility, PO Box 76, Epping, NSW 1710, Australia}
\affiliation{Research School of Astronomy and Astrophysics, Australian National University, Canberra, ACT 2611, Australia}
\affiliation{Kavli Institute for Astronomy and Astrophysics, Peking University, Beijing 100871, China}
\correspondingauthor{Li Shao}
\email{lishao@pku.edu.cn}

\author{B\"{a}rbel S. Koribalski}
\affiliation{CSIRO Astronomy \& Space Science, Australia Telescope National Facility, PO Box 76, Epping, NSW 1710, Australia}

\author[0000-0002-6593-8820]{Jing Wang}
\affiliation{Kavli Institute for Astronomy and Astrophysics, Peking University, Beijing 100871, China}
\affiliation{CSIRO Astronomy \& Space Science, Australia Telescope National Facility, PO Box 76, Epping, NSW 1710, Australia}

\author{Luis C. Ho}
\affiliation{Kavli Institute for Astronomy and Astrophysics, Peking University, Beijing 100871, China}
\affiliation{Department of Astronomy, School of Physics, Peking University, Beijing 100871, China}
 
\author[0000-0002-8057-0294]{Lister Staveley-Smith}
\affiliation{International Centre for Radio Astronomy Research, University of Western Australia, 35 Stirling Highway, Crawley, WA 6009, Australia}
\affiliation{ARC Centre of Excellence for All-Sky Astrophysics (CAASTRO), Australia}

\begin{abstract}
In this paper we measure the far-infrared (FIR) and radio flux densities of a sample of 82 local gas-rich galaxies, including 70 ``dwarf'' galaxies ($M_*<10^9\,\mathrm{M_\odot}$), from the Local Volume \hi\ Survey (LVHIS), which is close to volume limited. It is found that LVHIS galaxies hold a tight linear FIR-radio correlation (FRC) over four orders of magnitude ($F_\mathrm{1.4GHz} \propto F_\mathrm{FIR}^{1.00\pm0.08}$). However, for detected galaxies only, a trend of larger FIR-to-radio ratio with decreasing flux density is observed. We estimate the star formation rate by combining UV and mid-IR data using empirical calibration. It is confirmed that both FIR and radio emission are strongly connected with star formation but with significant non-linearity. Dwarf galaxies are found radiation deficient in both bands, when normalized by star formation rate. It urges a ``conspiracy'' to keep the FIR-to-radio ratio generally constant. By using partial correlation coefficient in Pearson definition, we identify the key galaxy properties associated with the FIR and radio deficiency. Some major factors, such as stellar mass surface density, will cancel out when taking the ratio between FIR and radio fluxes. The remaining factors, such as \hi-to-stellar mass ratio and galaxy size, are expected to cancel each other due to the distribution of galaxies in the parameter space. Such cancellation is probably responsible for the ``conspiracy'' to keep the FRC alive.
\end{abstract}


\section{Introduction}
\label{sec:intro}

It is well known that for normal star forming galaxies, the far-infrared (FIR) and radio continuum fluxes are correlated \citep{1985A&A...147L...6D,1985ApJ...298L...7H,1992ARA&A..30..575C,2001ApJ...554..803Y,2003ApJ...586..794B}. This FIR-radio correlation (FRC) is one of the tightest relation in astronomy. It is partly understood with both FIR and radio emission being connected to massive stars. The major fraction of FIR emission comes from the re-emitted ultra-violet (UV) radiation by dust in star forming regions. It is often used as star formation rate (SFR) indicator \citep{1998ARA&A..36..189K}. The radio continuum emission in our wavelength range has two potential major components. The thermal component is considered to be bremsstrahlung emission from hot star forming regions and the non-thermal component is believed to be associated with supernova remnants \citep{1992ARA&A..30..575C}. They are both used as SFR indicators, separately or combined together \citep{2002AJ....124..675C,2011ApJ...737...67M,2017ApJ...836..185T}. However, it is known that both FIR and non-thermal radio continuum are far from perfect (i.e. linear) SFR indicators \citep[see, e.g][]{2003ApJ...586..794B}. The FIR emission does not only depends on star formation but also on the dust opacity. The dust mass and spatial distribution (e.g. clumpiness) are important factors. The non-thermal radio emission is even more complicated. It depends on many different factors, such as supernova rate, magnetic field and cosmic ray escaping. Only the supernova rate is directly linked with SFR, with a fixed initial mass function\footnote{The strength of magnetic field is found to be coupled with gas density \citep[][assuming energy equipartition between cosmic ray and magnetic field]{1997A&A...320...54N,2011A&A...529A..94C,2013A&A...552A..19T}. The gas density is known to be correlated with star formation rate via the well known Kennicutt-Schmidt law \citep{1959ApJ...129..243S,1998ApJ...498..541K}. So magnetic field strength is correlated with star formation rate \citep{2013A&A...552A..19T}. But the direct physical link between them is not clear yet.}. The non-linearity of both FIR-SFR and radio-SFR relations, combined with the fact that FIR and radio correlate linearly very well, leads to an intriguing ``conspiracy'' \citep{2003ApJ...586..794B}, which is not yet well understood. The physics behind FRC is still an interesting field to investigate.

Numerous theoretical models have been proposed to explain this correlation. Almost all the models use recent star formation as the basis, yet they differ on details of how to ``convert'' star formation rate to FIR and radio luminosities, while keeping a constant FIR-to-radio flux ratio. One way is to assume that both far-UV (FUV) star light and cosmic ray energy are fully converted to FIR and radio emission \citep[``calorimetry'', see][]{1989A&A...218...67V,1994InPhT..35..527V,1996A&A...306..677L}. However, the assumption is over-simplified. Only the most dense regions are expected to be ``optically thick'' for both UV photon and cosmic ray electrons \citep[see e.g.][]{2010ApJ...717....1L}. Therefore, a more detailed tuning of the model is needed. \citet{1993ApJ...415...93H} make an assumption that the magnetic field and gas are well coupled locally, leading to similar optical depths for UV photons and cosmic rays. Their argument depends on a simplified condition, a roughly constant effective escape length (the distance travelled within galaxy before escape) for cosmic rays. But it is unlikely to be universal, especially for dwarf galaxies due to its small size. \citet{2010ApJ...717....1L} propose a more detailed model, suggesting that some ``conspiracy'' is necessary to solve the problem for low gas surface density case. In galaxies with low gas surface density, which is the case for many dwarf galaxies, the lower radio emission due to cosmic ray escaping is somehow compensated by lower FIR emission due to low dust content. They also predict in their ``standard'' model that the FRC will eventually break at very low gas mass surface density \citep[see also][]{2003ApJ...586..794B,2010ApJ...717..196L,2016A&A...593A..77S}.

Dwarf galaxies, with their typical lower stellar mass surface density, are important for testing the theoretical models. However, due to observational difficulties, only a few studies were done in this low luminosity regime. \citet{2001ApJ...554..803Y} investigated a large sample of galaxies with the NRAO VLA Sky Survey \citep[NVSS,][]{1998AJ....115.1693C} and the Infrared Astronomical Satellite \citep[IRAS,][]{1984ApJ...278L...1N} data, suggesting the FRC spanned for 4 orders of magnitude. They spotted a marginal tendency for higher FIR-to-radio flux ratios for dwarf galaxies. However, as their sample is FIR selected, it is potentially biased towards strong FIR sources, leaving the issue unresolved. \citet{2008ApJ...676..970W} also used the NVSS data but focusing on a sample of 28 Spitzer \citep{2004ApJS..154....1W} 24\mum\ detected blue compact dwarf galaxies. They found that the dwarf galaxies followed the correlation of normal star forming galaxies. \citet{2011A&A...529A..94C} carried out a focused study on dwarf galaxies with a sample of 12 galaxies within the Local Group. They were unable to make any solid conclusion as only 3 of their galaxies were detected at 2.6 GHz, though the detections did not clearly deviate from the correlation. \citet{2012MNRAS.423L.127R} used stacking techniques to overcome this low detection rate problem. They stacked the radio and FIR images of 24 dwarf galaxies, also suggesting the FRC still held for dwarf galaxies.

In this paper, we re-visit this problem by investigating individually detected dwarf galaxies with our broadband radio continuum images from the Local Volume \hi\ Survey \citep[LVHIS,][]{2018MNRAS.tmp..467K}. The LVHIS sample is a collection of gas rich galaxies, suitable for studying the star formation properties of dwarf galaxies. The LVHIS sample is close to volume limited (in sense of \hi\ mass), making it easier to interpret the statistical results of this sample. We will describe our galaxy sample and data reduction in Section \ref{sec:data}. In Section \ref{sec:emission}, we will study FRC itself and its star formation origin. In Section \ref{sec:galprop} we will study the relation between FRC and various galaxy properties.

\section{Data}
\label{sec:data}

\subsection{LVHIS and dwarf galaxy sample}

LVHIS is a southern sky survey for all \hi\ rich nearby galaxies carried out with the Australia Telescope Compact Array (ATCA). ATCA is a radio telescope array of six $22\,\mathrm{m}$ diameter dishes, located near Narrabri in New South Wales, Australia. The primary beam size is 33.6 arcmin at 1.4 GHz. The longest baseline is $6\,\mathrm{km}$, giving highest spatial resolution around 7 arcsec at 1.4 GHz.

The LVHIS sample is a collection of all galaxies with Local Group velocities $v_{LG}<550\kms$ or distance $D<10\mpc$ detected by \hi\ Parkes All Sky Survey \citep[HIPASS,][]{2001MNRAS.322..486B}. There are 82 galaxies in this \hi\ selected sample (see Table \ref{tab:name}). In this paper, we define ``dwarf galaxies'' as galaxies with stellar mass smaller than $10^9\,\msun$ or galaxies without reliable stellar mass estimation due to too low brightness. The stellar mass is estimated from the WISE \citep[the Wide-field Infrared Survey Explorer][]{2010AJ....140.1868W} 3.6\mum\ luminosity or near-infrared H band luminosity \citep{2017MNRAS.472.3029W}. It is worth mentioning that this dwarf classification is generally consistent with morphological classification retrieved from the NASA/IPAC Extragalactic Database (NED). There are totally 70 dwarf galaxies in our LVHIS sample, including 9 \hi\ galaxies without identified optical counterpart. According to NED data, there are 3 galaxies (NGC 4945, NGC 5128 and the Circinus Galaxy) identified as potential active galactic nucleus (AGN) hosts. We exclude them from our analysis to minimize AGN contamination.

\startlongtable
\begin{deluxetable}{lccrrrrrr}
\tabletypesize{\footnotesize}
\tablecaption{Basic properties of the LVHIS sample.\label{tab:name}}
\tablehead{\multicolumn{1}{c}{LVHIS ID} & \multicolumn{1}{c}{Optical name} & \multicolumn{1}{c}{HIPASS ID} & \multicolumn{1}{c}{RA} & \multicolumn{1}{c}{Dec} & \multicolumn{1}{c}{Distance} & \multicolumn{1}{c}{Morph} & \multicolumn{1}{c}{$\log(M_*/\mathrm{M_\odot})$} & \multicolumn{1}{c}{$\log(M_\mathrm{HI}/\mathrm{M_\odot})$} \\
 & & & \multicolumn{1}{c}{deg} & \multicolumn{1}{c}{deg} & \multicolumn{1}{c}{Mpc} &  & & \\
\multicolumn{1}{c}{(1)} & \multicolumn{1}{c}{(2)} & \multicolumn{1}{c}{(3)} & \multicolumn{1}{c}{(4)} & \multicolumn{1}{c}{(5)} & \multicolumn{1}{c}{(6)} & \multicolumn{1}{c}{(7)} & \multicolumn{1}{c}{(8)} & \multicolumn{1}{c}{(9)} }
\startdata
LVHIS 001 & ESO 349--G031 & J0008--34 & $2.0557$ & $-34.57833$ & $3.21$ & IBm & $7.29$ & $7.07$ \\
LVHIS 002 & ESO 294--G010 &  & $6.6390$ & $-41.85530$ & $1.92$ & dS0/Im & $6.71$ & \nodata \\
LVHIS 003 & ESO 410--G005 & J0015--32 & $3.8815$ & $-32.17994$ & $1.92$ & dS0-a & $6.23$ & $5.91$ \\
LVHIS 004 & NGC 55 & J0015--39 & $3.7233$ & $-39.19664$ & $2.13$ & SBm & $9.36$ & $9.34$ \\
LVHIS 005 & NGC 300 & J0054--37 & $13.7228$ & $-37.68439$ & $2.15$ & Sd & $9.49$ & $9.36$ \\
LVHIS 006 & NGC 253 & J0047--25 & $11.8880$ & $-25.28822$ & $3.94$ & SABc & $10.43$ & $9.43$ \\
LVHIS 007 & NGC 247 & J0047--20 & $11.7856$ & $-20.76039$ & $3.65$ & SABd & $9.74$ & $9.32$ \\
LVHIS 008 & NGC 625 & J0135--41 & $23.7693$ & $-41.43619$ & $3.89$ & SBm & $8.70$ & $7.94$ \\
LVHIS 009 & ESO 245--G005 & J0145--43 & $26.2656$ & $-43.59804$ & $4.43$ & IBm & $8.31$ & $8.59$ \\
LVHIS 010 & ESO 245--G007 & J0150--44 & $27.7764$ & $-44.44469$ & $0.44$ & Im & $6.77$ & $4.92$ \\
LVHIS 011 & ESO 115--G021 & J0237--61 & $39.4470$ & $-61.33669$ & $4.99$ & SBdm & $8.06$ & $8.81$ \\
LVHIS 012 & ESO 154--G023 & J0256--54 & $44.2099$ & $-54.57142$ & $5.76$ & SBm & $8.74$ & $9.01$ \\
LVHIS 013 & ESO 199--G007 & J0258--49 & $44.5174$ & $-49.38305$ & $6.60$ & Sd & $7.00$ & $7.22$ \\
LVHIS 014 & NGC 1313 & J0317--66 & $49.5669$ & $-66.49825$ & $4.07$ & SBd & $9.49$ & $9.28$ \\
LVHIS 015 & NGC 1311 & J0320--52 & $50.0290$ & $-52.18553$ & $5.20$ & SBm & $8.23$ & $7.94$ \\
LVHIS 016 & AM 0319--662 & J0321--66 & $50.2600$ & $-66.31917$ & $3.98$ & dIrr & $7.11$ & $5.96$ \\
LVHIS 017 & IC 1959 & J0333--50 & $53.3025$ & $-50.41425$ & $6.05$ & SBm & $8.54$ & $8.35$ \\
LVHIS 018 & NGC 1705 & J0454--53 & $73.5563$ & $-53.36106$ & $5.11$ & S0 & $8.24$ & $7.87$ \\
LVHIS 019 & ESO 252--IG001 & J0457--42 & $74.2446$ & $-42.80389$ & $7.20$ & dIrr & $7.32$ & $8.09$ \\
LVHIS 020 & ESO 364--G?029 & J0605--33 & $91.4384$ & $-33.08084$ & $7.60$ & IBm & $7.58$ & $8.48$ \\
LVHIS 021 & AM 0605--341 & J0607--34 & $91.8321$ & $-34.20444$ & $7.40$ & SBdm & $7.90$ & $8.08$ \\
LVHIS 022 & NGC 2188 & J0610--34 & $92.5397$ & $-34.10621$ & $7.40$ & SBm & $8.93$ & $8.65$ \\
LVHIS 023 & ESO 121--G020 & J0615--57 & $93.9758$ & $-57.72546$ & $6.05$ & Im & $7.22$ & $7.80$ \\
LVHIS 024 & ESO 308--G022 & J0639--40 & $99.8863$ & $-40.72083$ & $7.70$ & dIrr & $7.82$ & $7.79$ \\
LVHIS 025 & AM 0704--582 & J0705--58 & $106.3283$ & $-58.52028$ & $4.90$ & SBm & $6.18$ & $8.27$ \\
LVHIS 026 & ESO 059--G001 & J0731--68 & $112.8258$ & $-68.18799$ & $4.57$ & IBm & $8.08$ & $7.92$ \\
LVHIS 027 & NGC 2915 & J0926--76 & $141.5480$ & $-76.62633$ & $3.78$ & I0 & $8.40$ & $8.50$ \\
LVHIS 028 & ESO 376--G016 & J1043--37 & $160.8627$ & $-37.04260$ & $7.10$ & dIrr & $8.01$ & $8.09$ \\
LVHIS 029 & ESO 318--G013 & J1047--38 & $161.9233$ & $-38.85367$ & $6.50$ & SBd & $7.36$ & $7.96$ \\
LVHIS 030 & ESO 215--G?009 & J1057--48 & $164.3746$ & $-48.17861$ & $5.25$ & dIrr & $7.96$ & $8.85$ \\
LVHIS 031 & NGC 3621 & J1118--32 & $169.5688$ & $-32.81406$ & $6.70$ & Sd & $9.67$ & $9.96$ \\
LVHIS 032 &  & J1131--31 & $172.8942$ & $-31.67453$ & $6.70$ & dIrr & \nodata & $7.03$ \\
LVHIS 033 &  & J1132--32 & $173.2954$ & $-32.96328$ & $6.70$ & dIrr & $7.73$ & $7.03$ \\
LVHIS 034 & ESO 320--G014 & J1137--39 & $174.4718$ & $-39.22034$ & $6.08$ & dIrr & $8.08$ & $7.23$ \\
LVHIS 035 & ESO 379--G007 & J1154--33 & $178.6812$ & $-33.55999$ & $5.22$ & dIrr & $7.30$ & $7.49$ \\
LVHIS 036 & ESO 379--G024 & J1204--35 & $181.2361$ & $-35.74303$ & $4.90$ & dIrr & $7.54$ & $7.16$ \\
LVHIS 037 & ESO 321--G014 & J1214--38 & $183.4567$ & $-38.23137$ & $3.18$ & IBm & $7.02$ & $7.08$ \\
LVHIS 038 & IC 3104 & J1219--79 & $184.6919$ & $-79.72607$ & $2.27$ & IBm & $8.18$ & $6.99$ \\
LVHIS 039 & ESO 381--G018 & J1244--35 & $191.1766$ & $-35.96672$ & $5.32$ & dIrr & $7.21$ & $7.24$ \\
LVHIS 040 & ESO 381--G020 & J1246--33 & $191.5028$ & $-33.83694$ & $5.45$ & IBm & $7.41$ & $8.36$ \\
LVHIS 041 &  & J1247--77 & $191.8850$ & $-77.58164$ & $3.16$ & Im & $7.07$ & $6.99$ \\
LVHIS 042 & [CFC97] CEN 06 & J1305--40 & $196.2587$ & $-40.08278$ & $5.78$ & dIrr & $7.75$ & $7.55$ \\
LVHIS 043 & NGC 4945 & J1305--49 & $196.3645$ & $-49.46821$ & $3.80$ & SBcd & $10.15$ & $9.14$ \\
LVHIS 044 & ESO 269--G058 & J1310--46 & $197.6372$ & $-46.99091$ & $3.80$ & I0 & $8.47$ & $7.26$ \\
LVHIS 045 &  & J1321--31 & $200.2892$ & $-31.53367$ & $5.22$ & dIrr & \nodata & $7.53$ \\
LVHIS 046 & NGC 5102 & J1321--36 & $200.4900$ & $-36.63024$ & $3.40$ & S0 & $9.48$ & $8.36$ \\
LVHIS 047 & AM 1321--304 & J1324--30 & $201.1510$ & $-30.97181$ & $4.63$ & dIrr & $7.88$ & $6.93$ \\
LVHIS 048 & NGC 5128 & J1324--42 & $201.3651$ & $-43.01911$ & $3.75$ & S0 & $11.02$ & $8.48$ \\
LVHIS 049 & IC 4247 & J1326--30 & $201.6851$ & $-30.36242$ & $4.97$ & S? & $7.49$ & $7.28$ \\
LVHIS 050 & ESO 324--G024 & J1327--41 & $201.9057$ & $-41.48068$ & $3.73$ & Im & $8.04$ & $8.23$ \\
LVHIS 051 & ESO 270--G017 & J1334--45 & $203.6971$ & $-45.54750$ & $6.95$ & SBm & $8.74$ & $8.84$ \\
LVHIS 052 & UGCA 365 & J1336--29 & $204.1296$ & $-29.23489$ & $5.25$ & Im & $7.18$ & $7.43$ \\
LVHIS 053 & NGC 5236 & J1337--29 & $204.2540$ & $-29.86542$ & $4.92$ & Sc & $10.50$ & $9.91$ \\
LVHIS 054 &  & J1337--39 & $204.3542$ & $-39.89631$ & $4.83$ & Im & \nodata & $7.57$ \\
LVHIS 055 & NGC 5237 & J1337--42 & $204.4127$ & $-42.84697$ & $3.40$ & I0 & $8.28$ & $7.48$ \\
LVHIS 056 & ESO 444--G084 & J1337--28 & $204.3333$ & $-28.04500$ & $4.61$ & Im & $6.91$ & $7.91$ \\
LVHIS 057 & NGC 5253 & J1339--31 & $204.9832$ & $-31.64011$ & $3.56$ & S0 & $8.61$ & $7.96$ \\
LVHIS 058 & IC 4316 & J1340--28 & $205.0767$ & $-28.89222$ & $4.41$ & IBm & $7.91$ & $7.08$ \\
LVHIS 059 & NGC 5264 & J1341--29 & $205.4028$ & $-29.91308$ & $4.53$ & IBm & $8.66$ & $7.69$ \\
LVHIS 060 & ESO 325--G?011 & J1345--41 & $206.2521$ & $-41.86111$ & $3.40$ & IBm & $7.14$ & $7.86$ \\
LVHIS 061 &  & J1348--37 & $207.1421$ & $-37.96889$ & $5.75$ & dIrr & \nodata & $7.10$ \\
LVHIS 062 & ESO 174--G?001 & J1348--53 & $206.9921$ & $-53.34747$ & $3.60$ & Im? & $7.64$ & $8.20$ \\
LVHIS 063 & ESO 383--G087 & J1349--36 & $207.3229$ & $-36.06344$ & $3.45$ & SBdm & $8.94$ & $7.85$ \\
LVHIS 064 &  & J1351--47 & $207.8383$ & $-46.99806$ & $5.73$ & dIrr & \nodata & $7.44$ \\
LVHIS 065 & NGC 5408 & J1403--41 & $210.8371$ & $-41.37769$ & $4.81$ & IBm & $8.25$ & $8.44$ \\
LVHIS 066 & Circinus Galaxy & J1413--65 & $213.2915$ & $-65.33922$ & $4.20$ & Sb & $10.09$ & $9.83$ \\
LVHIS 067 & UKS 1424--460 & J1428--46 & $217.0154$ & $-46.30167$ & $3.58$ & IBm & $7.53$ & $7.70$ \\
LVHIS 068 & ESO 222--G010 & J1434--49 & $218.7608$ & $-49.42064$ & $5.80$ & dIrr & $7.79$ & $7.70$ \\
LVHIS 069 &  & J1441--62 & $220.4258$ & $-62.76783$ & $6.00$ & dIrr & $7.77$ & $7.31$ \\
LVHIS 070 & ESO 272--G025 & J1443--44 & $220.8563$ & $-44.70525$ & $5.90$ & dIrr & $8.04$ & $7.11$ \\
LVHIS 071 & ESO 223--G009 & J1501--48 & $225.2861$ & $-48.29046$ & $6.49$ & Im & $8.86$ & $8.98$ \\
LVHIS 072 & ESO 274--G001 & J1514--46 & $228.5577$ & $-46.80794$ & $3.09$ & Sd & $9.01$ & $8.49$ \\
LVHIS 073 &  & J1526--51 & $231.5933$ & $-51.17506$ & $5.70$ & dIrr & $6.92$ & $7.58$ \\
LVHIS 074 & ESO 137--G018 & J1620--60 & $245.2433$ & $-60.49103$ & $6.40$ & Sc & $8.43$ & $8.62$ \\
LVHIS 075 & IC 4662 & J1747--64 & $266.7869$ & $-64.64176$ & $2.44$ & IBm & $8.01$ & $8.16$ \\
LVHIS 076 & ESO 461--G036 & J2003--31 & $300.9891$ & $-31.68161$ & $7.83$ & dIrr & $8.15$ & $8.03$ \\
LVHIS 077 & IC 5052 & J2052--69 & $313.0232$ & $-69.20164$ & $6.03$ & SBd & $9.09$ & $8.89$ \\
LVHIS 078 & IC 5152 & J2202--51 & $330.6730$ & $-51.29644$ & $1.97$ & Im & $8.15$ & $7.95$ \\
LVHIS 079 & UGCA 438 & J2326--32 & $351.6147$ & $-32.38875$ & $2.18$ & IBm & $6.87$ & $6.62$ \\
LVHIS 080 & UGCA 442 & J2343--31 & $355.9398$ & $-31.95678$ & $4.27$ & SBm & $7.34$ & $8.35$ \\
LVHIS 081 & ESO 149--G003 & J2352--52 & $358.0117$ & $-52.57772$ & $5.90$ & IBm & $7.41$ & $7.77$ \\
LVHIS 082 & NGC 7793 & J2357--32 & $359.4576$ & $-32.59103$ & $3.91$ & Sd & $9.47$ & $8.95$ \\
\enddata
\tablecomments{Columns: (1): LVHIS ID; (2) optical name; (3) HIPASS ID; (4) right accension; (5) declination; (6) distance; (7) morphology classification; (8) stellar mass; (9) \hi\ mass.\\A machine readable version of this table is available as online supplementary data.}
\end{deluxetable}

\subsection{Continuum observations and data reduction}
\label{sec:data_reduction}

The 1.4 GHz (20\cm) radio continuum observations were taken with three configurations of the ATCA with baselines up to 6 km, including at least one compact configuration. See \citet{2018MNRAS.tmp..467K} for a description of these observations. Most galaxies were observed in two frequency bands, one narrow band for high resolution \hi\ observation and one broad band (128 MHz, divided into 32 channels) centred at either 1384 or 1376 MHz. The integration time is around 30 hours for each galaxy. We only use the broad band data to produce continuum maps. The narrow band continuum images are too shallow to make any meaningful contribution to our analysis.

We use \textsc{miriad} v1.5 to reduce the data. An automatic flagging for radio frequency interference based on Stokes V properties is executed right after applying calibration tables. Manual flagging is also applied if necessary. We take several iterations to self-calibrate the data to improve the phase and/or amplitude calibration. All the visibility data from different configurations are combined together to produce dirty maps, using $\mathrm{robust}=0$ weighting, to provide a compromise between sensitivity and resolution. We clean the dirty map in an area of $1.25^{\circ}\times1.25^{\circ}$ around our galaxy. If there are strong sources at the edge of the beam, we expand our clean area to make sure they are also cleaned, or subtract those strong sources directly from visibility data to improve image quality. The rms threshold to stop cleaning is set around $3\times10^{-5}\jy$, but could be manually adjusted according to actual data quality. The restored radio continuum images of LVHIS galaxies are displayed in Appendix \ref{sec:radio_image}. The size of the synthesis beam varies significantly, due to many factors, such as object latitude, array configuration and data flagging. The typical major and minor axes are 18.3 and 9.0 arcsec, respectively.

We estimate the noise level of the continuum image ($\sigma_\mathrm{bkg}$) by computing the standard deviation of 3-$\sigma$ clipped pixel values in a $10'\times10'$ region centred at the galaxy. Within this region, any 3-$\sigma$ detected source is masked out to avoid source contamination. We note that the detail of the mask is not important because the original pixel histogram in the region is noise dominated. The size of the region is selected to be sufficiently large to get meaningful background noise, but small enough to represent the local noise level at the position of our galaxy (the rms is usually worse at the edge of the image due to the primary beam correction). The typical rms of our continuum image is $\sim0.05$ mJy/beam. Our continuum images of NGC 253 and NGC 4945 are discarded due to strong cleaning residuals, which significantly degrade the image quality. We use literature values instead for these two galaxies and any other galaxy without LVHIS broad band continuum data. Polarization data are not used for this study.

\subsection{1.4 GHz flux density measurement}

We use two different methods to measure the 1.4 GHz flux densities of our LVHIS galaxies. Here we briefly describe the methods and discuss the reliability of the measurements.

\subsubsection{SFIND flux density}

We use the \textsc{miriad} task \textsc{sfind} to detect and measure the radio continuum flux densities of our galaxies. We use the default false discovery rate (FDR) algorithm \citep{2002AJ....123.1086H} for source detection. We set the ``smoothing'' box size (parameter ``rmsbox'') as 200 pixels (300 arcsec), to make sure the background noise can be determined from a sufficiently large source-free area. We set the default percentage of acceptable false pixels (parameter ``alpha'') as 2.0\%, corresponding to a $\sim4$-$\sigma$ threshold for our images. We also compare the result with the output of a previous \textsc{sfind} algorithm based on simple sigma clipping\footnote{see http://www.atnf.csiro.au/computing/software/miriad/doc/sfind.html}. The flux density difference between the two methods is negligible for our galaxies.

We match the detections with optical/\hi\ positions. We check the matching by eye to remove false matching, which is more likely to happen when the radio source is clearly extended or the target galaxy is not detected at all. The flux densities and their uncertainties are directly extracted from the \textsc{sfind} output for detections. It works fine for point source or barely resolved source. For extended radio source that is detected as multiple objects by \textsc{sfind}, we add up the flux densities of individual components as total flux density. We combine the background uncertainty and the Gaussian fitting uncertainty calculated by \textsc{sfind}.

\subsubsection{Aperture flux density}

As a large fraction of our galaxies are extended sources, it is possible that \textsc{sfind} misses some flux density. For comparison, we measure the total flux density within a given aperture, if available, to better account for the extended emission. The aperture is based on the B band 25 $\mathrm{mag/arcsec}^2$ isophotal ellipse, taken from the surface photometry catalogue of the ESO-Uppsala galaxies \citep{1989spce.book.....L}. If the B band aperture is not available, we use WISE 3.4\mum\ 1-$\sigma$ isophotal ellipse \citep[see][]{2017MNRAS.472.3029W}. There are 62 LVHIS galaxies with available B band aperture, and 17 with a WISE aperture (see Table \ref{tab:flux}). As the synthesis beam size varies dramatically from galaxy to galaxy, we convolve the ellipse with the beam shape (2 times the major and minor axes, respectively) in order to enlarge the aperture to make sure we do not miss radio flux (therefore, the actual photometry aperture is slightly larger than the optical aperture listed in Table \ref{tab:flux}). We subtract nearby sources within the aperture but not associated with our target galaxy. We manually identify the deconvolution model components of these sources, subtract models from the visibility data, remake the radio image and measure the aperture flux density. This minimizes the neighbour contamination to our flux density measurement.

\begin{figure*}
  \centering \includegraphics[width=8cm]{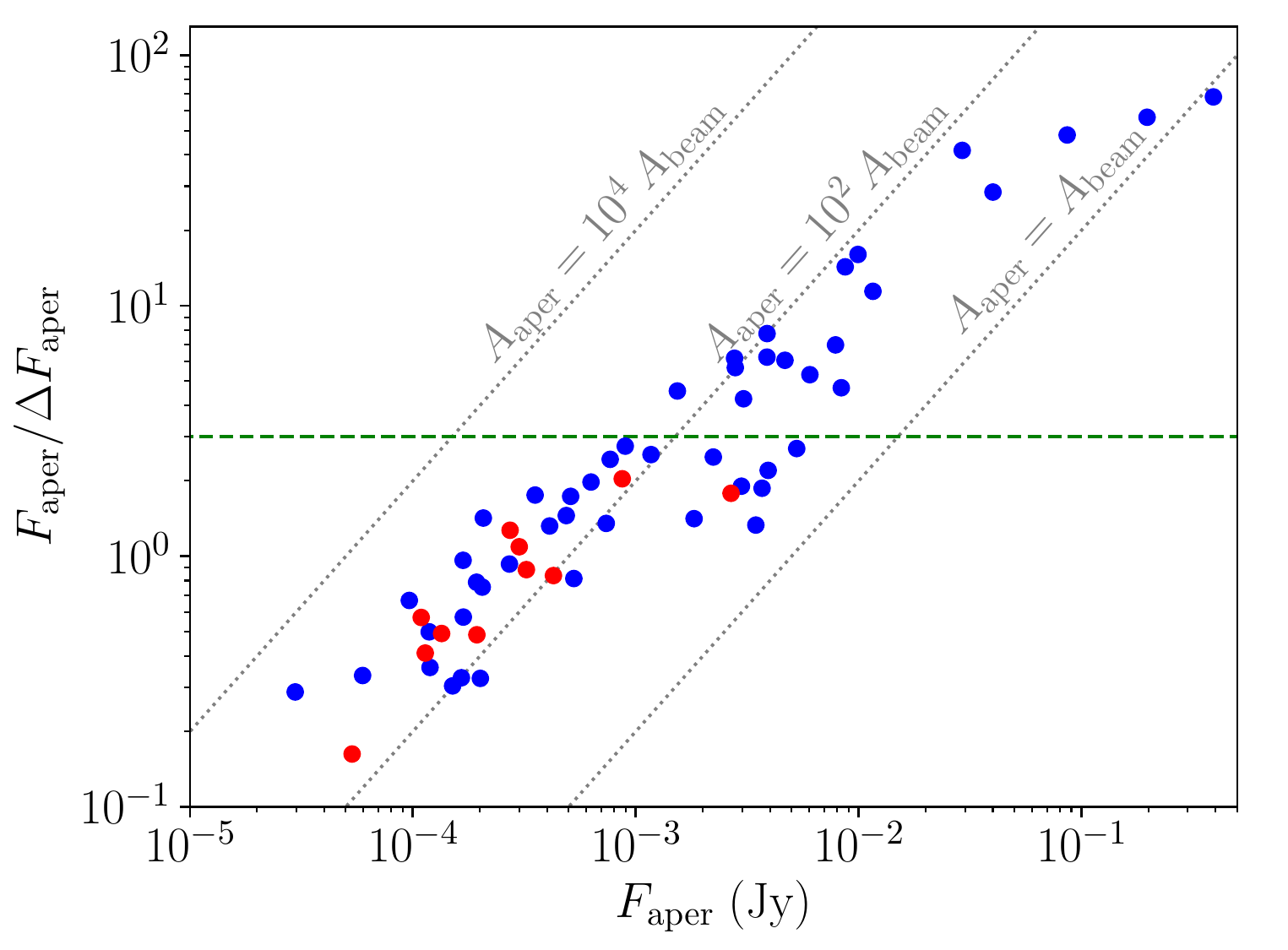}
  \caption{Signal-to-noise ratio as a function of integrated aperture flux density. The red symbols are the objects with negative flux densities, but shown here as absolute values. The dashed line indicates $F_\mathrm{aper}/\Delta F_\mathrm{aper} = 3$. The dotted lines indicate the expected relation with given aperture area (1, $10^2$, $10^4$ times the beam area, respectively) and typical image rms (0.05\mjy/beam).}
  \label{fig:aper_snr}
\end{figure*}

The aperture flux density uncertainty is estimated as $\sigma = \sigma_\mathrm{bkg} \sqrt{2A_\mathrm{aper} / A_\mathrm{beam}}$. Here $\sigma_\mathrm{bkg}$ is the rms value of the background, as described above. $A_\mathrm{aper}$ is the size of the aperture in unit of pixels. $A_\mathrm{beam}$ is the size of the synthesis beam in unit of pixels\footnote{The uncertainty of the aperture flux can be estimated by $\sigma_\mathrm{measure} = \sqrt{f_\mathrm{corr} \sum\sigma_i^2} = \sqrt{f_\mathrm{corr} A_\mathrm{aper} \sigma_\mathrm{bkg}^2}$. $f_\mathrm{corr}$ is the correlated noise correction factor. For 2D Gaussian beam, $f_\mathrm{corr} = 2A_\mathrm{beam}$. To convert the unit from $\jy/\mathrm{beam}$ to \jy, $\sigma = \sigma_\mathrm{measure} / A_\mathrm{beam}$. Therefore, $\sigma = \sigma_\mathrm{bkg} \sqrt{2A_\mathrm{aper} / A_\mathrm{beam}}$.}. Figure \ref{fig:aper_snr} shows the measured signal-to-noise ratio as a function of measured aperture flux density. As aperture flux density is an integration of all pixel values within the aperture, it is expected to get negative ones simply due to statistical fluctuation for weak sources. The objects with negative flux densities are shown as red symbols with their absolute values. Negative flux densities are all well below 3-$\sigma$. Therefore we use a minimum signal-to-noise ratio of 3 to define a detection.

\subsubsection{Comparison}

\begin{figure*}
  \centering \includegraphics[width=8cm]{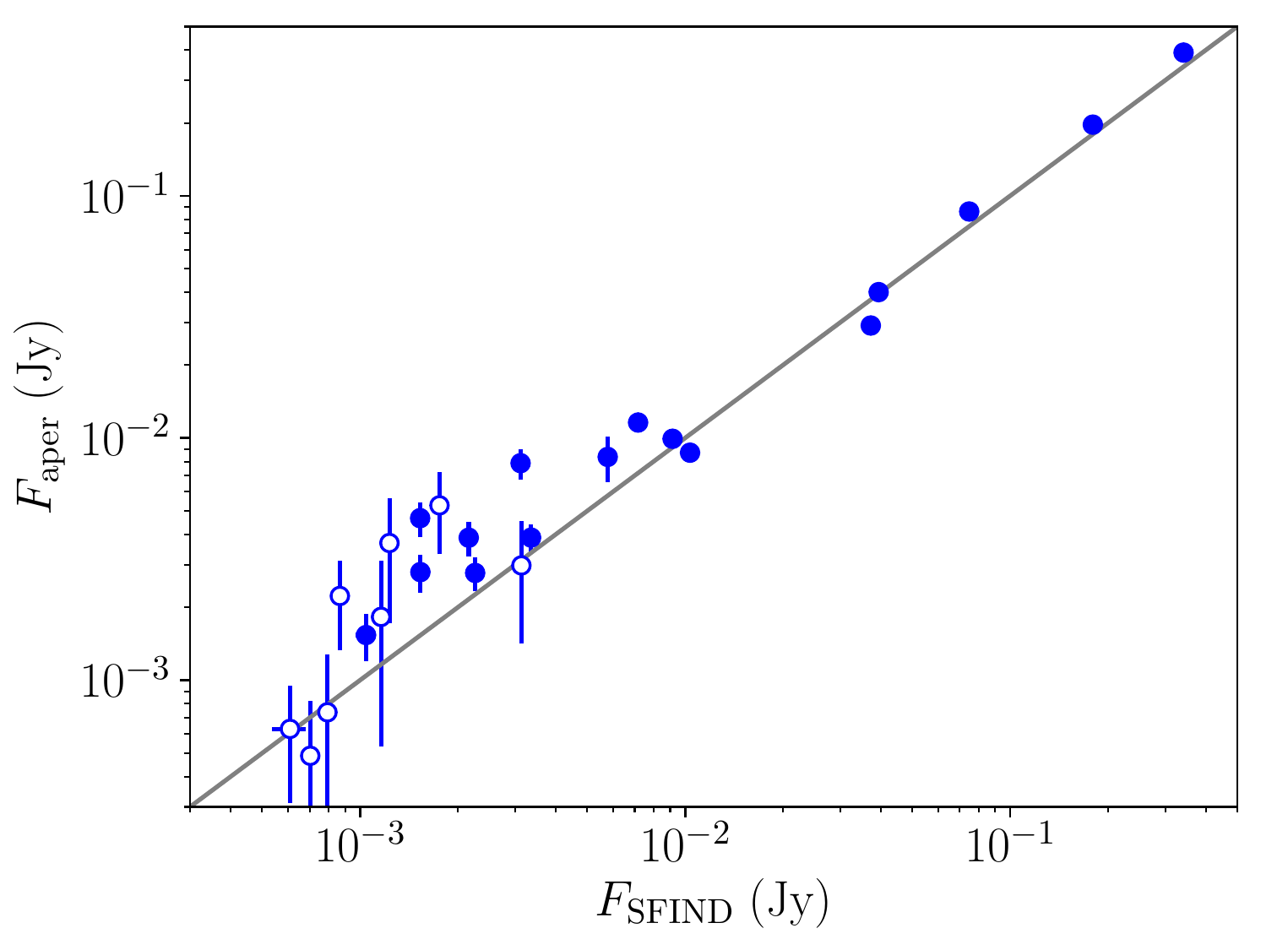}
  \caption{Comparison of \textsc{sfind} and aperture flux densities for galaxies detected with \textsc{sfind} method. The solid and empty symbols indicate the signal-to-noise ratio of aperture flux density greater and less than 3, respectively. The 1-$\sigma$ error bars are displayed.}
  \label{fig:sfind_aper}
\end{figure*}

In Figure \ref{fig:sfind_aper} we compare flux densities obtained with the two methods mentioned above. Above 10\mjy\, they are generally consistent with each other. However, the aperture flux densities seem systematically higher for sources below 10\mjy. This is expected because the aperture can pick up flux densities that are: 1) not well cleaned; 2) buried in noise due to low surface brightness. The aperture is probably superior to \textsc{sfind} for measuring total flux density. The aperture flux density is also better for large extended source as de-blending is not necessary. However, the aperture is not optimized on the radio image itself so that it may include too many empty sky pixels, reducing the signal-to-noise ratio. This has strong impact on detecting weak point sources. The aperture flux density is also more prone to image artefacts, such as cleaning residuals of nearby strong sources. We are not able to totally rule out the possibility that the aperture flux density excess in low flux density region is due to image artefacts. However, it will not affect the main result of this paper.

For final flux density catalogue (Table \ref{tab:flux}), we adopt aperture flux densities first if above 3-$\sigma$. There are 18 galaxies with aperture flux densities and 8 galaxies with \textsc{sfind} flux densities. For these detections, we add additional 1\% calibration error\footnote{http://www.atnf.csiro.au/observers/memos/d96783$\sim$1.pdf}. For non-detections, we provide 3-$\sigma$ upper limits estimated from the aperture photometry if the aperture is available (37 galaxies). Otherwise we provide 3-$\sigma$ upper limit assuming an underlying point source as a rough estimation (3 galaxies). For sources without LVHIS broad band continuum image (16 galaxies, including NGC 253 and NGC 4945), we use literature flux densities (8 galaxies) close to our frequency ($1.4\,\mathrm{GHz}$) and preferentially from extended photometry. If no error is given in literature, we assume 5\% uncertainty. The total detection rate we achieve is 41.5\% (34 galaxies). The results are summarized in Table \ref{tab:flux}, in which the ``Flag'' column indicates the source of the radio flux density we use.

\subsection{FIR flux density}

We measure FIR flux densities from IRAS all-sky survey data \citep{1984ApJ...278L...1N}. The angular resolution of IRAS is poor, 60 and 100 arcsec at 60\mum\ and 100\mum\ (diameter of 80\% encircled energy, diffraction limited). However, some of LVHIS galaxies are still extended. Fluxes from pipeline produced IRAS catalogues, i.e. the Point Source Catalog (PSC) and the Faint Source Catalog (FSC), are expected to be systematically underestimated (for more details of PSC and FSC, please read IRAS Explanatory Supplement\footnote{http://irsa.ipac.caltech.edu/IRASdocs/exp.sup/toc.html}). As flux density is the key parameter of our study, we use the IRAS Scan Processing and Integration tool\footnote{http://irsa.ipac.caltech.edu/applications/Scanpi/} (Scanpi, version 6.0) to get better measurements. Scanpi is an interactive set of tools to extract, display and measure flux density from calibrated scan data (the basic data format of IRAS). It provides tools to perform more detailed photometry, which offers several advantages. First, it is possible to measure the flux density of highly extended sources, up to around 15 arcmin in radius, suitable for LVHIS galaxies. Second, the sensitivity is a factor of 2-5 better than that of PSC. It allows us to measure lower flux densities. Third, de-blending nearby sources and removing foreground or background emission can be done interactively. This again helps to detect weaker objects and improve signal-to-noise ratio.

Usually we use the default parameter set in Scanpi. For data extraction, we use all the scans with ``cross-scan distance'' (the distance of scan from the source) less than 1.7 arcmin and the median coadding method, which is robust to outliers. For background fitting, we use all the data points within 30 arcmin radius but excluding the central $\pm4$ arcmin (60\mum) or $\pm6$ arcmin (100\mum) region. A 2nd-order polynomial is fit to the background region and extrapolated to the source region. A point source template based on IRAS PSF is fit to the data within the central $\pm3.2$ arcmin (60\mum) or $\pm6.4$ arcmin (100\mum) region, to estimate point source flux density. Readers are referred to Scanpi website for more details of the input parameters.

Scanpi provides three different flux density measurements: model flux density, zero-point flux density and centre flux density. The model flux density is based on template fitting to the central region, as mentioned above. It is the best estimator when the target is point source and the fitting quality is high. However, the fitting is not always reliable due to data quality and/or numerical stability. It is not useful when the fitting quality is poor. It also misses fluxes for extended sources. The zero-point flux density is the integration of the averaged scan between two nearest zero-crossings (the position where the flux density reaches zero, after background subtraction). It is a good flux density estimator for highly extended sources. But it is strongly affected by the reliability of background estimation. If the background is poorly determined, especially when there are blended sources or foreground cirrus structures, the estimated flux density will be dramatically wrong. The centre flux density is the integration of averaged scan within the central $\pm2.5$ arcmin (60\mum) or $\pm4$ arcmin (100\mum) region. It is a compromise of model flux density and zero-point flux density. It gives a reasonable alternative estimation for point-like sources when the template fitting fails. It is less dependant on background than the zero-point flux density but not as useful when the source extends beyond the integration region. For each galaxy, we manually choose the optimal method.

\begin{figure*}
  \centering \includegraphics[width=8cm]{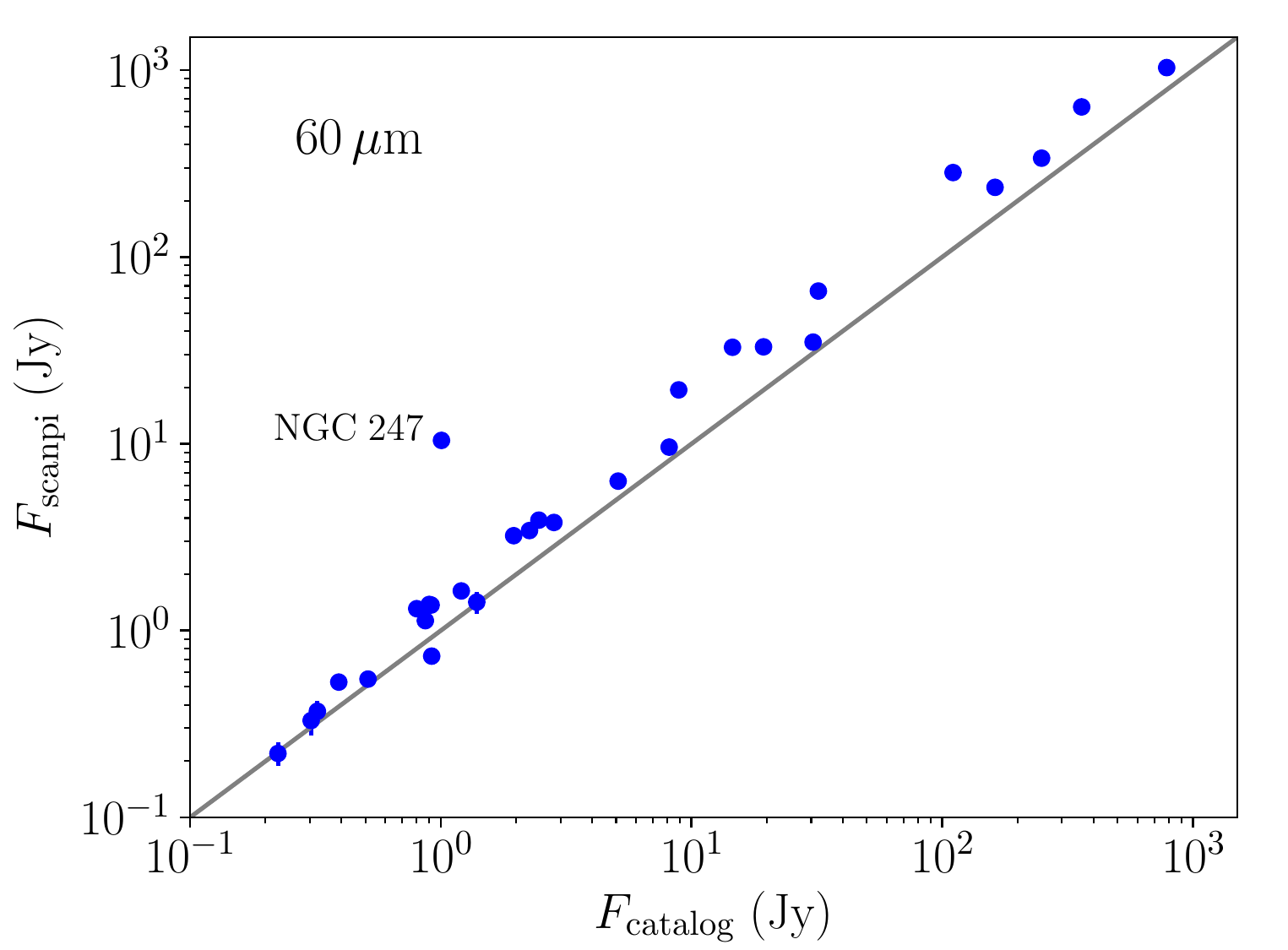} \includegraphics[width=8cm]{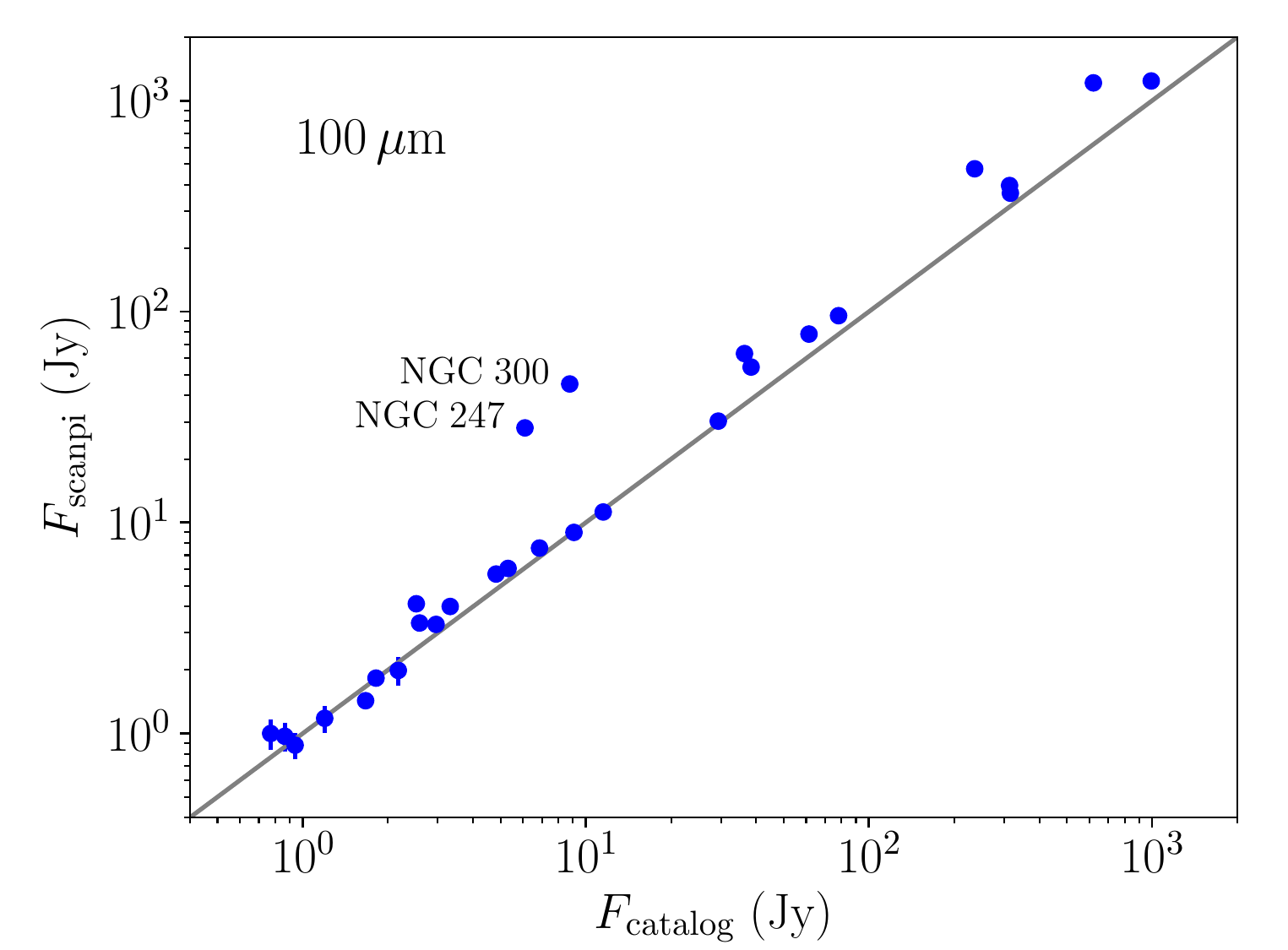}
  \caption{Direct comparison of IRAS Scanpi flux density and catalogue flux density (from PSC or FSC) at 60 and 100\mum. Clear outliers are labelled (see text for details).}
  \label{fig:fir_flux}
\end{figure*}

\begin{figure*}
  \centering \includegraphics[width=8cm]{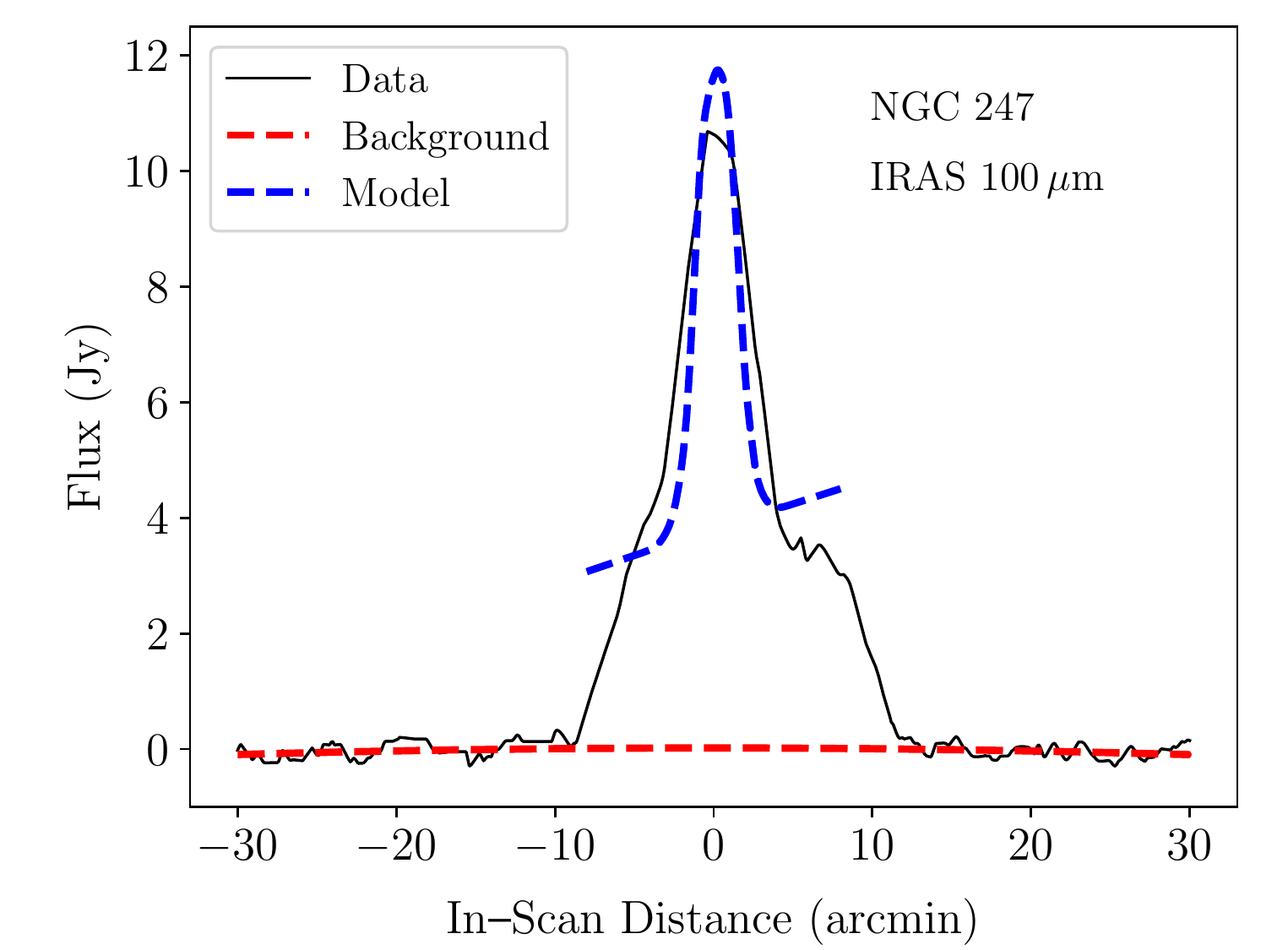} \includegraphics[width=8cm]{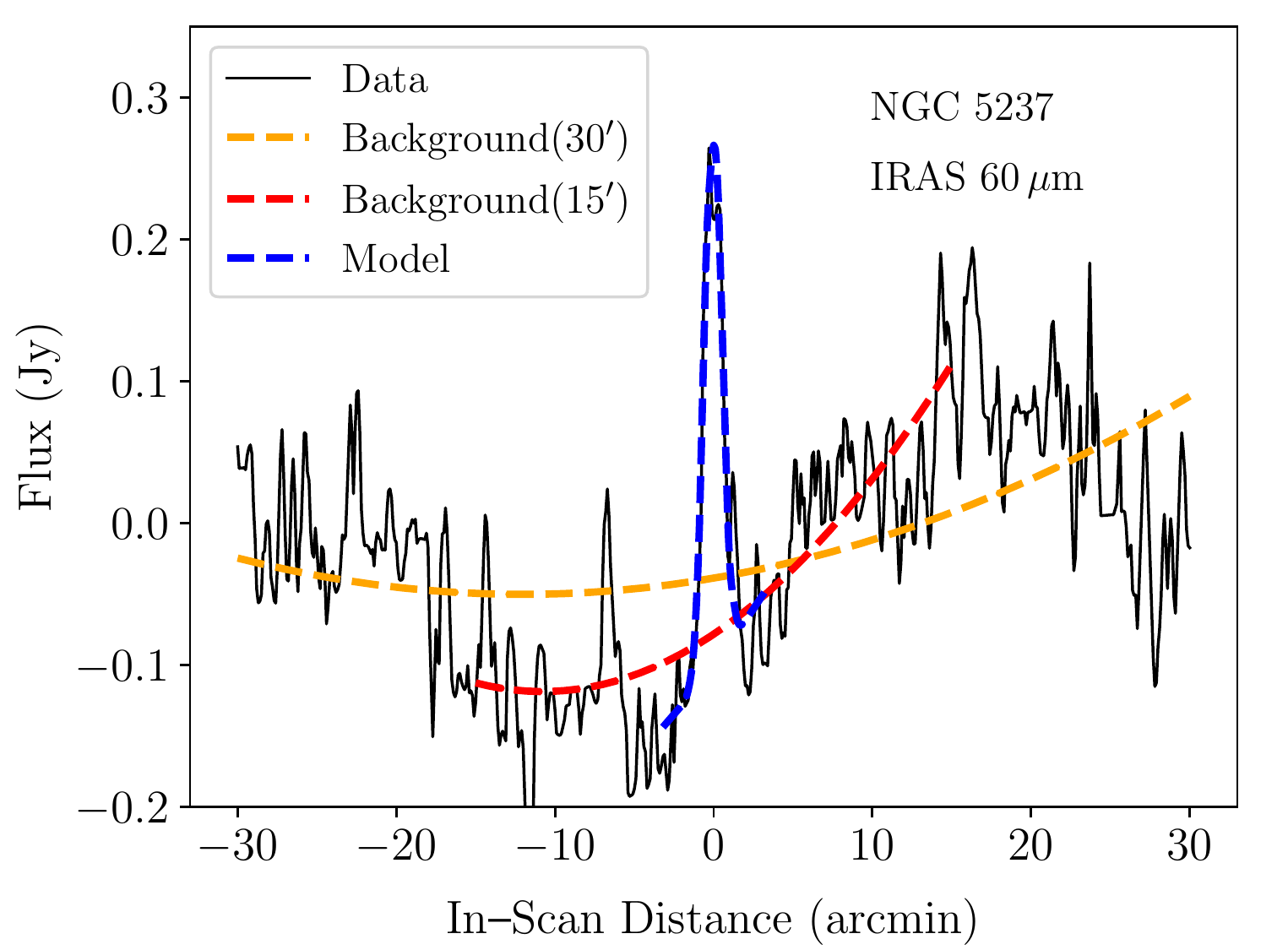}
  \caption{Examples of Scanpi scan data. Red and orange lines show the background calculated by Scanpi (using different region parameters). Blue line shows the result of automatic model fitting by Scanpi.}
  \label{fig:scanpi}
\end{figure*}

Figure \ref{fig:fir_flux} shows a comparison between the flux density we measured with Scanpi and that from PSC or FSC. It is clear that the Scanpi flux density is systematically larger than the catalogue one. The offset is larger when the source is brighter. It is consistent with our expectation. The brighter sources are usually more massive galaxies, more likely to be extended sources. The offset is stronger in 60\mum\ than in 100\mum. It can be understood as the PSF FWHM of 100\mum\ is much larger than that of 60\mum. We also notice two clear outliers, NGC 247 and NGC 300. They have much larger Scanpi flux densities at the same catalogue flux density level than other objects. In the left panel of Figure \ref{fig:scanpi}, we show the scan data of NGC 247, an extreme example of extended source. In such case, we use zero-point flux density. But it is necessary to manually adjust the background fitting region (especially the exclusion region) to make sure the background is not overestimated.

We estimate the flux density uncertainties based on background rms. It is sensitive to the background fitting region we choose, and the quality of background fitting. Usually smaller background fitting region gives a better background solution (by avoiding close neighbours and large scale fluctuations) and leads to smaller rms. However, the Scanpi data points are not totally independent of each other due to over-sampling. If the background fitting region is too small, the rms will be underestimated. We keep the background region larger than 12 arcmin to minimize such problem. However it is difficult to quantify the impact of background subtraction to the flux density uncertainty we get. We use flux densities with signal-to-noise ratio larger than 3 as detections. For detections, we add 10\% additional calibration uncertainty \citep{1984ApJ...278L...1N}. For non-detections, we give 3-$\sigma$ upper limits based on background rms.

There are 15 (at 60\mum) and 12 (at 100\mum) galaxies detected with Scanpi but without counterparts in PSC or FSC. We find that most of these objects show complicated background, implying that poor background subtraction is the main reason for PSC/FSC non-detection. The right panel of Figure \ref{fig:scanpi} shows the scan data of NGC 5237, as an example. The galaxy itself is relatively weak hence it is buried in the large scale fluctuation of the background when performing blind source finding. We also show the strong impact of background estimation to flux density measurement for such weak objects. The red and orange dashed lines are the estimated background when using fitting region radius of 15 arcmin and 30 arcmin, respectively. The 30 arcmin background does not well represent the local background at the position of galaxy, due to the complex background fluctuation, and will cause significant flux missing if we use zero-point flux density or centre flux density. Similar situation may occur even for a moderately bright object, if it is close to some bright cirrus filament or a bright galaxy. Our manual fine-tuning avoids such background problem.

\startlongtable
\begin{deluxetable}{cccrrcrrr}
\tablecaption{Flux measurements of LVHIS galaxies.\label{tab:flux}}
\tablehead{\multicolumn{1}{c}{LVHIS ID} & \multicolumn{1}{c}{$\mathrm{RMS_{1.4\,GHz}}$} & \multicolumn{1}{c}{$R_\mathrm{opt}$} & \multicolumn{1}{c}{$\mathrm{PA}_\mathrm{opt}$} & \multicolumn{1}{c}{$F_\mathrm{1.4\,GHz}$} & \multicolumn{1}{c}{Flag} & \multicolumn{1}{c}{$F_\mathrm{60\mum}$} & \multicolumn{1}{c}{$F_\mathrm{100\mum}$} & \multicolumn{1}{c}{$F_\mathrm{FIR}$}\\
 & \multicolumn{1}{c}{$\mathrm{mJy/beam}$} & \multicolumn{1}{c}{arcsec} & \multicolumn{1}{c}{degree} & \multicolumn{1}{c}{$\mathrm{mJy}$} &  & \multicolumn{1}{c}{$\mathrm{Jy}$} & \multicolumn{1}{c}{$\mathrm{Jy}$} & \multicolumn{1}{c}{$\mathrm{10^{-14}\,W\,m^{-2}}$}\\
\multicolumn{1}{c}{(1)} & \multicolumn{1}{c}{(2)} & \multicolumn{1}{c}{(3)} & \multicolumn{1}{c}{(4)} & \multicolumn{1}{c}{(5)} & \multicolumn{1}{c}{(6)} & \multicolumn{1}{c}{(7)} & \multicolumn{1}{c}{(8)} & \multicolumn{1}{c}{(9)} }
\startdata
LVHIS 001 & 0.041 & $24.8\times24.8$ & 0 & $<0.98$ & 3 & $<0.17$ & $<0.38$ & $<0.74$ \\
LVHIS 002 & \nodata & $25.6\times19.7$ & 6 & \nodata & 0 & $<0.16$ & $<0.46$ & $<0.78$ \\
LVHIS 003 & 0.046 & $34.2\times28.3$ & 54 & $<1.5$ & 3 & $<0.12$ & $<0.37$ & $<0.61$ \\
LVHIS 004 & \nodata & $879.0\times237.6$ & 108 & $381\pm19$ & 5\tablenotemark{a} & $66\pm7$ & $96\pm10$ & $334\pm25$ \\
LVHIS 005 & \nodata & $580.7\times387.1$ & 111 & \nodata & 0 & $17\pm2$ & $45\pm5$ & $113\pm8$ \\
LVHIS 006 & 0.976 & $899.5\times224.9$ & 52 & $5970\pm90$ & 5\tablenotemark{b} & $1031\pm103$ & $1240\pm124$ & $4912\pm370$ \\
LVHIS 007 & \nodata & $574.1\times172.4$ & 174 & \nodata & 0 & $10\pm1$ & $28\pm3$ & $69\pm5$ \\
LVHIS 008 & \nodata & $194.5\times59.8$ & 92 & $10\pm1$ & 5\tablenotemark{c} & $6.3\pm0.6$ & $9.0\pm0.9$ & $32\pm2$ \\
LVHIS 009 & 0.067 & $97.5\times93.8$ & 122 & $1.16\pm0.07$ & 2 & $0.22\pm0.04$ & $0.3\pm0.1$ & $1.2\pm0.2$ \\
LVHIS 010 & \nodata & $70.7\times63.6$ & 90 & \nodata & 0 & $<0.14$ & $<0.29$ & $<0.57$ \\
LVHIS 011 & 0.047 & $147.6\times24.3$ & 44 & $3.0\pm0.7$ & 1 & $0.11\pm0.03$ & $0.4\pm0.1$ & $0.8\pm0.2$ \\
LVHIS 012 & 0.072 & $233.9\times68.3$ & 38 & $1.23\pm0.08$ & 2 & $0.37\pm0.06$ & $0.9\pm0.2$ & $2.3\pm0.3$ \\
LVHIS 013 & 0.046 & $21.9\times8.7$ & 4 & $<0.60$ & 3 & $<0.083$ & $<0.28$ & $<0.45$ \\
LVHIS 014 & 0.063 & $326.5\times226.8$ & 39 & $391\pm7$ & 1 & $33\pm3$ & $63\pm6$ & $186\pm13$ \\
LVHIS 015 & 0.042 & $95.2\times27.2$ & 40 & $2.8\pm0.5$ & 1 & $0.53\pm0.06$ & $1.0\pm0.2$ & $2.9\pm0.3$ \\
LVHIS 016 & 0.038 & $22.6\times18.6$ & 121 & $<0.83$ & 3 & $<0.087$ & $<0.40$ & $<0.57$ \\
LVHIS 017 & 0.038 & $98.6\times20.8$ & 147 & $3.9\pm0.5$ & 1 & $0.55\pm0.07$ & $1.2\pm0.2$ & $3.3\pm0.3$ \\
LVHIS 018 & 0.038 & $53.6\times40.6$ & 28 & $9.9\pm0.6$ & 1 & $1.1\pm0.1$ & $1.8\pm0.2$ & $6.0\pm0.5$ \\
LVHIS 019 & 0.090 & $41.1\times10.8$ & 56 & $<1.9$ & 3 & $<0.15$ & $<0.40$ & $<0.70$ \\
LVHIS 020 & 0.129 & $66.7\times47.6$ & 52 & $<1.8$ & 3 & $0.10\pm0.03$ & $0.34\pm0.09$ & $0.8\pm0.2$ \\
LVHIS 021 & 0.059 & $40.6\times20.3$ & 92 & $0.61\pm0.07$ & 2 & $0.14\pm0.03$ & $<0.35$ & $<1.0$ \\
LVHIS 022 & 0.047 & $163.6\times38.1$ & 175 & $12\pm1$ & 1 & $3.4\pm0.3$ & $5.7\pm0.6$ & $18\pm1$ \\
LVHIS 023 & 0.063 & $23.6\times20.0$ & 49 & $<0.82$ & 3 & $<0.068$ & $<0.31$ & $<0.44$ \\
LVHIS 024 & 0.048 & $14.9\times14.9$ & 0 & $<0.82$ & 3 & $<0.084$ & $0.22\pm0.06$ & $<0.63$ \\
LVHIS 025 & 0.050 & $10.0\times6.6$ & 43 & $<0.52$ & 3 & $<0.11$ & $<0.35$ & $<0.57$ \\
LVHIS 026 & 0.046 & $55.5\times42.7$ & 83 & $0.70\pm0.04$ & 2 & $0.17\pm0.03$ & $<0.33$ & $<1.1$ \\
LVHIS 027 & 0.044 & $72.2\times36.1$ & 129 & $2.8\pm0.5$ & 1 & $1.4\pm0.1$ & $1.4\pm0.2$ & $6.3\pm0.5$ \\
LVHIS 028 & 0.059 & $23.1\times18.5$ & 129 & $1.5\pm0.3$ & 1 & $<0.16$ & $<0.41$ & $<0.73$ \\
LVHIS 029 & 0.054 & $63.7\times10.0$ & 75 & $<0.86$ & 3 & $0.10\pm0.03$ & $<0.46$ & $<0.99$ \\
LVHIS 030 & 0.036 & $18.4\times9.2$ & 72 & $<0.53$ & 3 & $0.15\pm0.05$ & $<2.3$ & $<3.5$ \\
LVHIS 031 & 0.101 & $291.0\times145.5$ & 159 & $197\pm4$ & 1 & $33\pm3$ & $78\pm8$ & $206\pm15$ \\
LVHIS 032 & 0.057 & \nodata & \nodata & $<0.17$ & 4 & $<0.14$ & $<0.73$ & $<1.0$ \\
LVHIS 033 & 0.102 & $19.5\times15.3$ & 137 & $<2.2$ & 3 & $<0.15$ & $<0.96$ & $<1.3$ \\
LVHIS 034 & 0.079 & $20.6\times16.9$ & 86 & $<1.3$ & 3 & $<0.27$ & $<1.4$ & $<2.0$ \\
LVHIS 035 & 0.061 & $15.4\times13.1$ & 90 & $<0.64$ & 3 & $<0.11$ & $<0.28$ & $<0.49$ \\
LVHIS 036 & 0.079 & $17.4\times13.0$ & 30 & $<1.1$ & 3 & $<0.15$ & $<0.68$ & $<0.98$ \\
LVHIS 037 & 0.051 & $41.1\times17.6$ & 20 & $<0.74$ & 3 & $<0.11$ & $<0.47$ & $<0.69$ \\
LVHIS 038 & \nodata & $70.7\times40.1$ & 45 & \nodata & 0 & $0.31\pm0.05$ & $1.3\pm0.3$ & $2.6\pm0.4$ \\
LVHIS 039 & 0.047 & $24.2\times14.2$ & 83 & $<0.88$ & 3 & $<0.17$ & $<0.60$ & $<0.93$ \\
LVHIS 040 & 0.047 & $71.4\times28.6$ & 138 & $<0.98$ & 3 & $<0.14$ & $<0.36$ & $<0.63$ \\
LVHIS 041 & \nodata & $15.4\times15.4$ & 29 & \nodata & 0 & $<0.22$ & $<0.89$ & $<1.3$ \\
LVHIS 042 & 0.049 & $31.4\times16.5$ & 94 & $<1.2$ & 3 & $<0.14$ & $0.5\pm0.1$ & $<1.3$ \\
LVHIS 043 & 0.492 & $698.2\times161.2$ & 43 & $6450\pm100$ & 5\tablenotemark{d} & $635\pm64$ & $1214\pm122$ & $3594\pm258$ \\
LVHIS 044 & 0.041 & $103.2\times60.7$ & 77 & $0.87\pm0.06$ & 2 & $0.60\pm0.06$ & $1.1\pm0.1$ & $3.3\pm0.3$ \\
LVHIS 045 & 0.044 & \nodata & \nodata & $<0.13$ & 4 & $<0.17$ & $<0.27$ & $<0.64$ \\
LVHIS 046 & 0.043 & $291.0\times121.3$ & 48 & $1.76\pm0.05$ & 2 & $1.3\pm0.1$ & $4.1\pm0.5$ & $9.4\pm0.7$ \\
LVHIS 047 & 0.058 & $31.2\times24.0$ & 118 & $<0.82$ & 3 & $<0.18$ & $<0.78$ & $<1.1$ \\
LVHIS 048 & \nodata & $920.4\times708.0$ & 32 & $4520\pm226$ & 5\tablenotemark{e} & $236\pm24$ & $396\pm40$ & $1265\pm91$ \\
LVHIS 049 & 0.064 & $43.5\times17.4$ & 158 & $<1.5$ & 3 & $<0.16$ & $<0.50$ & $<0.82$ \\
LVHIS 050 & 0.063 & $89.9\times68.1$ & 50 & $<4.5$ & 3 & $0.17\pm0.05$ & $0.7\pm0.1$ & $1.4\pm0.2$ \\
LVHIS 051 & \nodata & $342.0\times40.2$ & 118 & \nodata & 0 & $1.4\pm0.1$ & $3.3\pm0.4$ & $8.7\pm0.7$ \\
LVHIS 052 & 0.048 & $35.0\times14.4$ & 31 & $<0.93$ & 3 & $<0.16$ & $<0.37$ & $<0.70$ \\
LVHIS 053 & 0.985 & $461.3\times390.9$ & 45 & $2445\pm122$ & 5\tablenotemark{a} & $283\pm28$ & $475\pm48$ & $1518\pm110$ \\
LVHIS 054 & 0.057 & $6.7\times6.7$ & 11 & $<0.44$ & 3 & $<0.15$ & $<0.71$ & $<1.0$ \\
LVHIS 055 & 0.046 & $65.2\times50.1$ & 103 & $0.79\pm0.05$ & 2 & $0.39\pm0.06$ & $0.8\pm0.1$ & $2.3\pm0.3$ \\
LVHIS 056 & \nodata & $39.2\times24.5$ & 126 & \nodata & 0 & $<0.14$ & $<0.41$ & $<0.68$ \\
LVHIS 057 & 0.082 & $154.5\times66.6$ & 42 & $86\pm2$ & 1 & $35\pm3$ & $30\pm3$ & $152\pm12$ \\
LVHIS 058 & 0.058 & $18.4\times11.5$ & 52 & $<0.57$ & 3 & $0.14\pm0.04$ & $0.25\pm0.03$ & $0.8\pm0.1$ \\
LVHIS 059 & 0.141 & $102.1\times62.3$ & 65 & $<5.3$ & 3 & $0.33\pm0.07$ & $1.0\pm0.2$ & $2.3\pm0.3$ \\
LVHIS 060 & 0.066 & $81.1\times40.5$ & 128 & $6\pm1$ & 1 & $0.73\pm0.09$ & $2.0\pm0.4$ & $4.9\pm0.5$ \\
LVHIS 061 & 0.112 & $9.9\times7.4$ & 39 & $<1.1$ & 3 & $<0.25$ & $<0.95$ & $<1.5$ \\
LVHIS 062 & 0.047 & $55.9\times19.1$ & 2 & $<0.94$ & 3 & $0.21\pm0.04$ & $<1.3$ & $<2.4$ \\
LVHIS 063 & 0.063 & $171.4\times137.1$ & 93 & $3.1\pm0.2$ & 2 & $1.6\pm0.2$ & $4.0\pm0.4$ & $10.3\pm0.8$ \\
LVHIS 064 & 0.039 & $8.1\times6.5$ & 20 & $<0.31$ & 3 & $<0.12$ & $<0.71$ & $<0.97$ \\
LVHIS 065 & 0.072 & $83.9\times51.5$ & 62 & $40\pm1$ & 1 & $3.8\pm0.4$ & $3.3\pm0.3$ & $16\pm1$ \\
LVHIS 066 & \nodata & $259.4\times129.7$ & 40 & $1500\pm75$ & 5\tablenotemark{f} & $338\pm34$ & $364\pm37$ & $1556\pm119$ \\
LVHIS 067 & 0.041 & $29.9\times15.0$ & 122 & $<0.87$ & 3 & $<0.17$ & $<0.74$ & $<1.1$ \\
LVHIS 068 & 0.085 & $11.7\times11.7$ & 18 & $<0.99$ & 3 & $<0.22$ & $<0.88$ & $<1.3$ \\
LVHIS 069 & 0.608 & $9.2\times9.2$ & 7 & $<7.8$ & 3 & $<1.8$ & $<3.3$ & $<7.3$ \\
LVHIS 070 & 0.043 & $48.3\times29.2$ & 65 & $<0.88$ & 3 & $<0.17$ & $<0.69$ & $<1.0$ \\
LVHIS 071 & 0.038 & $75.3\times67.8$ & 134 & $4.7\pm0.8$ & 1 & $1.8\pm0.2$ & $6.3\pm0.7$ & $14\pm1$ \\
LVHIS 072 & 0.061 & $294.4\times49.1$ & 38 & $8\pm1$ & 1 & $1.4\pm0.2$ & $<4.4$ & $<11$ \\
LVHIS 073 & 0.088 & \nodata & \nodata & $<0.27$ & 4 & $<0.71$ & $<3.3$ & $<4.8$ \\
LVHIS 074 & 0.045 & $117.2\times43.9$ & 30 & $3.9\pm0.6$ & 1 & $2.5\pm0.3$ & $4.5\pm0.6$ & $14\pm1$ \\
LVHIS 075 & 0.048 & $65.2\times34.7$ & 126 & $29.2\pm0.8$ & 1 & $9.6\pm1.0$ & $11\pm1$ & $45\pm3$ \\
LVHIS 076 & 0.046 & $16.4\times8.2$ & 22 & $<0.43$ & 3 & $<0.15$ & $<0.37$ & $<0.68$ \\
LVHIS 077 & 0.044 & $213.3\times39.4$ & 143 & $8.7\pm0.6$ & 1 & $3.2\pm0.3$ & $6.1\pm0.6$ & $18\pm1$ \\
LVHIS 078 & 0.073 & $147.6\times98.4$ & 100 & $8\pm2$ & 1 & $3.9\pm0.4$ & $7.6\pm0.8$ & $22\pm2$ \\
LVHIS 079 & 0.038 & $50.6\times44.0$ & 138 & $<1.5$ & 3 & $<0.18$ & $<0.53$ & $<0.90$ \\
LVHIS 080 & 0.057 & $113.2\times22.6$ & 48 & $<1.4$ & 3 & \nodata & \nodata & \nodata \\
LVHIS 081 & 0.043 & $63.0\times10.5$ & 148 & $<0.71$ & 3 & $0.22\pm0.04$ & $0.4\pm0.1$ & $1.3\pm0.2$ \\
LVHIS 082 & \nodata & $311.9\times178.2$ & 98 & $103\pm5$ & 5\tablenotemark{a} & $19\pm2$ & $55\pm5$ & $132\pm9$ \\
\enddata
\tablecomments{Columns: (1): LVHIS ID; (2) rms of 1.4 GHz image; (3) size of the ESO-Uppsala/WISE aperture; (4) position angle of the aperture; (5) 1.4 GHz flux density; (6) source flag of 1.4 GHz data; (7) 60\mum\ flux density; (8) 100\mum\ flux density; (9) far-IR flux density.\\Flag meanings: 0 -- no broad band data; 1 -- aperture flux density; 2 -- \textsc{sfind} flux density; 3 -- aperture upper limit; 4 -- 3-$\sigma$ upper limit (assuming point source); 5 -- literature. References are labeled as below.\\A machine readable version of this table is available as online supplementary data.}
\tablenotetext{a}{\citealt{1996ApJS..103...81C}}
\tablenotetext{b}{\citealt{2010ApJ...710.1462W}}
\tablenotetext{c}{\citealt{2004ApJ...610..772C}}
\tablenotetext{d}{\citealt{1997MNRAS.284..830E}}
\tablenotetext{e}{\citealt{2003PASJ...55..351T}}
\tablenotetext{f}{\citealt{1990PKS...C......0W}}
\end{deluxetable}

\section{FIR and radio continuum emission in local volume galaxies}
\label{sec:emission}

\subsection{FIR-radio correlation}
\label{subsec:frc}

As in \citet{2001ApJ...554..803Y}, we define the FIR-to-radio ratio $q$ as:

\begin{equation}
q \equiv \log(\frac{F_\mathrm{FIR}}{3.75\times10^{12}\wmm}) - \log(\frac{F_\mathrm{1.4\,GHz}}{10^{26}\jy}).
\end{equation}

Here the total FIR flux density is calculated from 60\mum\ and 100\mum\ flux densities as below \citep{1988ApJS...68..151H}:

\begin{equation}
  F_{\mathrm{FIR}} = 1.26 \times 10^{-14} \times (2.58 \times \frac{f_\nu(60\mum)}{1\jy} + \frac{f_\nu(100\mum)}{1\jy})\wmm.
\end{equation}

\begin{figure*}
  \centering \includegraphics[width=16cm]{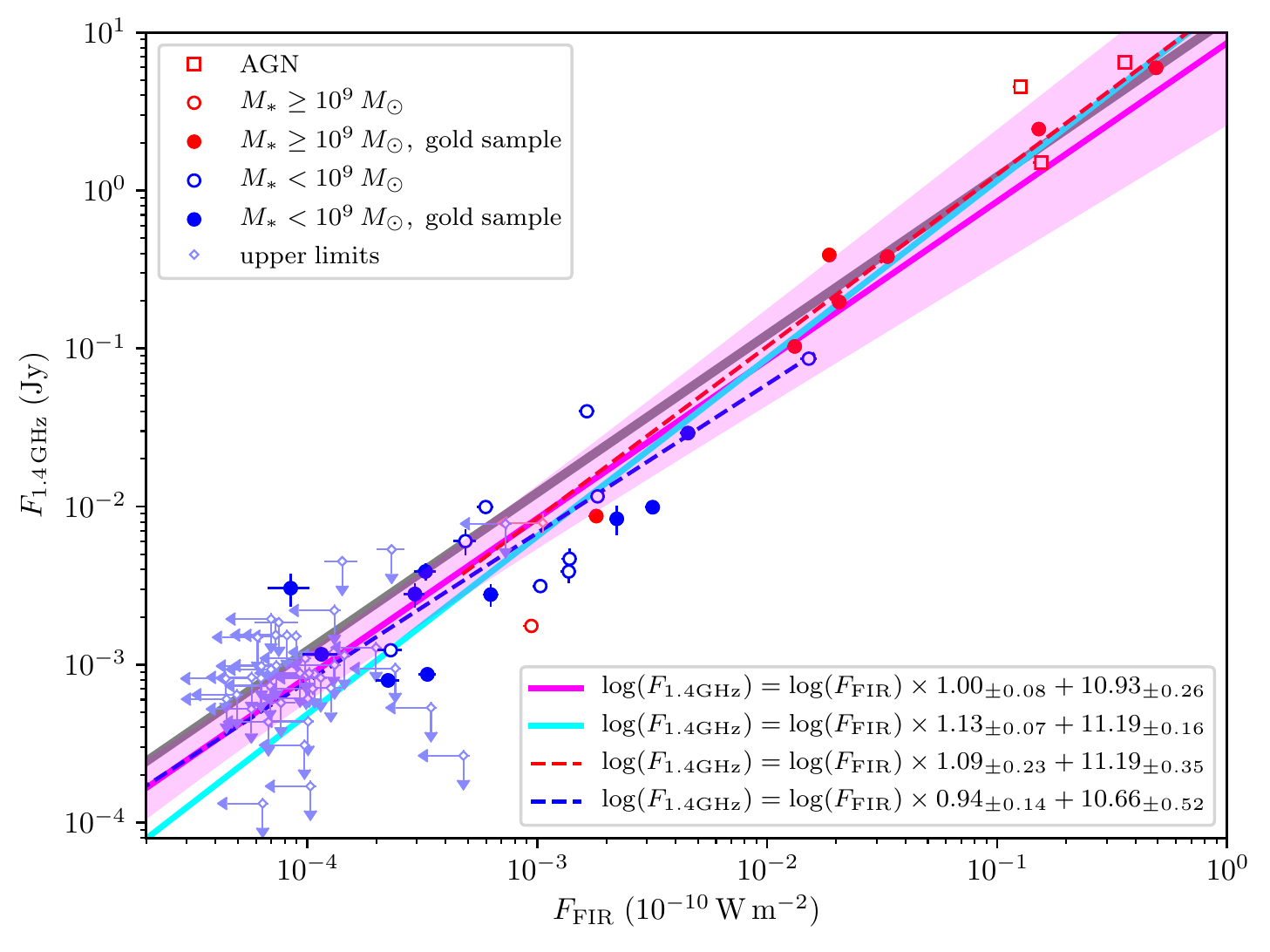}
  \caption{The FIR-radio correlation for LVHIS galaxies. The red and blue data points are LVHIS ``massive'' ($M_*\geq10^9\msun$) and ``dwarf'' ($M_*<10^9\msun$) galaxies, respectively. The square points are (potential) AGN host galaxies. The solid symbols are gold sample galaxies, described in Section \ref{subsec:corr_method}. Error bars are displayed for each object unless smaller than the symbol. The magenta and cyan lines are power law model fits to full sample (excluding AGN), with and without censored data, respectively. 95\% confidence band is colour-shaded for the magenta line. The red and blue dashed lines are fits to massive and dwarf galaxies, with censored data, respectively. The solid grey line indicates $q=2.34$ \citep{2001ApJ...554..803Y}}.
  \label{fig:frc}
\end{figure*}

Figure \ref{fig:frc} shows the distribution of LVHIS galaxies on the FIR-radio plane. There are totally 30 galaxies detected in both wavelengths (3 of them are potentially contaminated by nuclear emission). We confirm strong correlation between FIR and radio continuum flux densities, for full sample, massive galaxies and dwarf galaxies, respectively, by using the Spearman rank correlation coefficient (see Table \ref{tab:stats}).

The data are fit with power law model ($\log(y)=a\log(x)+b$) by using orthogonal distance regression method in log-log space (throughout this paper, ``linear'' refers $a=1$). In order to properly estimate the uncertainty of the fitting results, bootstrapping method is used. 10000 random runs are performed to resample the galaxies. The flux density of each simulated source is the observed value added by a zero-centred and normally distributed random noise with flux density uncertainty as the standard deviation. For non-detections, the observed flux density is assumed to be zero. As the data fitting is done in log-log space, all simulated flux densities with negative value are sign-flipped (it does not change the flux density probability distribution for non-detections and only has negligible impact on detected objects as the detection threshold is 3-$\sigma$). We find the index of best-fit power law model for the full sample including all upper limits (magenta line) is consistent with unity. The scatter of detected objects around the best-fit model is about 0.34 dex (y axis direction), confirming a good correlation over four orders of magnitude. We note that we get similar result by using the Schmitt's binning method \citep{1985ApJ...293..178S}.

We find no significant difference between massive and dwarf galaxies (red and blue lines in Figure \ref{fig:frc}). They are both consistent with the full sample relation. Yet the result may be slightly different if the upper limits are excluded from analysis. We notice that the index of best-fit power law model for detection-only (cyan line in Figure \ref{fig:frc}) is slightly larger than one with $\sim2$-$\sigma$ significance. We compare the mean or median $q$ of detected massive and dwarf galaxies, finding marginal difference (Table \ref{tab:stats}). However, a two-sample Kolmogorov-–Smirnov (K--S) test to the $q$ distributions of massive and dwarf galaxies gives the statistics of 0.38 with $p=0.30$, not sufficient to rule out the null hypothesis.

\begin{figure}
  \centering \includegraphics[width=8cm]{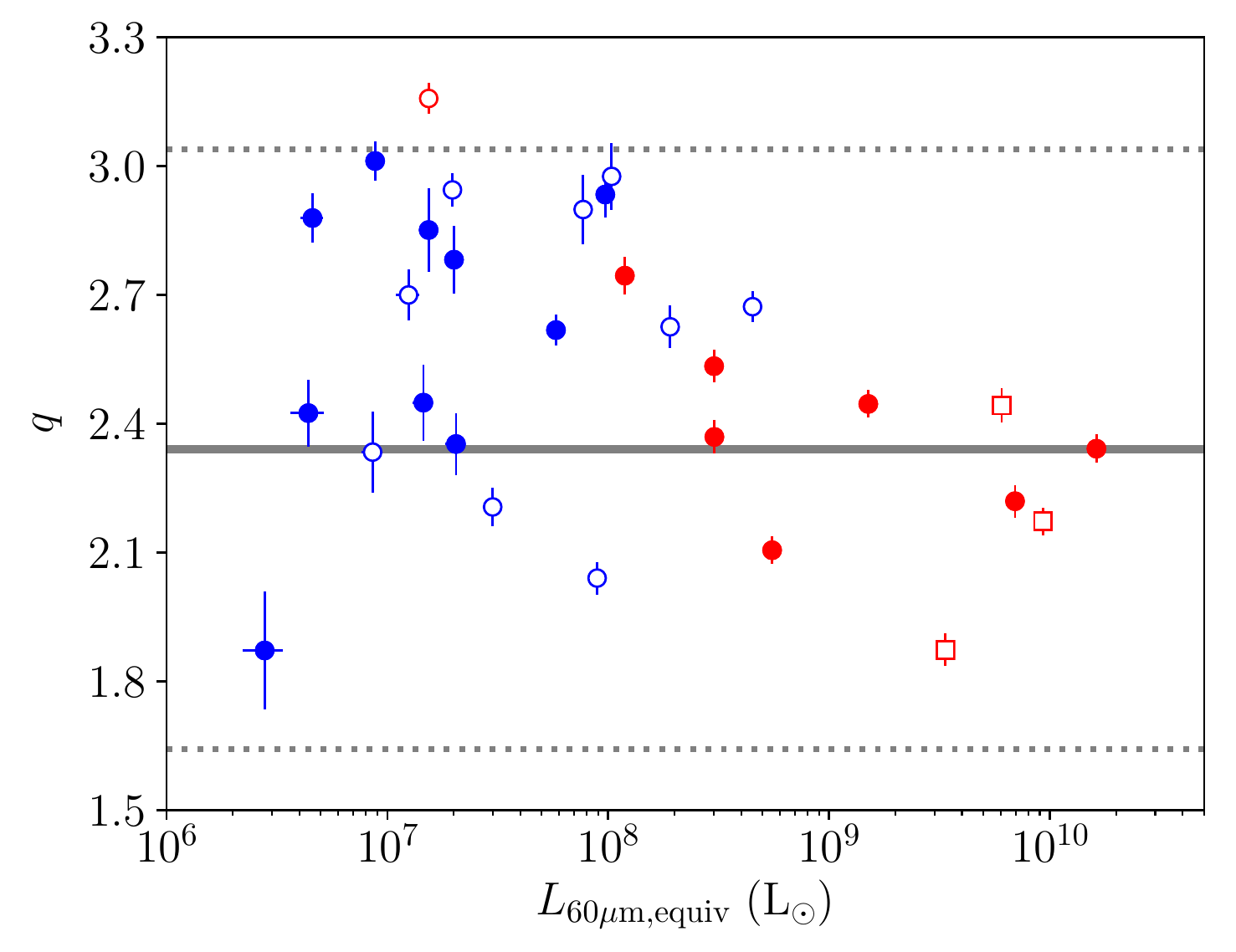}
  \caption{The $q$ value as a function of equivalent 60\mum\ luminosity $L_\mathrm{60\mu m,equiv}$ for detections. The symbol scheme is the same as Figure \ref{fig:frc}. The grey lines are the same as in Figure 6 of \citet{2001ApJ...554..803Y}: the solid line is $q=2.34$ and the dotted lines show the value corresponding to 5 times larger radio and FIR flux density than expected.}
  \label{fig:lir_q}
\end{figure}

Figure \ref{fig:lir_q} shows the $q$ as a function of equivalent 60\mum\ luminosity, which is the luminosity contribution from IRAS 60\mum\ band to the total FIR luminosity, defined in Equation 3 of \citet{2001ApJ...554..803Y}. It provides a better view of the comparison between massive and dwarf galaxies, for detections. It is clear that the low luminosity systems, mostly dwarf galaxies, have larger $q$. If we split the detection sample at $L_\mathrm{60\mu m,equiv}=3\times10^8\,\mathrm{L_\odot}$, we find that luminous and faint galaxies have different $q$ ($2.38\pm0.07$ versus $2.64\pm0.08$; two-sample K-S test statistics 0.55 with $p=0.05$). The lack of significant $q$ difference between massive and dwarf galaxies, as reported above, is probably caused by the presence of an outlier with exceptionally large $q$ (above the top dotted line) and low FIR luminosity in the massive galaxy sample\footnote{This specific object, NGC 5102, is catalogued with the \textsc{sfind} flux density, which may miss the extended emission, causing a unreal small radio flux density. Its aperture flux density is indeed about 3 times larger. Unfortunately, the optical aperture for this object is too large, probably including too many empty pixels, causing an overestimation of flux uncertainty. The signal-to-noise ratio of its aperture flux density is slightly smaller than 3, failed to pass the detection threshold. We note, however, it is not the typical situation for our galaxies (see Figure \ref{fig:cont_1}).}.

Our flux densities are compared with the empirical relation (grey solid line in Figure \ref{fig:frc} and \ref{fig:lir_q}, $q=2.34$) established with the IRAS and NVSS catalogues \citep{2001ApJ...554..803Y}. The $q=2.34$ line is slightly offset with our best-fit power law model for full sample (with upper limits) by $\sim0.15$ dex. It still lies within the 95\% confidence band, but only marginally at the faint end ($F_\mathrm{FIR}\lesssim3\times10^{-14}\mathrm{W\,m^{-2}}$). The massive galaxies follow the empirical relation better than dwarf galaxies. For detections, the best-fit power law model implies the deviation from the constant $q$ at faint end. The median or average $q$ of dwarf galaxies is found to be statistically different from the empirical value (see Table \ref{tab:stats}). By directly comparing our Figure \ref{fig:lir_q} with Figure 6 of \citet{2001ApJ...554..803Y}, we find they are highly consistent, in sense of clustering of low-luminosity galaxies ($L_\mathrm{60\mu m,equiv}<10^9\,\mathrm{L_\odot}$) above the $q=2.34$ line. But our result extends to much lower luminosity regime ($>10^6\,\mathrm{L_\odot}$ versus $>10^8\,\mathrm{L_\odot}$).

The slightly different results from using or not using upper limits in the analysis can be understood as our overall detection rate is lower than 50\%. Therefore we emphasize that the faint end deviation from empirical relation with high statistical significance level is only observed in the sample of detected objects, which may potentially be biased in a subtle way. The discrepancy between the empirical relation and the full sample result (i.e. with censored data) is only marginal at the faint end. We are not able to make solid conclusion whether the empirical FRC is already broken or not.

In summary, LVHIS galaxies are consistent with a tight linear FRC, though a marginal offset from the empirical relation \citep[$q=2.34$,][]{2001ApJ...554..803Y} is observed at faint end. Our detections suggest a slightly super-linear FIR-radio relation, consistent with measured larger $q$ for faint galaxies (mostly dwarf galaxies). Our detected dwarf galaxies clearly deviate from the empirical relation.

\begin{longrotatetable}
\begin{deluxetable*}{l|ccc|ccc|ccc}
\tablewidth{0.85\linewidth}
\tablecaption{Statistical properties of detections.\label{tab:stats}}
\tablehead{Sample & \multicolumn{2}{c}{$q$} & $r_s(F_\mathrm{FIR}, F_\mathrm{1.4GHz})$ & \multicolumn{2}{c}{$L_\mathrm{FIR}/\mathrm{SFR}$} & $r_s(F_\mathrm{FIR}, \mathrm{SFR}/D^2)$ & \multicolumn{2}{c}{$L_\mathrm{1.4GHz}/\mathrm{SFR}$} & $r_s(F_\mathrm{1.4GHz}, \mathrm{SFR}/D^2)$ \\
 & & & & \multicolumn{2}{c}{$10^9\,\mathrm{L_\odot\,M_\odot^{-1}\,yr}$} & & \multicolumn{2}{c}{$10^{-6}\,\mathrm{L_\odot\,Hz^{-1}\,M_\odot^{-1}\,yr}$} & \\
 & mean & median & & mean & median & & mean & median & }
\startdata
full & $2.57_{\pm0.06}$ & $2.62_{\pm0.11}$ & 0.90 ($2\!\times\!10^{-10}$) & $2.29_{\pm0.36}$ & $1.82_{\pm0.21}$ & 0.90 ($2\!\times\!10^{-10}$) & $2.15_{\pm0.41}$ & $1.69_{\pm0.28}$ & 0.92 ($9\!\times\!10^{-9}$) \\
massive & $2.49_{\pm0.12}$ & $2.41_{\pm0.11}$ & 0.93 ($8\!\times\!10^{-4}$) & $3.05_{\pm0.48}$ & $2.50_{\pm0.63}$ & 0.85 ($3\!\times\!10^{-3}$) & $4.01_{\pm0.76}$ & $3.88_{\pm1.13}$ & 0.93 ($2\!\times\!10^{-3}$) \\
dwarf & $2.61_{\pm0.07}$ & $2.67_{\pm0.12}$ & 0.81 ($2\!\times\!10^{-5}$) & $1.89_{\pm0.44}$ & $1.67_{\pm0.21}$ & 0.70 ($1\!\times\!10^{-3}$) & $1.15_{\pm0.17}$ & $1.03_{\pm0.31}$ & 0.79 ($1\!\times\!10^{-3}$) \\
empirical & \multicolumn{2}{c}{2.34} & & \multicolumn{2}{c}{2.7} & & \multicolumn{2}{c}{3.3} & \\
\enddata
\tablecomments{The uncertainty is estimated with bootstrap method. $r_s$ is the Spearman rank correlation coefficient. The probability $p$ for null hypothesis is listed in bracket.}
\end{deluxetable*}
\end{longrotatetable}

\subsection{FIR/radio flux densities and star formation rate}
\label{subsec:sfr}

It is known that both FIR and radio luminosities are correlated with star formation rates for massive non-AGN galaxies \citep[see e.g.][]{1998ARA&A..36..189K,2002AJ....124..675C}. And indeed, it is the basis for FRC studies on galactic scale. For LVHIS galaxies, we calculate the star formation rate from the GALEX FUV flux density, calibrated with stellar synthesis models \citep{2013seg..book..419C}, and WISE mid-IR flux density, calibrated with infrared bright galaxies \citep{2013AJ....145....6J}\footnote{If the galaxy is not detected in mid-IR, we assume it is mainly caused by low dust obscuration, therefore the UV flux density alone is sufficient to give reasonable star formation rate estimation. In dwarf galaxies, low dust obscuration is common.}. By adding up the obscured (mid-IR) and unobscured (FUV) SFR, we get the attenuation-corrected SFR \citep[Equation \ref{eqn:sfr}, see Section 3.3.2 of][for more details]{2017MNRAS.472.3029W}. It provides reliable SFR estimation independent of FIR or radio calibrations. Throughout the paper, we use Kroupa initial mass function \citep{2001MNRAS.322..231K}.

\begin{equation}
\label{eqn:sfr}
(\mathrm{SFR/1\,M_\odot\,yr^{-1}}) = 1.76\times10^{-10} (L_\mathrm{FUV}/\mathrm{L_\odot}) + 7.50\times10^{-10} (L_\mathrm{22\mum}/\mathrm{L_\odot}).
\end{equation}

\begin{figure*}
  \centering
  \includegraphics[width=8cm]{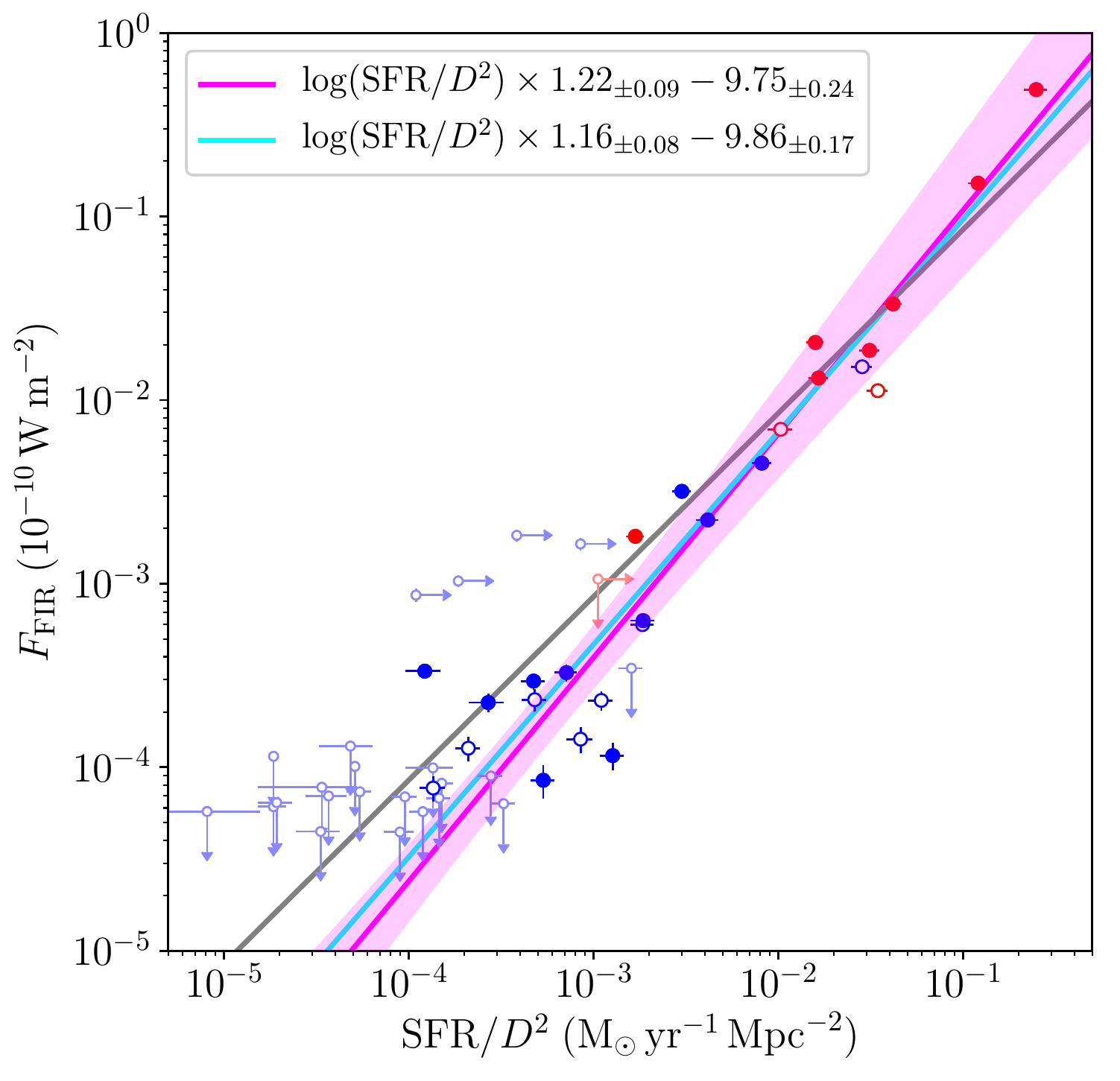} \includegraphics[width=8cm]{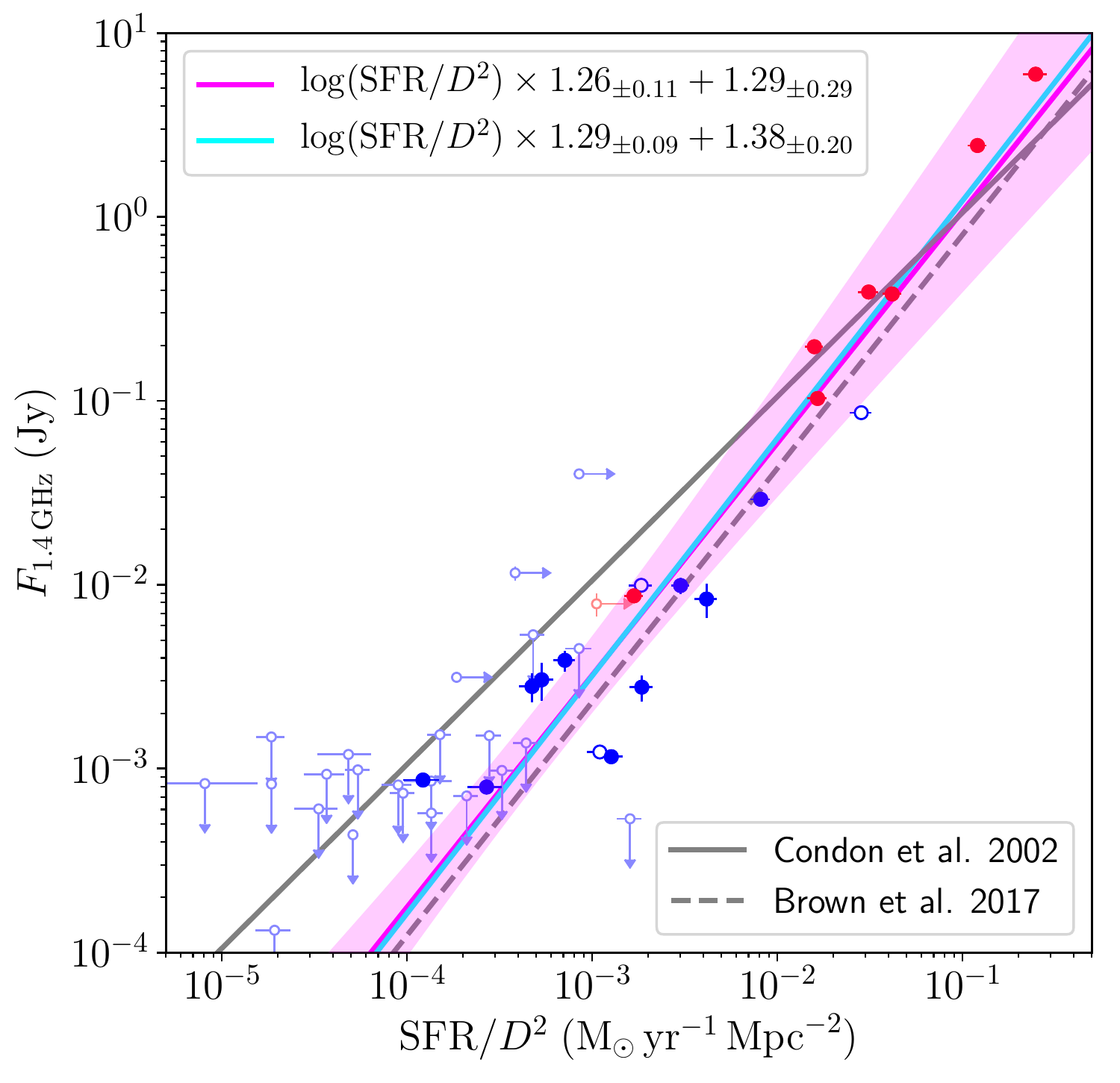}
  \caption{The correlation between FIR/radio flux densities and distance-normalized star formation rates for non-AGN LVHIS galaxies. The symbol and line scheme is the same as Figure \ref{fig:frc}. The grey solid line is the empirical linear SFR calibration (Equation \ref{eqn:sfr_fir} for left panel and Equation \ref{eqn:sfr_1.4_condon} for right panel). The grey dashed line in right panel is a super-linear correlation adopted from literature (Equation \ref{eqn:sfr_1.4_brown}).}
  \label{fig:sfr}
\end{figure*}

We confirm the strong correlation between the distance-normalized star formation rates and the FIR or radio flux densities for LVHIS galaxies, with large Spearman rank correlation coefficient (see Table \ref{tab:stats}). As shown in Figure \ref{fig:sfr}, again in log-log space we use the power law model to fit the full sample, with (magenta line) and without upper limits (cyan line). The fitting results suggest that the flux densities do not increase linearly with the distance normalized SFR.

For convenience, here we define the luminosity-to-SFR ratio ($\mathrm{eff}_\mathrm{FIR} \equiv L_\mathrm{FIR}/\mathrm{SFR}$; $\mathrm{eff}_\mathrm{1.4\,GHz} \equiv L_\mathrm{1.4\,GHz}/\mathrm{SFR}$) as ``radiation efficiency'' (FIR efficiency and radio efficiency, respectively). The physical motivation of this will be mentioned in Section \ref{subsec:radi_eff}. The super-linear best fit model implies a lower radiation efficiency in low SFR regime. For detections, we calculate the median or mean value of the radiation efficiency in each sample (listed in Table \ref{tab:stats}). It is indeed that the dwarf galaxies have lower radiation efficiency than massive galaxies, for both radio continuum and FIR. Two-sample K-–S tests give statistics of 0.65 ($p=0.005$) for FIR and 0.69 ($p=0.006$) for radio continuum, suggesting a significant difference between massive and dwarf galaxies.

For FIR, we compare our data with the FIR star formation rate calibration based on starburst models \citep{1998ARA&A..36..189K}, but assuming only half of the radiation power is dust absorbed to better reflect the condition in normal galaxies \citep[see][]{2006A&A...451..417D,2013seg..book..419C}, and the total IR to IRAS FIR ratio as 1.75 \citep{2000ApJ...533..682C}.

\begin{equation}
\label{eqn:sfr_fir}
(L_\mathrm{FIR}/1\,\mathrm{W}) \approx 1.0\times10^{36}(\mathrm{SFR}/1\,\mathrm{M_\odot\,yr^{-1}}).
\end{equation}

For radio continuum, we compare with empirical linear star formation rate calibration at 1.4 GHz \citep[Equation 27 in][rescaled to Kroupa initial mass function]{2002AJ....124..675C}:

\begin{equation}
\label{eqn:sfr_1.4_condon}
(L_\mathrm{1.4GHz}/1\,\mathrm{W\,Hz^{-1}}) \approx 1.3\times10^{21}(\mathrm{SFR}/1\,\mathrm{M_\odot\,yr^{-1}}).
\end{equation}

We note that this calibration is based on total 1.4 GHz luminosity (thermal plus non-thermal). By combining Equation \ref{eqn:sfr_fir} and \ref{eqn:sfr_1.4_condon}, we get $q\approx2.33$, highly consistent with empirical value. We find that the radiation efficiency of massive galaxies is consistent with the expectation from empirical linear star formation rate calibration, both for radio and FIR (Table \ref{tab:stats}). But dwarf galaxies have clearly smaller radiation efficiency.

We also compare with a recent non-linear calibration on 1.4 GHz luminosity \citep{2017ApJ...847..136B}, by converting \ha\ luminosity to SFR using empirical calibration \citep{2013seg..book..419C}:

\begin{equation}
\label{eqn:sfr_1.4_brown}
(L_\mathrm{1.4GHz}/1\,\mathrm{W\,Hz^{-1}}) \approx 4.5\times10^{19}(\mathrm{SFR}/0.055\,\mathrm{M_\odot\,yr^{-1}})^{1.27}.
\end{equation}

It is clear that our power law fitting result for radio continuum is much more consistent with literature non-linear calibration \citep{2017ApJ...847..136B}, especially the index value (our 1.26 versus their 1.27). The scatter around our best fit model is around 0.2 dex (orthogonal distance), equivalent to $\sim0.3$ dex on y-axis direction, larger than their scatter 0.2 dex though.

As both FIR and radio continuum show similar trend of radiation deficiency at low SFR end, the effects will be partially, if not completely, cancelled out when calculating $q$, causing a tighter FRC. We note that the indices of best-fit power law models for FIR and radio are similar, especially when including censored data. This is consistent with the observed linear FRC, and with the so-called ``conspiracy'' to keep FRC tight \citep{2003ApJ...586..794B}.

\section{FIR-radio correlation and galaxy properties}
\label{sec:galprop}

In previous section, we find that radiation deficiency happens at low SFR end for both FIR and radio. Here we will study the physical drive of such behaviour to ultimately improve our knowledge about the nature of FRC.

\subsection{Understanding radiation efficiency}
\label{subsec:radi_eff}

As mentioned in Section \ref{sec:intro}, both FIR and radio continuum emission essentially come from massive stars in star forming galaxies. The energy of the FIR and thermal radio emission is converted from UV photons, which are emitted by massive stars and absorbed by interstellar medium\footnote{Old stars can also heat up interstellar medium. But they are much less efficient, leading to lower dust temperature and little bremsstrahlung photons. It is negligible in star forming galaxies, such as our galaxies, and in shorter wavelength FIR bands, such as our IRAS 60 and 100\mum\ bands.}. The energy of the non-thermal radio emission is mainly converted from cosmic rays, which are accelerated by supernova remnants and propagated within the galaxy. If the physical condition does not change, one would expect the output luminosity simply scales up linearly with the number of young stars, which is effectively SFR by assuming a fixed initial mass function and a certain timescale \citep[see, for example, Section 6.2 of][for normal galaxies]{1992ARA&A..30..575C}. Therefore the radiation efficiency is an indicator of the capability of converting injected energy (in form of UV photon and cosmic ray) to FIR or radio photon energy. Higher radiation efficiency means larger fraction of injected energy converted into observed emission. Here we emphasize, however, the radiation efficiency itself is a model-free observational quantity without any assumption of detailed radiation processes.

The FIR efficiency is likely an indicator of the dust opacity. Higher dust opacity leads to more absorbed UV photons and more energy converted into dust thermal energy. The observed drop of FIR efficiency in dwarf galaxies is indeed a clear sign of decreased dust opacity in these smaller systems. It is consistent with previous studies. It was found that the UV-to-FIR flux ratio of galaxy decreases with FIR or bolometric luminosities \citep[see e.g.][]{1996ApJ...457..645W,2005ApJ...619L..51B} and star formation rate \citep[see e.g.][]{2006ApJ...643..173S,2010ApJ...714.1256C}. Linear calibration, such as the popular one provided by \citet{1998ARA&A..36..189K}, can only be applied to highly obscured starburst galaxies, as stated in the paper (the grey solid line in the left panel of our Figure \ref{fig:sfr} already assumes 50\% UV photon leakage). It was demonstrated by \citet{2010ApJ...714.1256C} that the spatially resolved FIR-SFR relation deviated from linear relation at $\lesssim10^{-2}\,\mathrm{M_\odot\,yr^{-1}\,kpc^{-2}}$. Most of our dwarf galaxies have SFR surface density lower than this value. Therefore a super-linear correlation between FIR luminosity and SFR for our sample is not surprising.

The observed radio deficiency in dwarf galaxies is consistent with previous studies \citep[see, e.g.][]{2003ApJ...586..794B,2014AJ....147..103H,2015A&A...579A.102B,2017ApJ...847..136B}. But it is more difficult to understand, since there are two major components with significantly different physical mechanism. The thermal component is probably non-negligible in dwarf galaxies at 1.4 GHz \citep[e.g.][]{2006MNRAS.370..363H,2017MNRAS.471..337B}. However, the thermal component is probably linearly correlated with SFR \citep{2011ApJ...737...67M}, on different scales \citep{2013A&A...557A.129T}, due to its nature of bremsstrahlung emission directly from the ionized gas in star forming regions \citep{1992ARA&A..30..575C}. As we do not have proper uniform multi-band radio data to perform spectral decomposition \citep[see e.g.][]{2011ApJ...737...67M}, based on studies mentioned above, here we assume that the radio thermal component luminosity is proportional to star formation rate ($L_\mathrm{1.4\,GHz}^\mathrm{th} = c_\mathrm{th}\times\mathrm{SFR}$, where $c_\mathrm{th}$ is a constant). Then the radio efficiency can be written as:

\begin{equation}
\mathrm{eff}_\mathrm{1.4\,GHz} = (L_\mathrm{1.4\,GHz}^\mathrm{sync} + L_\mathrm{1.4\,GHz}^\mathrm{th}) / \mathrm{SFR} = L_\mathrm{1.4\,GHz}^\mathrm{sync} / \mathrm{SFR} + c_\mathrm{th}.
\end{equation}

Here $L_\mathrm{1.4\,GHz}^\mathrm{sync}$ is the non-thermal synchrotron component of the radio continuum. It is clear that when we use the quantity $\mathrm{eff}_\mathrm{1.4\,GHz}$, the effect of the thermal component is represented by a constant. Therefore the observed radio deficiency is likely caused by a non-thermal radiation deficiency. Many factors can cause such deficiency. We will discuss the most possible explanation based on further analysis below.

\subsection{Correlation analysis}
\label{subsec:corr_method}

To understand the underlying physics, we need to know what drives such deficiency behaviour. Instead of direct comparison with somehow degenerated theoretical models, we focus on the observables themselves. Here we use the correlation coefficient between galaxy property and radiation efficient, positive for correlation and negative for anti-correlation, to indicate the relevance of the specific property. A correlation coefficient with large absolute value implies that the specific galaxy property is an important factor, probably responsible, for the radiation deficiency.

To avoid non-detections in this analysis, here we define a sub-sample of galaxies, named as ``gold sample'', with the following criterion: 1) detected in both FIR and radio bands; 2) with stellar mass measurement; 3) with GALEX \citep[Galaxy Evolution Explorer,][]{2005ApJ...619L...1M} detection for reliable star formation rate estimation; 4) with ESO Uppsala B band diameter measurement; 5) not AGN. We will perform more detailed analysis on the gold sample below. There are totally 17 galaxies in this sample. The gold sample galaxies are shown in Figure \ref{fig:frc} as solid symbols. They are evenly populated by massive (7 galaxies) and dwarf galaxies (10 galaxies). Some important galaxy properties are listed in Table \ref{tab:prop}. We note that both calibration/system and statistical errors are included. The calibration uncertainties for stellar mass and star formation rate are assumed to be 25\% and 15\%, respectively \citep{2013seg..book..419C}. The uncertainty of \hi\ mass is around 15\% \citep{2017MNRAS.472.3029W}. Conservatively, the uncertainty of size measurement is assumed to be around 10\% \citep{1987A&A...184...86P}. Readers are referred to \citet{2017MNRAS.472.3029W} for more details about the these measurements.

\startlongtable
\begin{deluxetable}{lcrrrrrrrr}
\tablewidth{0.85\linewidth}
\tablecaption{Derived galaxy properties of gold sample.\label{tab:prop}}
\tablehead{\multicolumn{1}{c}{LVHIS ID} & \multicolumn{1}{c}{Optical name} & \multicolumn{1}{c}{$\log(M_*)$} & \multicolumn{1}{c}{$\log(\mathrm{SFR})$} & \multicolumn{1}{c}{$\log(M_\mathrm{HI})$} & \multicolumn{1}{c}{cons\tablenotemark{a}} & \multicolumn{1}{c}{$D25$} & \multicolumn{1}{c}{$R_\mathrm{UV}$} & \multicolumn{1}{c}{$\log(\mu_*)$} & \multicolumn{1}{c}{$\log(\mu_\mathrm{SFR})$}\\
 & & \multicolumn{1}{c}{[$\mathrm{M_\odot}$]} & \multicolumn{1}{c}{[$\mathrm{M_\odot\,yr^{-1}}$]} & \multicolumn{1}{c}{[$\mathrm{M_\odot}$]} &  & \multicolumn{1}{c}{kpc} & \multicolumn{1}{c}{kpc} & \multicolumn{1}{c}{[$\mathrm{M_\odot\,kpc^{-2}}$]} & \multicolumn{1}{c}{[$\mathrm{M_\odot\,yr^{-1}\,kpc^{-2}}$]}\\
\multicolumn{1}{c}{(1)} & \multicolumn{1}{c}{(2)} & \multicolumn{1}{c}{(3)} & \multicolumn{1}{c}{(4)} & \multicolumn{1}{c}{(5)} & \multicolumn{1}{c}{(6)} & \multicolumn{1}{c}{(7)} & \multicolumn{1}{c}{(8)} & \multicolumn{1}{c}{(9)} & \multicolumn{1}{c}{(10)} }
\startdata
LVHIS 004 & NGC 55 & $9.36_{\pm0.10}$ & $-0.72_{\pm0.05}$ & $9.34_{\pm0.07}$ & $3.6_{\pm0.5}$ & $18.2_{\pm1.8}$ & $1.16_{\pm0.12}$ & $2.01_{\pm0.12}$ & $-0.94_{\pm0.08}$ \\
LVHIS 006 & NGC 253 & $10.43_{\pm0.10}$ & $0.59_{\pm0.06}$ & $9.43_{\pm0.07}$ & $3.2_{\pm0.4}$ & $34.4_{\pm3.4}$ & $3.07_{\pm0.31}$ & $2.84_{\pm0.12}$ & $0.17_{\pm0.09}$ \\
LVHIS 008 & NGC 625 & $8.70_{\pm0.10}$ & $-1.34_{\pm0.05}$ & $7.94_{\pm0.09}$ & $3.6_{\pm0.5}$ & $7.3_{\pm0.7}$ & $0.06_{\pm0.01}$ & $2.00_{\pm0.12}$ & $-1.20_{\pm0.08}$ \\
LVHIS 009 & ESO 245-G005 & $8.31_{\pm0.10}$ & $-1.60_{\pm0.07}$ & $8.59_{\pm0.07}$ & $2.4_{\pm0.3}$ & $4.2_{\pm0.4}$ & $0.61_{\pm0.06}$ & $1.15_{\pm0.12}$ & $-2.41_{\pm0.09}$ \\
LVHIS 011 & ESO 115-G021 & $8.06_{\pm0.10}$ & $-1.88_{\pm0.07}$ & $8.81_{\pm0.07}$ & $2.8_{\pm0.4}$ & $7.1_{\pm0.7}$ & $0.92_{\pm0.09}$ & $1.28_{\pm0.12}$ & $-2.79_{\pm0.09}$ \\
LVHIS 014 & NGC 1313 & $9.49_{\pm0.10}$ & $-0.29_{\pm0.05}$ & $9.28_{\pm0.07}$ & $3.3_{\pm0.5}$ & $12.9_{\pm1.3}$ & $1.00_{\pm0.10}$ & $1.83_{\pm0.12}$ & $-1.10_{\pm0.08}$ \\
LVHIS 015 & NGC 1311 & $8.23_{\pm0.10}$ & $-1.89_{\pm0.07}$ & $7.94_{\pm0.08}$ & $3.1_{\pm0.4}$ & $4.8_{\pm0.5}$ & $0.33_{\pm0.03}$ & $1.69_{\pm0.12}$ & $-2.84_{\pm0.09}$ \\
LVHIS 017 & IC 1959 & $8.54_{\pm0.10}$ & $-1.58_{\pm0.06}$ & $8.35_{\pm0.07}$ & $2.5_{\pm0.3}$ & $5.8_{\pm0.6}$ & $0.80_{\pm0.08}$ & $2.02_{\pm0.12}$ & $-2.50_{\pm0.09}$ \\
LVHIS 027 & NGC 2915 & $8.40_{\pm0.10}$ & $-1.58_{\pm0.06}$ & $8.50_{\pm0.08}$ & $3.3_{\pm0.5}$ & $2.6_{\pm0.3}$ & $0.22_{\pm0.02}$ & $2.22_{\pm0.12}$ & $-2.01_{\pm0.09}$ \\
LVHIS 031 & NGC 3621 & $9.67_{\pm0.10}$ & $-0.15_{\pm0.05}$ & $9.96_{\pm0.07}$ & $3.0_{\pm0.4}$ & $18.9_{\pm1.9}$ & $4.02_{\pm0.40}$ & $2.28_{\pm0.13}$ & $-1.32_{\pm0.08}$ \\
LVHIS 044 & ESO 269-G058 & $8.47_{\pm0.10}$ & $-2.75_{\pm0.10}$ & $7.26_{\pm0.10}$ & $3.3_{\pm0.5}$ & $3.8_{\pm0.4}$ & $0.24_{\pm0.02}$ & $1.78_{\pm0.12}$ & $-3.36_{\pm0.11}$ \\
LVHIS 053 & NGC 5236 & $10.50_{\pm0.10}$ & $0.47_{\pm0.05}$ & $9.91_{\pm0.07}$ & $3.3_{\pm0.5}$ & $22.0_{\pm2.2}$ & $1.78_{\pm0.18}$ & $2.50_{\pm0.12}$ & $-0.65_{\pm0.08}$ \\
LVHIS 055 & NGC 5237 & $8.28_{\pm0.10}$ & $-2.51_{\pm0.09}$ & $7.48_{\pm0.16}$ & $3.1_{\pm0.4}$ & $2.1_{\pm0.2}$ & $0.17_{\pm0.02}$ & $2.11_{\pm0.12}$ & $-3.10_{\pm0.11}$ \\
LVHIS 075 & IC 4662 & $8.01_{\pm0.10}$ & $-1.31_{\pm0.05}$ & $8.16_{\pm0.07}$ & $2.8_{\pm0.4}$ & $1.5_{\pm0.2}$ & $0.27_{\pm0.03}$ & $1.92_{\pm0.12}$ & $-1.01_{\pm0.08}$ \\
LVHIS 077 & IC 5052 & $9.09_{\pm0.10}$ & $-1.21_{\pm0.05}$ & $8.89_{\pm0.09}$ & $2.8_{\pm0.4}$ & $12.5_{\pm1.2}$ & $1.03_{\pm0.10}$ & $2.02_{\pm0.12}$ & $-1.91_{\pm0.08}$ \\
LVHIS 078 & IC 5152 & $8.15_{\pm0.11}$ & $-1.79_{\pm0.06}$ & $7.95_{\pm0.08}$ & $3.0_{\pm0.4}$ & $2.8_{\pm0.3}$ & $0.21_{\pm0.02}$ & $1.88_{\pm0.13}$ & $-1.86_{\pm0.09}$ \\
LVHIS 082 & NGC 7793 & $9.47_{\pm0.10}$ & $-0.60_{\pm0.05}$ & $8.95_{\pm0.06}$ & $2.6_{\pm0.4}$ & $11.8_{\pm1.2}$ & $1.28_{\pm0.13}$ & $2.03_{\pm0.12}$ & $-1.34_{\pm0.08}$ \\
\enddata
\tablecomments{Columns: (1) LVHIS ID; (2) optical name; (3) stellar mass; (4) star formation rate; (5) \hi\ mass; (6) WISE 3.4\mum\ 90-to-50 percent light radius ratio; (7) B band $25\,\mathrm{mag/arcsec^2}$ diameter; (8) FUV 50\% light radius; (9) stellar mass surface density; (10) star formation rate surface density.\\A machine readable version of this table is available as online supplementary data.}
\end{deluxetable}

We estimate the uncertainty of the correlation coefficient by using bootstrapping method. The uncertainty can help us to distinguish intrinsically weak correlation with statistically insignificant correlation\footnote{A correlation with coefficient of 0.2 (usually regarded as ``weak'') but with uncertainty of 0.05 is statistically more significant than a correlation with coefficient of 0.5 (sometimes regarded as ``strong'') but with uncertainty of 0.5. For the former one, we are quite confident that there is some weak correlation. For the latter one, we are not confident if there is any correlation. The correlation may not be as ``strong'' as it appears, or even does not exist. A ``strong'' correlation driven by outliers will end up with large uncertainty and highly asymmetric distribution using the bootstrapping method.}. All the correlation coefficients are calculated for logarithmic (i.e. the logarithm of values are used instead of their original values, unless the physical quantity is already defined in logarithmic scale, such as $q$), with Pearson definition (we also test the Spearman definition, getting similar result).

The major caveat in measuring the correlation coefficient is that it reflects not only the physical correlation between parameters but also the sample bias. For example, if a sample of galaxies have wide distance distribution, distance modulation could cause strong apparent correlation between two different luminosities which are physically not connected. Our gold sample is probably highly biased, given the criteria we use to create it. It is almost impossible to quantify the bias (e.g. giving proper weighting) before the correlation analysis. Moreover, the physical parameters we are going to check, are correlated with each other as well (e.g. more massive galaxies are usually larger). It will cause artificial correlation, i.e. if A is physically correlated with B and C, then B and C will artificially correlate with each other without necessary physical link. It makes the interpretation of the correlation coefficient much more complicated. In order to overcome these problems, we use the partial correlation coefficient. The partial correlation coefficient between $a$ and $b$ with a third control parameter $c$, is defined as $\rho_{ab,c}=(\rho_{ab} - \rho_{ac}\rho_{bc})/\sqrt{(1-\rho_{ac}^2)(1-\rho_{bc}^2)}$, where $\rho_{ab}$ is the normal correlation coefficient between $a$ and $b$ (the same for $\rho_{ac}$ and $\rho_{bc}$). It measures the correlation strength between two parameters but nullifying the effects from a third parameter. The nullified parameter is also called ``controlled'' parameter. The partial correlation coefficient $\rho_{ab,c}$ can remove any apparent correlation caused by $c$, revealing the ``true'' physical correlation between $a$ and $b$.

We calculate both the normal and partial correlation coefficients. For partial correlation coefficients, we use many different parameters as the controlled parameter (one parameter each time). We compare these partial correlation coefficients. Two physical parameters are considered to be intrinsically correlated, if their partial correlation coefficients are always the same or change mildly, regardless of control parameter. On the other hand, two physical parameters are considered not correlated or sometimes artificially correlated, if their partial correlation coefficients change significantly (or even sign-flipped) when controlling different parameters.

The galaxy properties we are going to show are stellar mass ($M_*$), specific star formation rate ($\mathrm{SFR}/M_*$, sSFR), star formation efficiency ($\mathrm{SFR}/M_\mathrm{HI}$, SFe), \hi-to-stellar mass ratio ($f_\mathrm{HI}$), stellar concentration (90-to-50 percent light radius ratio, in WISE 3.4\mum\ band), B band $25\,\mathrm{mag/arcsec^2}$ diameter (D25), FUV 50\% light radius ($R_\mathrm{UV}$, a proxy of the size of the star forming region), stellar mass surface density ($\mu_*$), star formation rate surface density ($\mu_\mathrm{SFR}$) and FIR colour ($\mathrm{col_{IR}}$, 60-to-100\mum\ flux ratio). Many other properties, such as star formation rate, \hi\ mass ($M_\mathrm{HI}$), \hi\ diameter at $1\,\mathrm{M_\odot}/\mathrm{pc^2}$ ($D_\mathrm{HI}$) and \hi\ mass surface density ($\mu_\mathrm{HI}$) are also investigated but not shown in the figures for better readability.

\subsection{The origin of FIR deficiency}

\begin{figure*}
  \centering \includegraphics[width=16cm]{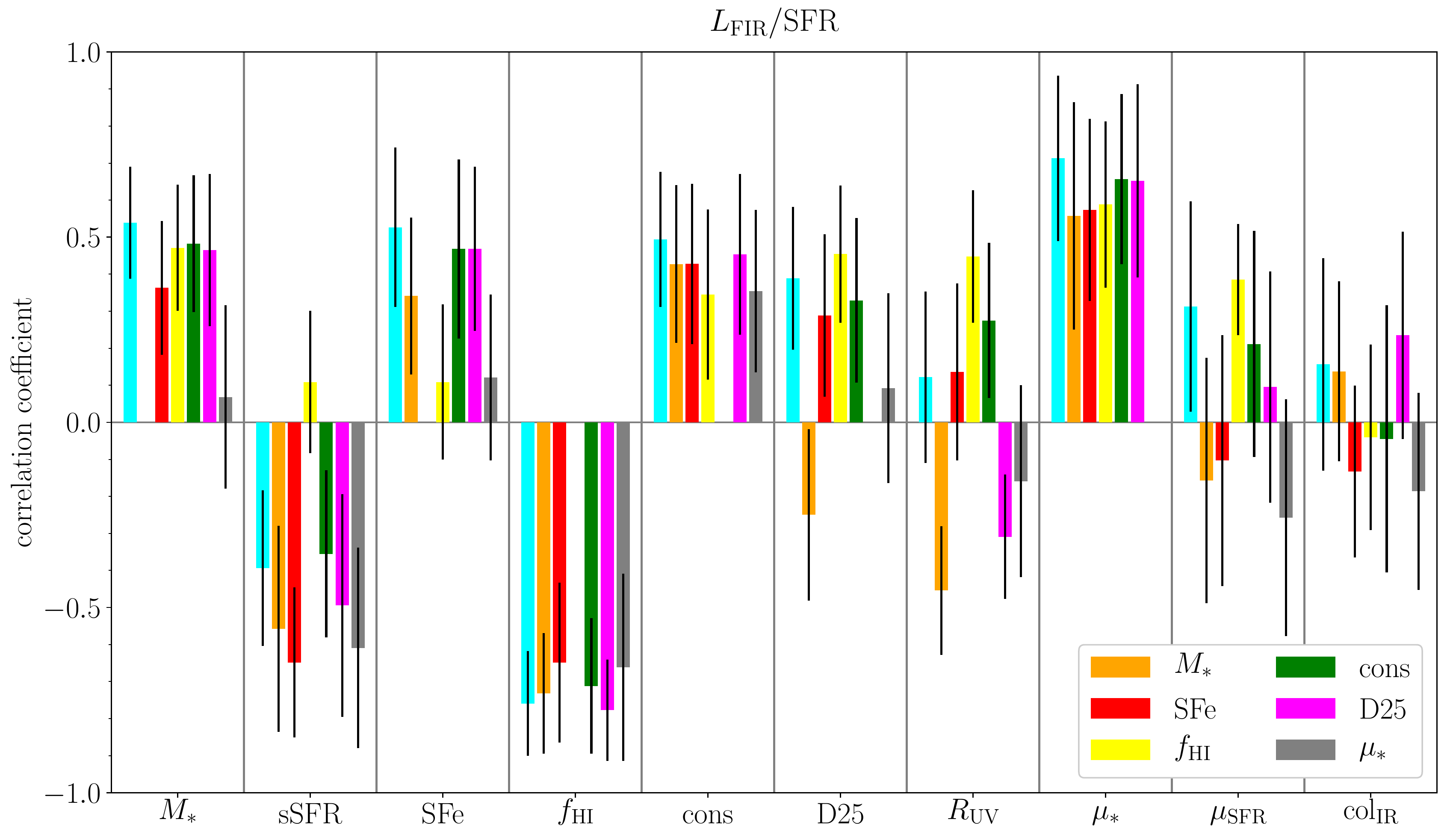}
  \caption{The correlation coefficients between FIR efficiency ($\mathrm{eff}_\mathrm{FIR} \equiv L_\mathrm{FIR}/\mathrm{SFR}$) and galaxy properties. The galaxy properties are stellar mass ($M_*$), specific star formation rate (sSFR), star formation efficiency (SFe), \hi-to-stellar mass ratio ($f_\mathrm{HI}$), stellar concentration (cons), B band $25\,\mathrm{mag/arcsec^2}$ diameter (D25), FUV 50\% light radius ($R_\mathrm{UV}$), stellar mass surface density ($\mu_*$), star formation rate surface density ($\mu_\mathrm{SFR}$) and FIR colour ($\mathrm{col_{IR}}$). Cyan bar is original correlation coefficient. Other colour bars are partial correlation coefficients with different control parameters, as shown in figure legend.}
  \label{fig:corr_fir}
\end{figure*}

The FIR efficiency ($\mathrm{eff}_\mathrm{FIR}$) has clear (anti-)correlation with \hi-to-stellar mass ratio and stellar mass surface density (Figure \ref{fig:corr_fir}), no matter which third quantity is controlled, indicating strong physical link between the FIR emission and these two quantities. Stellar concentration may also be weakly correlated, but is much less important than the two major factors. When we control the stellar mass surface density and \hi-to-stellar mass ratio simultaneously, all other partial correlation coefficients become zero.

For FIR emission, the stellar mass surface density and \hi-to-stellar mass ratio are found to be the most important physical parameters. The stellar mass surface density is known to be correlated with dust obscuration for more massive disk galaxies. \citet{2013ApJ...766...59G} reported a correlation between B band face-on central dust opacity and the stellar mass surface density for a sample of disk galaxies from the Galaxy and Mass Assembly survey \citep{2011MNRAS.413..971D}. They suggested that the correlation was driven by a linear relation between stellar mass and dust mass, combined with galaxy inclination \citep[see also earlier work, e.g.,][]{2007ApJ...659.1159S}. Our result supports their conclusion but with much lower typical stellar mass ($\sim3\times10^8\,\mathrm{M_\odot}$ versus their $\sim3\times10^{10}\,\mathrm{M_\odot}$) and larger morphological diversity (a large fraction of gold sample galaxies are irregular galaxies).

The anti-correlation between \hi-to-stellar mass ratio and FIR efficiency has not been noted previously. They are probably indirectly linked together. The \hi-to-stellar mass ratio or gas fraction is found to be anti-correlated with galaxy metallicity \citep[see e.g.][]{2013MNRAS.433L..35L, 2013MNRAS.433.1425B, 2013A&A...550A.115H, 2014ApJ...791..130Z, 2016MNRAS.455.1156B}, including on sub-galactic scales \citep{2015MNRAS.448.2126A}. It is probably a redshift independent, fundamental relation \citep{2014ApJ...791..130Z}. Galaxies without continuous gas supply, probably due to supernova feedback or lack of gas accretion, will build up metals due to past star formation \citep{2016MNRAS.455.1156B}. Metals can also be diluted, if a new gas supply is secured \citep[see e.g.][]{2014MNRAS.439.3817Y}. On the other hand, the dust obscuration is found to be strongly correlated with metallicity \citep[see e.g.,][]{1998ApJ...503..646H,2010ApJ...712.1070R,2011A&A...532A..56G,2012MNRAS.426.1782R,2016ApJ...822...29F}. It is expected as most of the metals will be contained by dust grains which are also attributed to the star light absorption. The anti-correlation can be interpreted as a natural consequence of these two fundamental correlations. In this work, we do not have galaxy gas metallicity measurements. In a recent paper by \citet{2017ApJ...846...68Q}, a positive correlation between metallicity and FIR-to-radio ratio $q$ was claimed, consistent with our interpretation.

For LVHIS sample, compared with massive galaxies, dwarf galaxies have lower stellar mass surface density ($\sim22\,\mathrm{M_\odot/pc^2}$ versus $\sim103\,\mathrm{M_\odot/pc^2}$, median value) and higher \hi-to-stellar mass ratio ($M_\mathrm{HI}/M_*\sim0.64$ versus $M_\mathrm{HI}/M_*\sim0.50$, median value). The FIR deficiency in dwarf regime mainly comes from these two effects.

\subsection{The origin of the radio deficiency}

\begin{figure*}
  \centering \includegraphics[width=16cm]{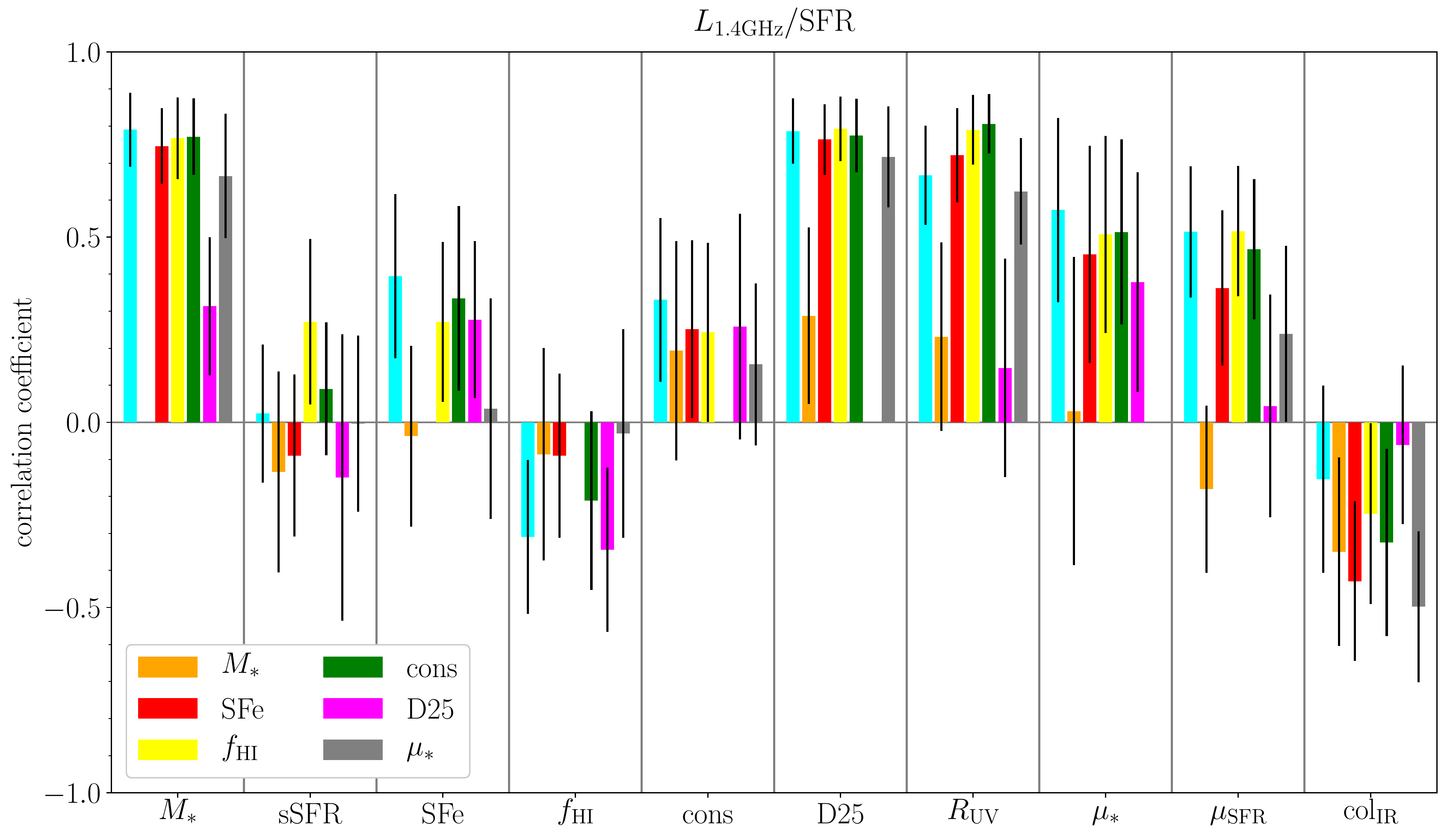}
  \caption{Same as Figure \ref{fig:corr_fir} but for the radio efficiency ($\mathrm{eff}_\mathrm{1.4\,GHz} \equiv L_\mathrm{1.4\,GHz}/\mathrm{SFR}$).}
  \label{fig:corr_radio}
\end{figure*}

The analysis for radio efficiency ($\mathrm{eff}_\mathrm{1.4\,GHz}$) is similar to what we did to FIR, but the result is less clear (Figure \ref{fig:corr_radio}). The most important physical associations are stellar mass, D25 and FUV size. Their correlation coefficients are large and do not significantly change unless controlling any of them. It suggests these three parameters have similar effects with high degree of degeneration, which is not broken by any of our control parameters. The stellar mass surface density and star formation rate surface density are probably also weakly correlated, given the pattern that they becomes zero only when controlling stellar mass or galaxy size. But they are less important.

The radio deficiency in dwarf galaxies is probably mainly due to cosmic rays escaping \citep{2010ApJ...717....1L,2011ApJ...739L..23H,2016A&A...593A..77S}, either diffusive or advective. To constrain the cosmic rays within the galaxy, it is necessary to keep the circular motion radius $r_\mathrm{CR}$ of a charged particle smaller than the effective height of galaxy $h_{g}$. For particles purely bound by magnetic field, $r_\mathrm{CR}\propto B^{-1}$. The strength of magnetic field is limited by the mass density due to the Parker instability \citep{1966ApJ...145..811P}. Observationally it is suggested that magnetic field strength is scaled with gas density \citep{2008ApJ...680..981R}. We expect the cosmic rays are less tightly bound in smaller galaxies with weaker magnetic field, causing higher diffusive loss rate. On the other hand, the required escape velocity is smaller for less massive galaxies. The high speed ejecta from star forming regions, together with cosmic rays, can leave smaller galaxies more easily. It will increase the advective cosmic ray loss \citep{2009ApJ...706..482M,2011ApJ...739L..23H}. These two effects both suggest a higher cosmic ray escape rate in dwarf galaxies. Indeed, Fermi observations suggest the cosmic ray density is $\sim70\%$ lower in Large Magellanic Cloud than in Milky Way \citep{2016A&A...586A..71A}. Furthermore, the synchrotron radiation power is directly correlated with magnetic field strength, $P_\mathrm{sync}\propto B^2$. The radio continuum emission is significantly enhanced with stronger magnetic field. Therefore, the radio emission is stronger if the galaxy is larger and more massive and the gas is denser. Yet, the star formation is also strongly affected by gas density \citep[according to the Kennicutt-Schmidt law, see e.g.][]{1998ApJ...498..541K}. The impact of gas density is partly (probably not totally though) cancelled, leaving the galaxy size as the only major factor.

The radio deficiency in dwarf regime is likely driven by the fact that dwarf galaxies are much smaller than massive galaxies (D25: $\sim2.8\,\mathrm{kpc}$ versus $\sim16\,\mathrm{kpc}$; $R_\mathrm{UV}$: $\sim0.3\,\mathrm{kpc}$ versus $\sim1.4\,\mathrm{kpc}$, median value).

\subsection{The ``conspiracy'' of FIR-radio correlation}

\begin{figure*}
  \centering \includegraphics[width=16cm]{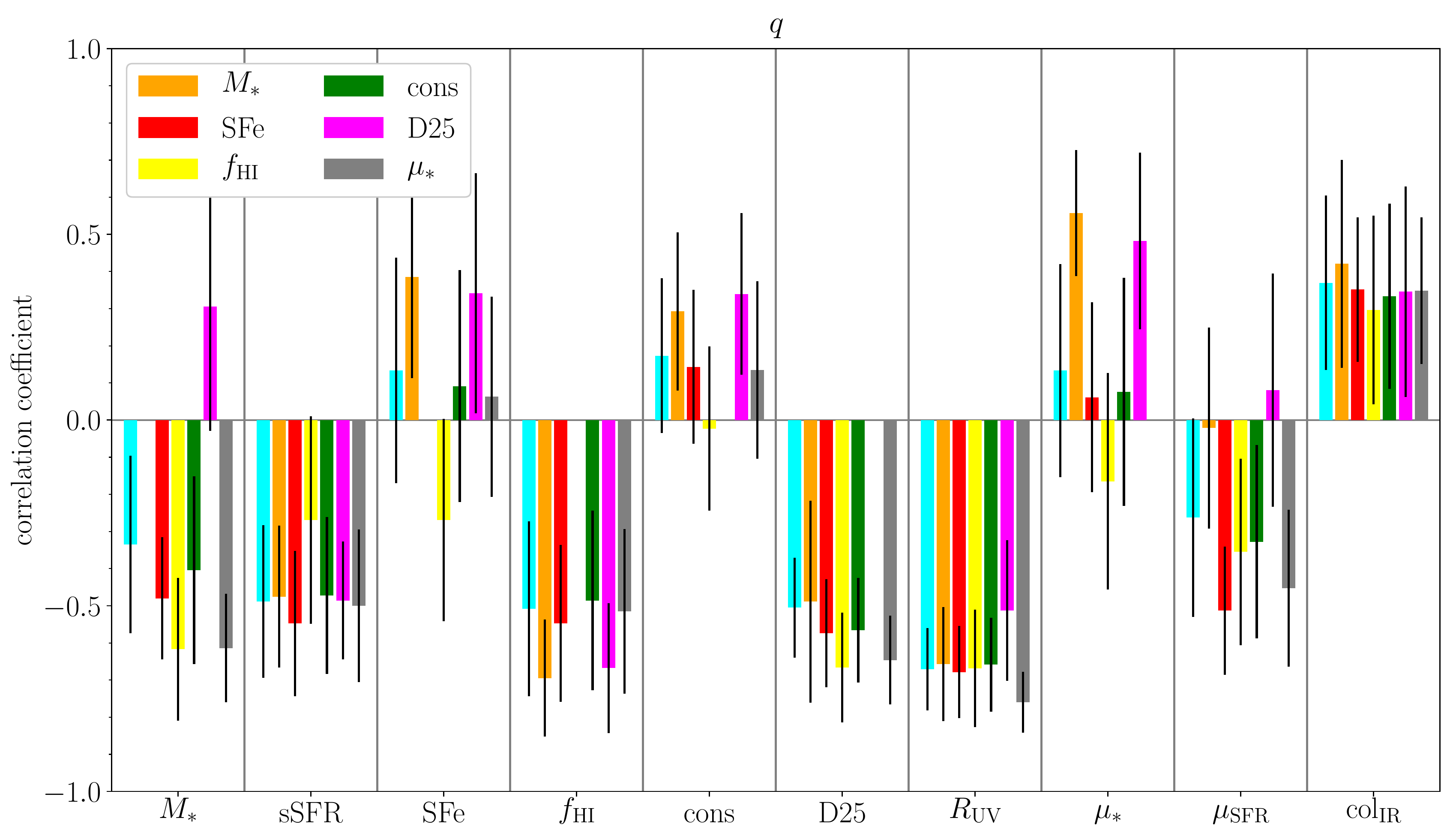}
  \caption{Same as Figure \ref{fig:corr_fir} but for the FIR-to-radio ratio $q$.}
  \label{fig:corr_q}
\end{figure*}

As $q$ differs $\mathrm{eff}_\mathrm{FIR}/\mathrm{eff}_\mathrm{1.4\,GHz}$ by only a constant factor, we would expect some of the most important physical parameters identified above to be identified here again. Indeed, we find $q$ is well anti-correlated with \hi-to-stellar mass ratio, D25 and FUV size. The sign of the correlation coefficient is consistent with the expectation from previous results (Figure \ref{fig:corr_q}). Specific star formation rate and FIR colour may also be (anti-)correlated with $q$.

The radiation deficiency is the key issue to break the FRC or keep FRC alive. The strength of deficiency in FIR and radio has to be the same to keep a constant $q$, i.e. the ``conspiracy'' of FRC\footnote{Here we note that ``conspiracy'' does not necessarily mean a strictly constant $q$ in practice. It is a characterized behaviour that both FIR and radio efficiencies vary significantly while the FIR-to-radio ratio is almost constant, i.e. a slow function of other properties such as luminosity, star formation rate, gas surface density, etc. \citep[see, e.g., Figure 1 of][]{2010ApJ...717....1L}.}. But as shown above, FIR and radio efficiencies are tightly correlated with somehow different galaxy properties. So the same deficiency strength for both FIR and radio is not automatically guaranteed. It seems we are not able to easily achieve the ``conspiracy''.

However, we notice they have some common feature: for some important physical parameters, such as stellar mass or stellar mass surface density, the signs of their correlation coefficients for FIR and radio are the same. It means when calculating $q$, the effects are (partly) cancelled out, i.e., $q$ becomes less sensitive to these physical quantities.

The remaining main contributors to the offset of $q$ are the \hi-to-stellar mass ratio, the size of galaxy and the size of star forming regions. They can however cancel each other out in a statistical way: more massive and larger galaxies statistically have lower \hi-to-stellar mass ratio. For example, our dwarf galaxies are smaller than massive galaxies but more \hi-rich than massive galaxies. Similar idea also works with other minor factors such as specific star formation rate. Therefore, in practice, one needs to sample a special part of the parameter space, where there are very few galaxies, to find the ``outlier'' to the FRC. This strengthens the tightness of the FRC. It also implies that in order to find galaxies that are not following FRC, one may need to focus on very large starburst galaxies \citep[see also][]{2003ApJ...593..733R} or small compact gas-poor galaxies.

Here we stress that because we are not able to make solid conclusion whether the FRC is broken or not at faint end (see our Section \ref{sec:emission}), the discussion about the origin of ``conspiracy'' presented above should be regarded as a general discussion, not specific for our sample. The non-detections of FIR and radio emission and the lack of proper galaxy property measurements for many local volume galaxies prevent us making more solid conclusion. It could be improved in future with deeper observation.

\section{Summary and perspectives}

In this paper, we use the LVHIS sample of gas-rich galaxies in local Universe. As the sample is close to volume limited, we are less affected by selection bias which is a big issue for many previous studies. We reduce the 20 cm radio continuum data and measure the flux density with two different methods. We use Scanpi to measure the FIR flux density and compare with catalogue values. Together with galaxy properties derived from supplementary data, we are able to test the FRC for galaxies with stellar mass as low as $\sim10^8\,\mathrm{M_\odot}$ and study the physics behind the FRC. Our main conclusions are:

\begin{enumerate}
\item LVHIS galaxies obey a tight linear FIR-radio correlation, generally consistent with empirical relation. But for detected galaxies, it is found that the FIR-to-radio ratio increases with decreasing FIR luminosity.
\item Both FIR and radio emission correlates with SFR, but deviating from a linear relation. FIR and radio emission both shows a deficiency in the low SFR regime, leading to small variance of the FIR-to-radio ratio.
\item The FIR efficiency is found to be well correlated with stellar mass surface density and anti-correlated with \hi-to-stellar mass ratio. The radio efficiency is mainly correlated with stellar mass, galaxy size and star forming region size.
\item Common parameter dependencies, such as stellar mass surface density, are found in both FIR and radio efficiency. By cancellation, the FIR-to-radio ratio is generally invariant. The remaining factors cancel with each other due to general galaxy properties, further strengthening the tightness of FRC.
\end{enumerate}

Beyond observables, we also provide possible physical explanations. But we emphasize that our study does not necessarily rule out other explanations.

\vspace{1em}

Despite of our efforts, there are several issues in this work, which could be improved in future:

\begin{itemize}
\item Sensitivity and statistics. Although we reach $\sim50\,\mathrm{\mu Jy}$ noise level, we still get detection rate less than 50\%. Our gold sample for detailed correlation analysis only contains 17 galaxies. To make more solid conclusions, we need more galaxies, especially low luminosity dwarf galaxies, with better coverage of parameter space. The Square Kilometre Array and their precursors, such as ASKAP (Australian Square Kilometre Array Pathfinder) and MeerKAT, will be very useful.

\item Analysis method. With correlation analysis, we may know the properties are correlated but not know how the correlations numerically arise. We are not able to directly compare with theoretical models. Further work is needed to incorporate these results better into theoretical framework. It for sure will require more data to break degeneracies in models.

\item Global physics only. In this work, the galaxies are always treated as whole. Future work could include a spatially resolved study to better understand the underlying physical processes. In the appendix we show some example galaxies with resolved radio/mid-IR/$\mathrm{H\alpha}$ data. It is however beyond the scope of this paper to do quantitative analysis on individual galaxies.
\end{itemize}

\acknowledgments
LCH thanks the support by the National Key R\&D Program of China (2016YFA0400702) and the National Science Foundation of China (11473002, 11721303).

This research has made use of the data products from the Australia Telescope Compact Array, which is part of the Australia Telescope National Facility, funded by the Australian Government for operation as a National Facility managed by CSIRO. This paper includes archived data obtained through the Australia Telescope Online Archive (http://atoa.atnf.csiro.au).

This publication makes use of data products from the Wide-field Infrared Survey Explorer, which is a joint project of the University of California, Los Angeles, and the Jet Propulsion Laboratory/California Institute of Technology, funded by the National Aeronautics and Space Administration.

This research has made use of the NASA/IPAC Infrared Science Archive, which is operated by the Jet Propulsion Laboratory, California Institute of Technology, under contract with the National Aeronautics and Space Administration.

This research has made use of the NASA/IPAC Extragalactic Database (NED) which is operated by the Jet Propulsion Laboratory, California Institute of Technology, under contract with the National Aeronautics and Space Administration.

\facilities{ATCA, IRAS, WISE, GALEX}

\clearpage

\appendix
\section{Appendix: radio continuum images of LVHIS galaxies}
\label{sec:radio_image}

The radio continuum images of all detected LVHIS galaxies are grouped according to the measured radio flux density into ``strong'' ($> 7\mathrm{mJy}$) and ``weak'' ($< 7\mathrm{mJy}$) sources. They are displayed in Figure \ref{fig:cont_1} and \ref{fig:cont_2} respectively. The 20 cm images are shown as contours on top of optical greyscale images, which are obtained from the Digitized Sky Survey (DSS2). The contour levels for strong galaxies are -3, 3, 5, 7, 10, 20, 40 and 100-$\sigma$. For weak galaxies, the contours are at -1.5, 1.5, 2, 3, 5, 7, 10, 20, 40 and 100-$\sigma$ levels. The absolute values of $\sigma$ for each map is listed in Table \ref{tab:flux}. The positive and negative contours are shown in red and cyan colours, respectively. The effective beam is displayed as yellow circle at the bottom left corner. For better visual effect, we smooth the radio image with 2-D Gaussian function to make the beam shape round. Hence the contours are less noisy than expected from the original images. We note that smoothing is only used for making contours. All flux densities are measured from the original images.

\renewcommand\thefigure{\thesection.\roman{figure}}
\begin{figure}[h]
\begin{center}
\includegraphics[width=0.47\linewidth]{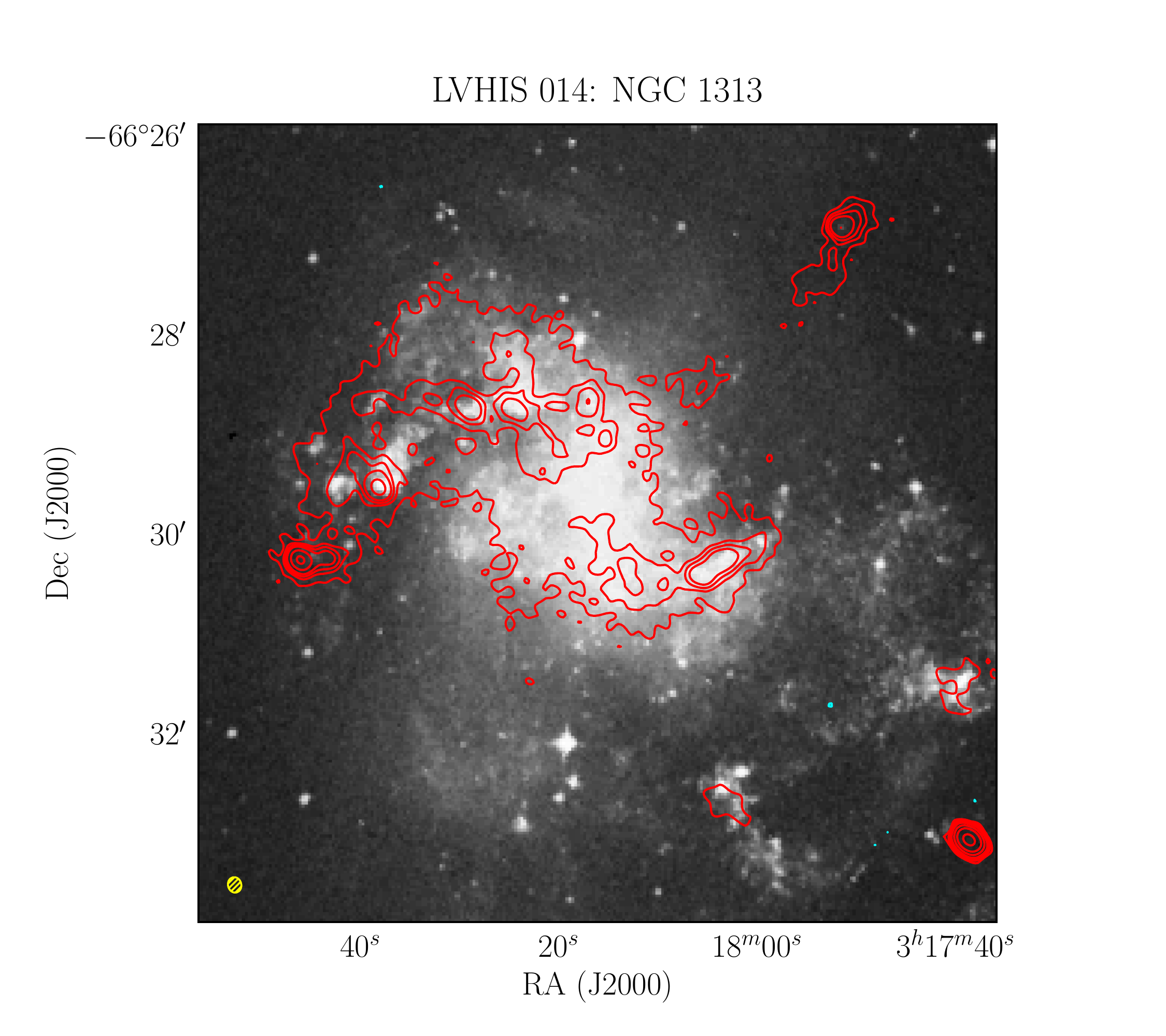} \includegraphics[width=0.47\linewidth]{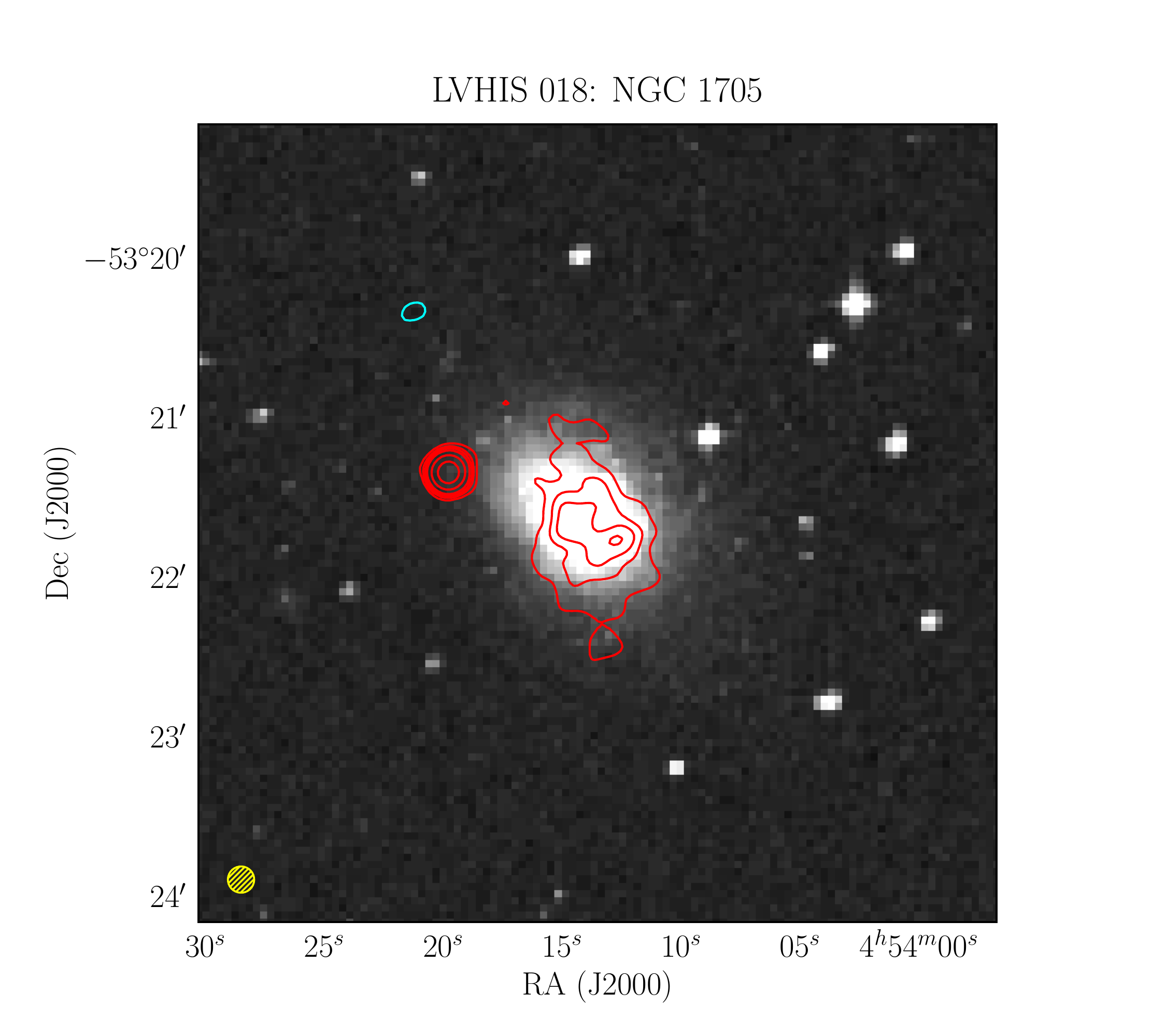} \\
\includegraphics[width=0.47\linewidth]{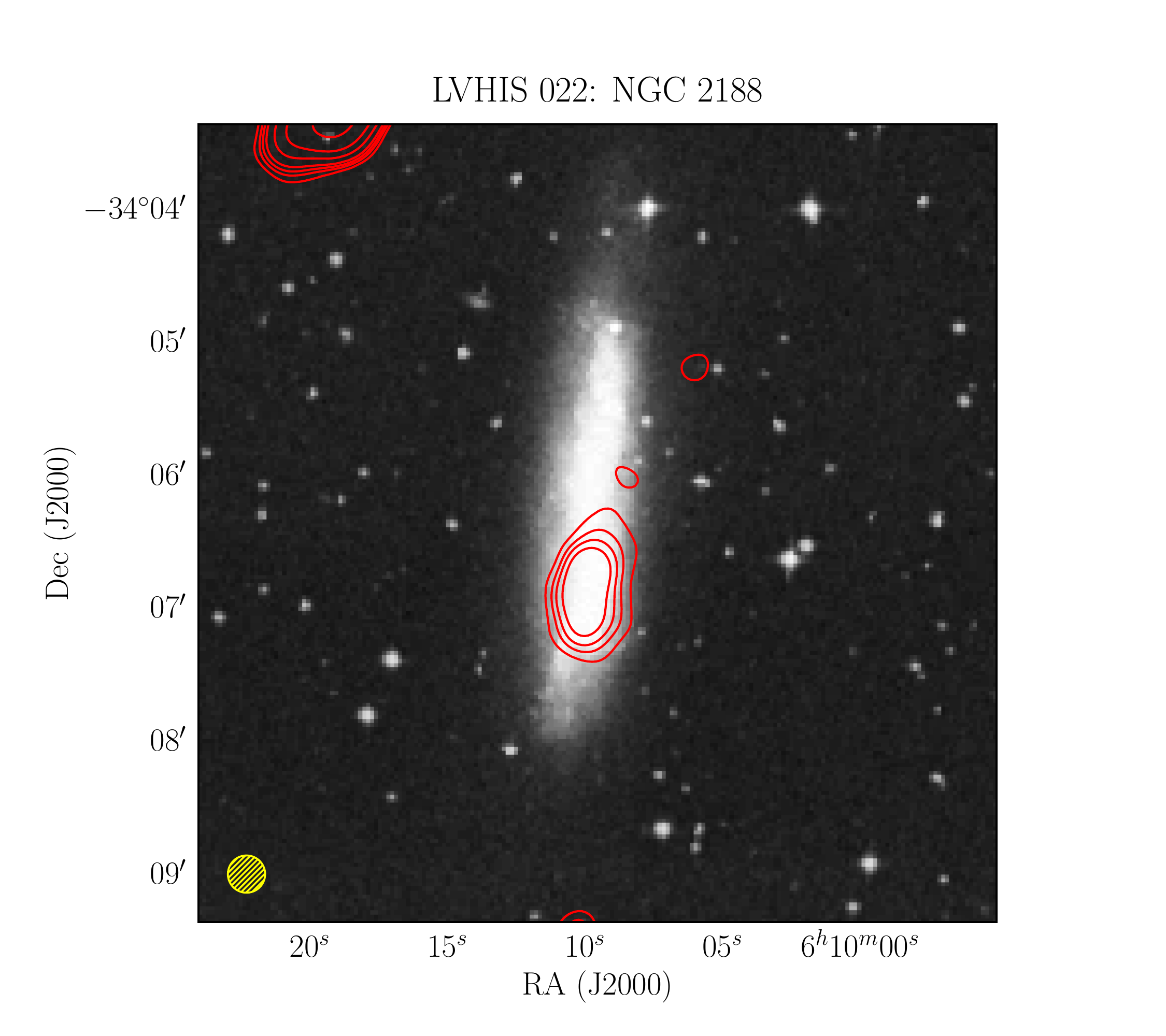} \includegraphics[width=0.47\linewidth]{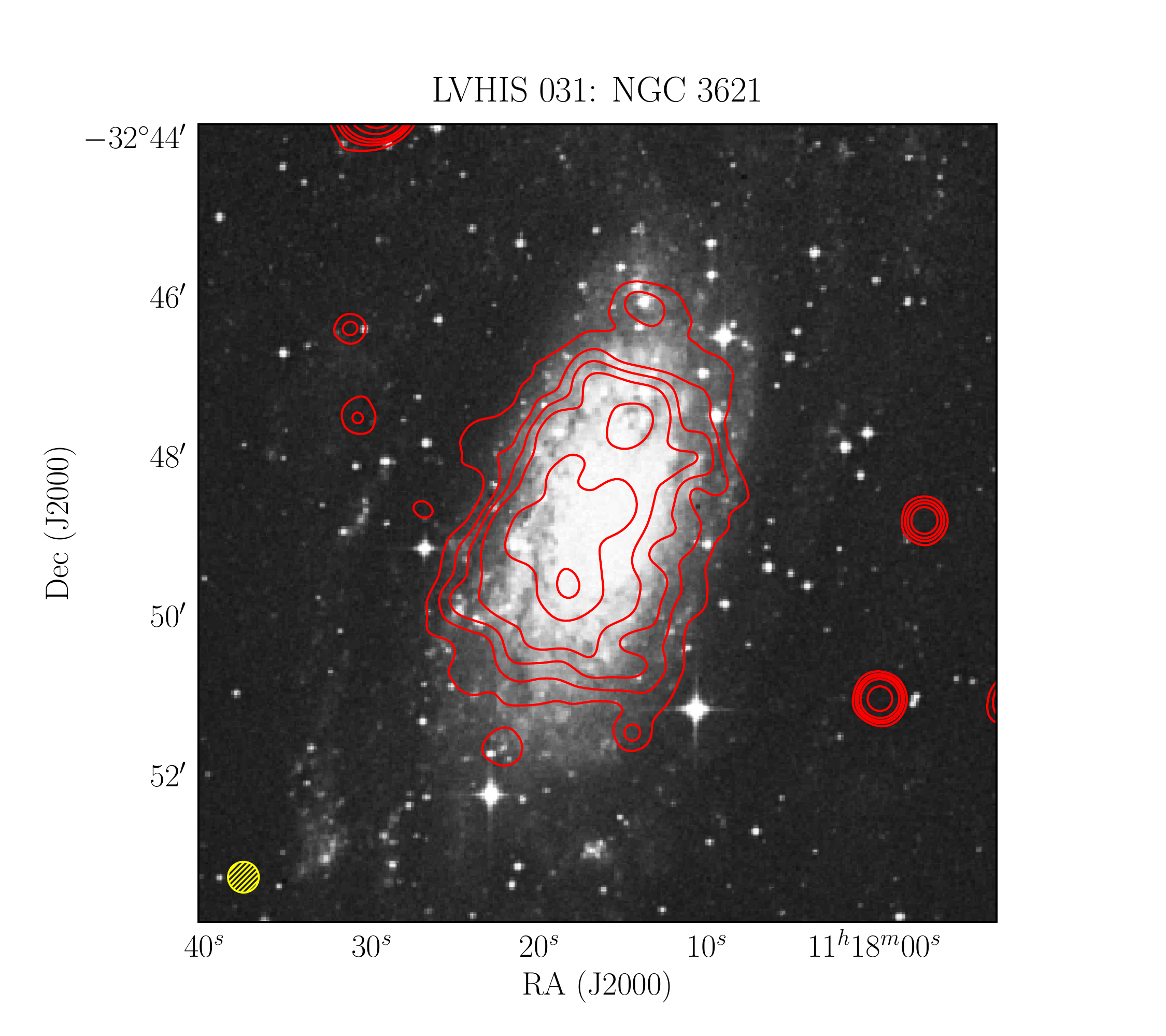} \\
\includegraphics[width=0.47\linewidth]{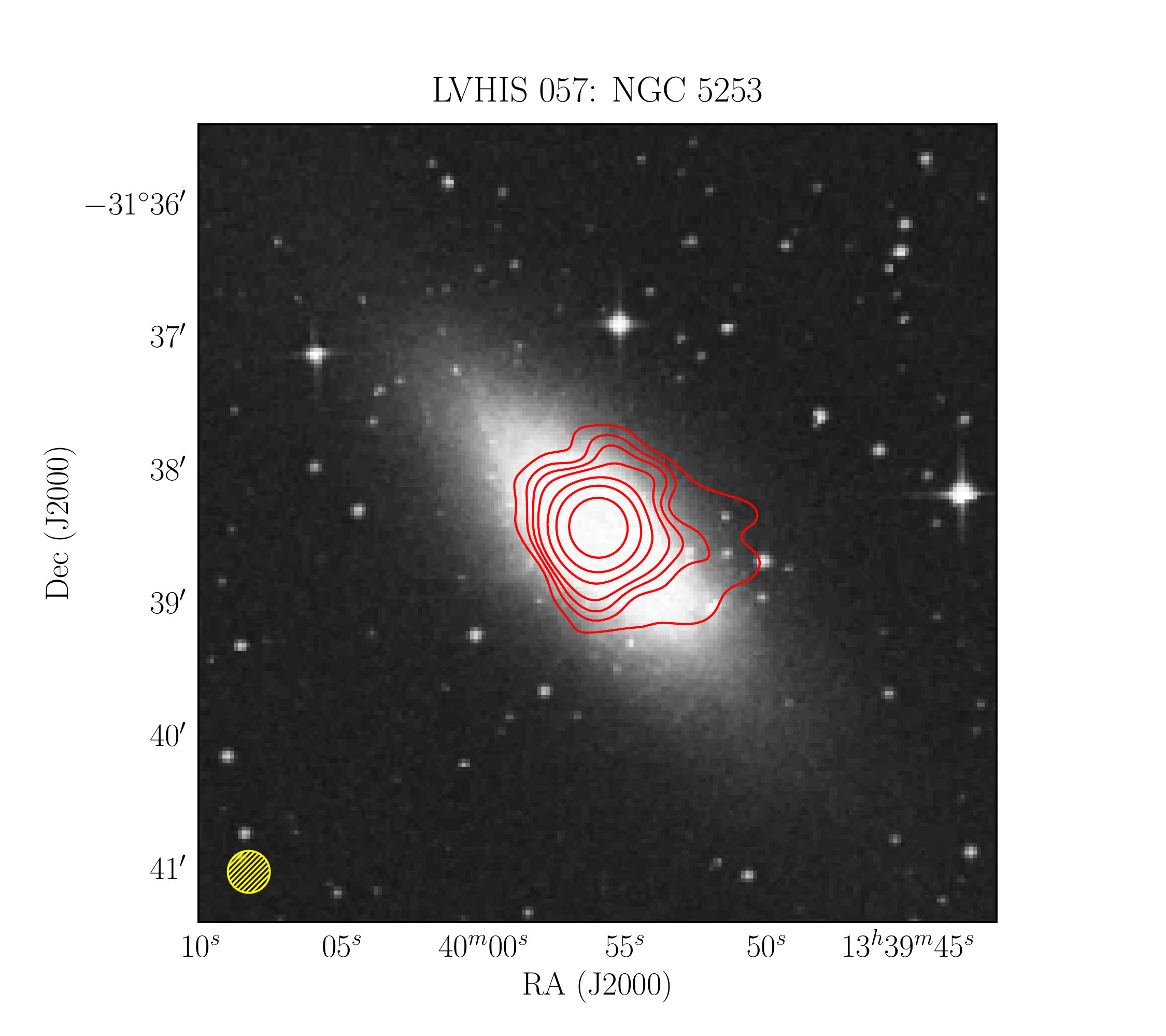} \includegraphics[width=0.47\linewidth]{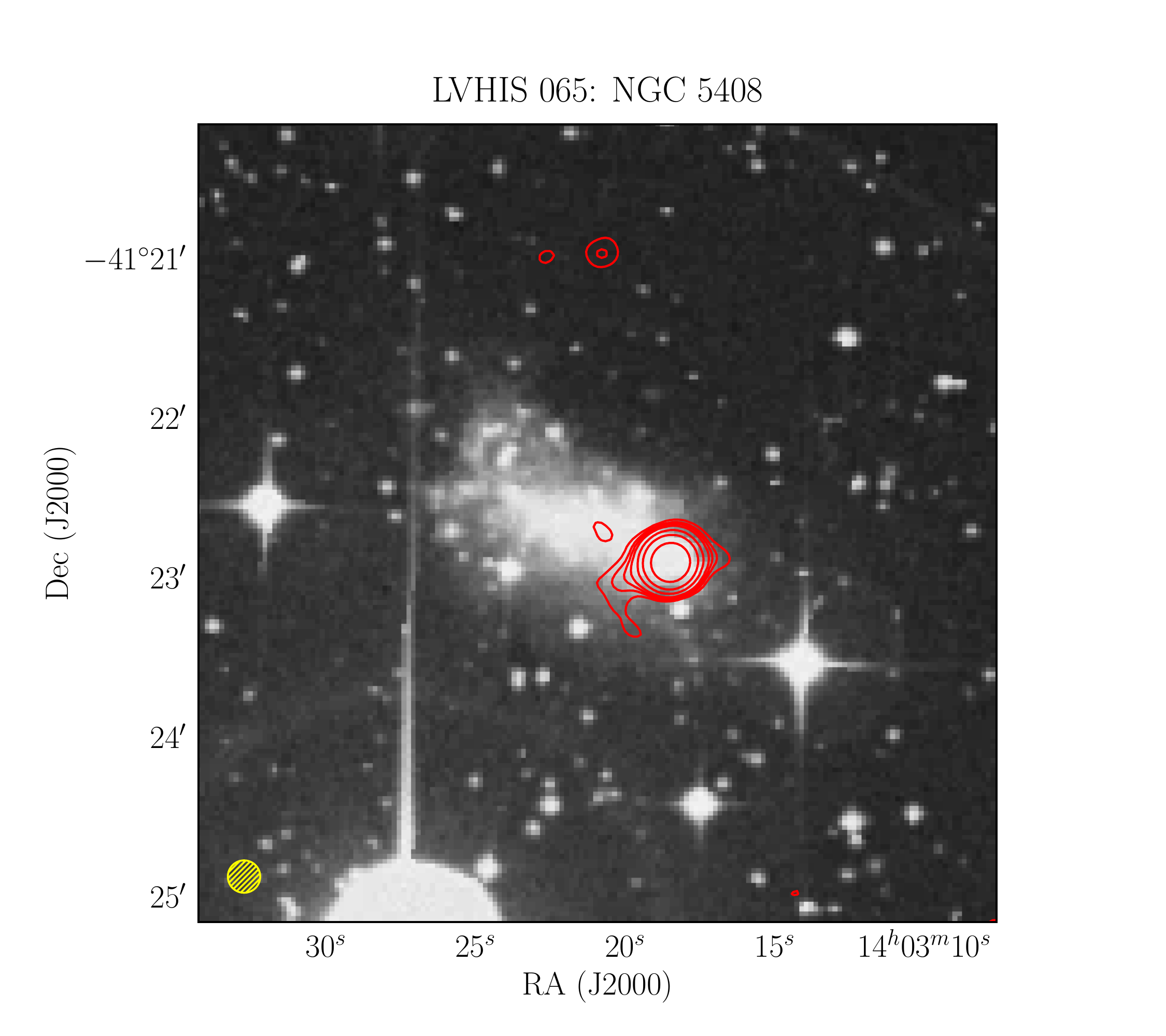} \\
\setcounter{figure}{0}
\caption{The images of 20 cm strong ($> 7\,\mathrm{mJy}$) LVHIS galaxies. \label{fig:cont_1}}
\end{center}
\end{figure}
\begin{figure}[h]
\begin{center}
\includegraphics[width=0.47\linewidth]{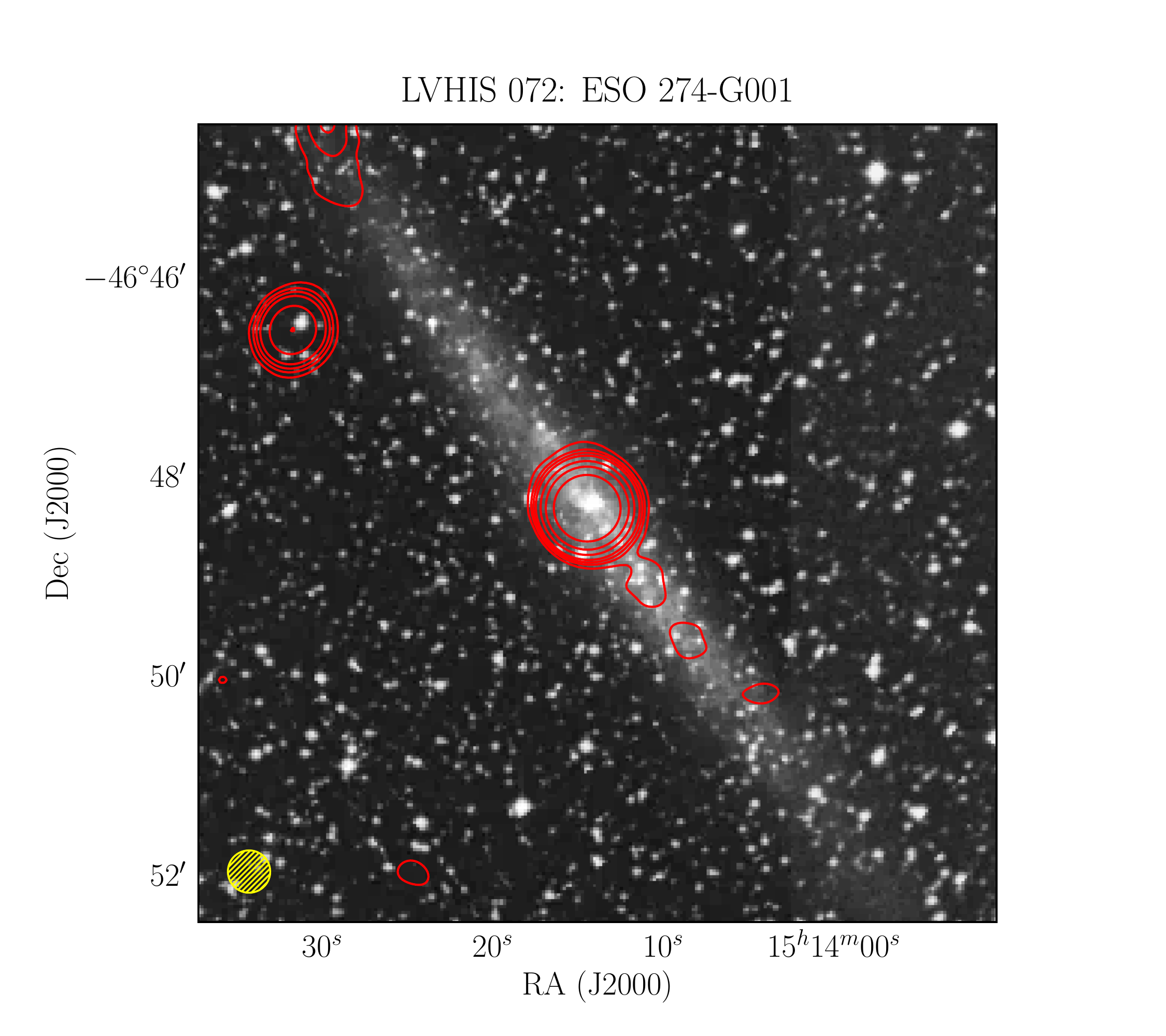} \includegraphics[width=0.47\linewidth]{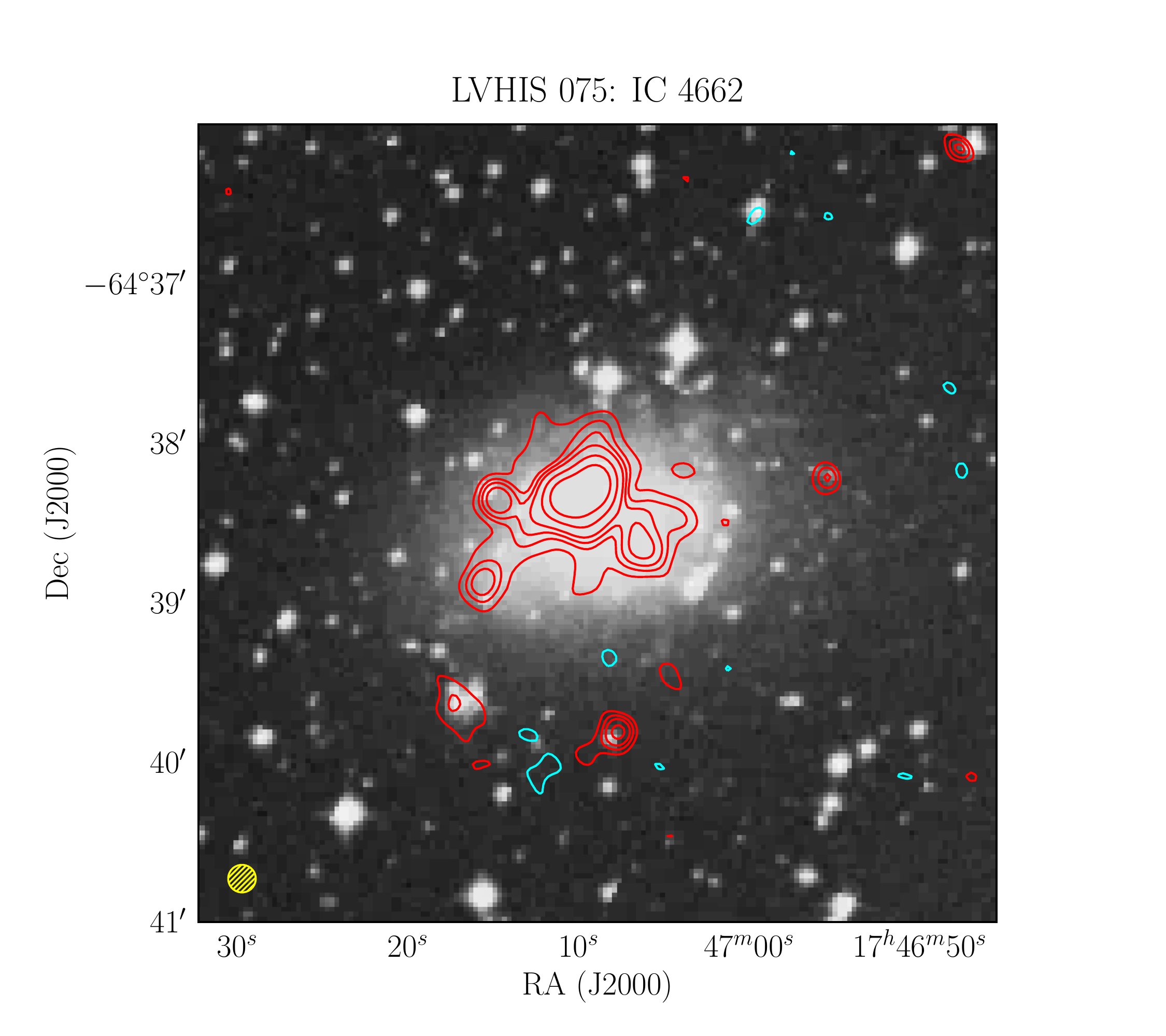} \\
\includegraphics[width=0.47\linewidth]{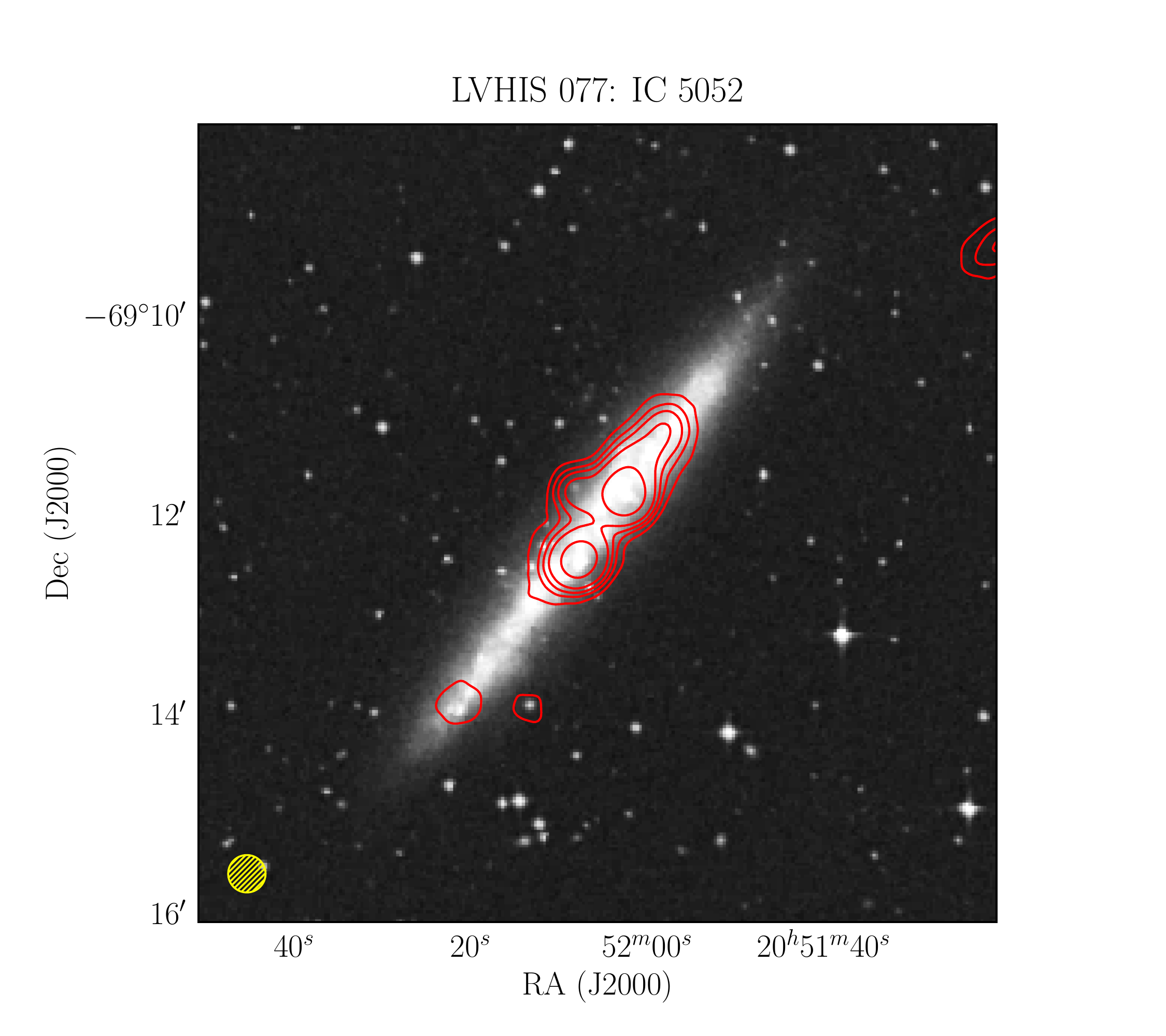} \includegraphics[width=0.47\linewidth]{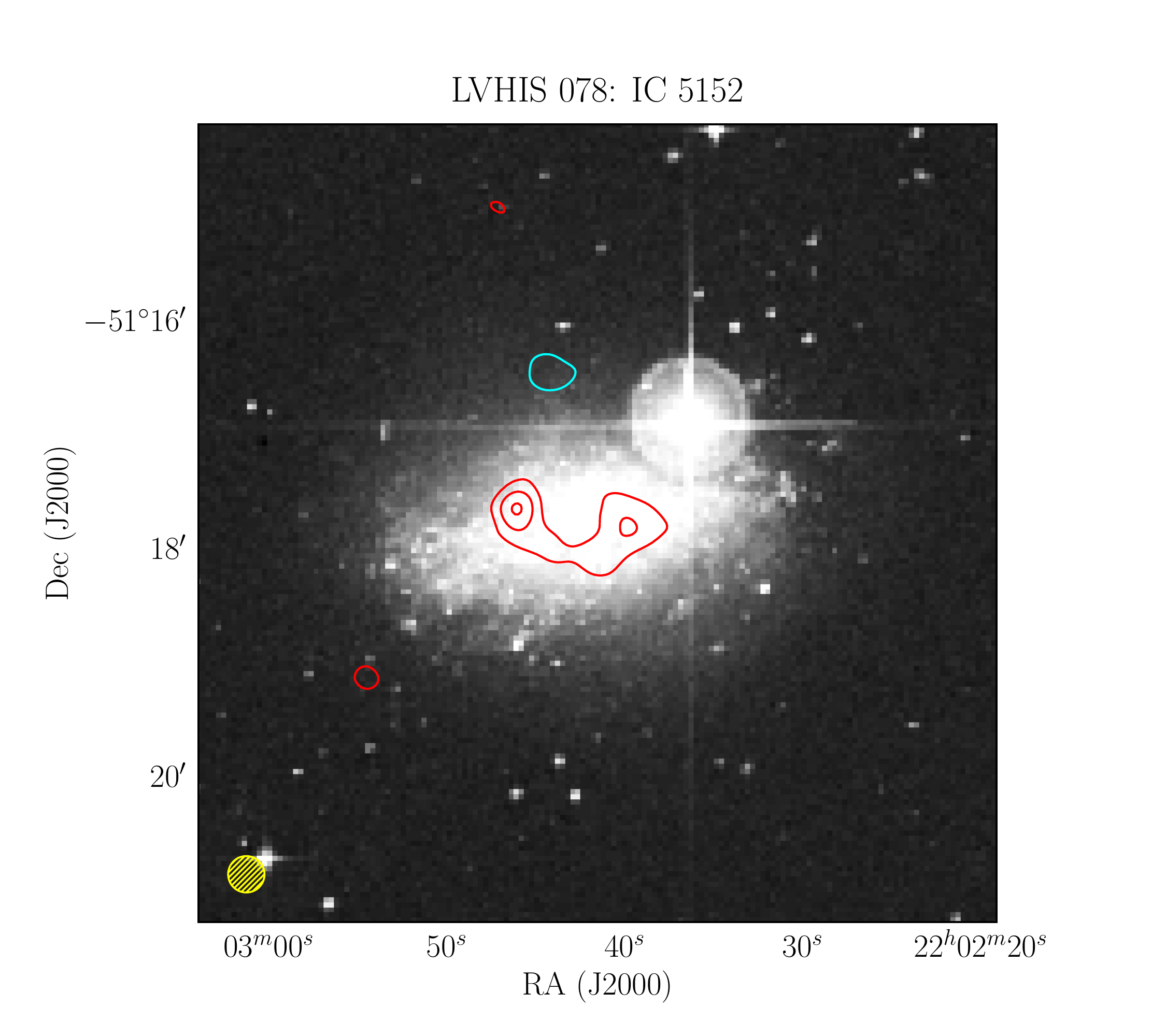} \\
\setcounter{figure}{0}
\caption{Continue: 20 cm strong LVHIS galaxies.}
\end{center}
\end{figure}
\begin{figure}[h]
\begin{center}
\includegraphics[width=0.47\linewidth]{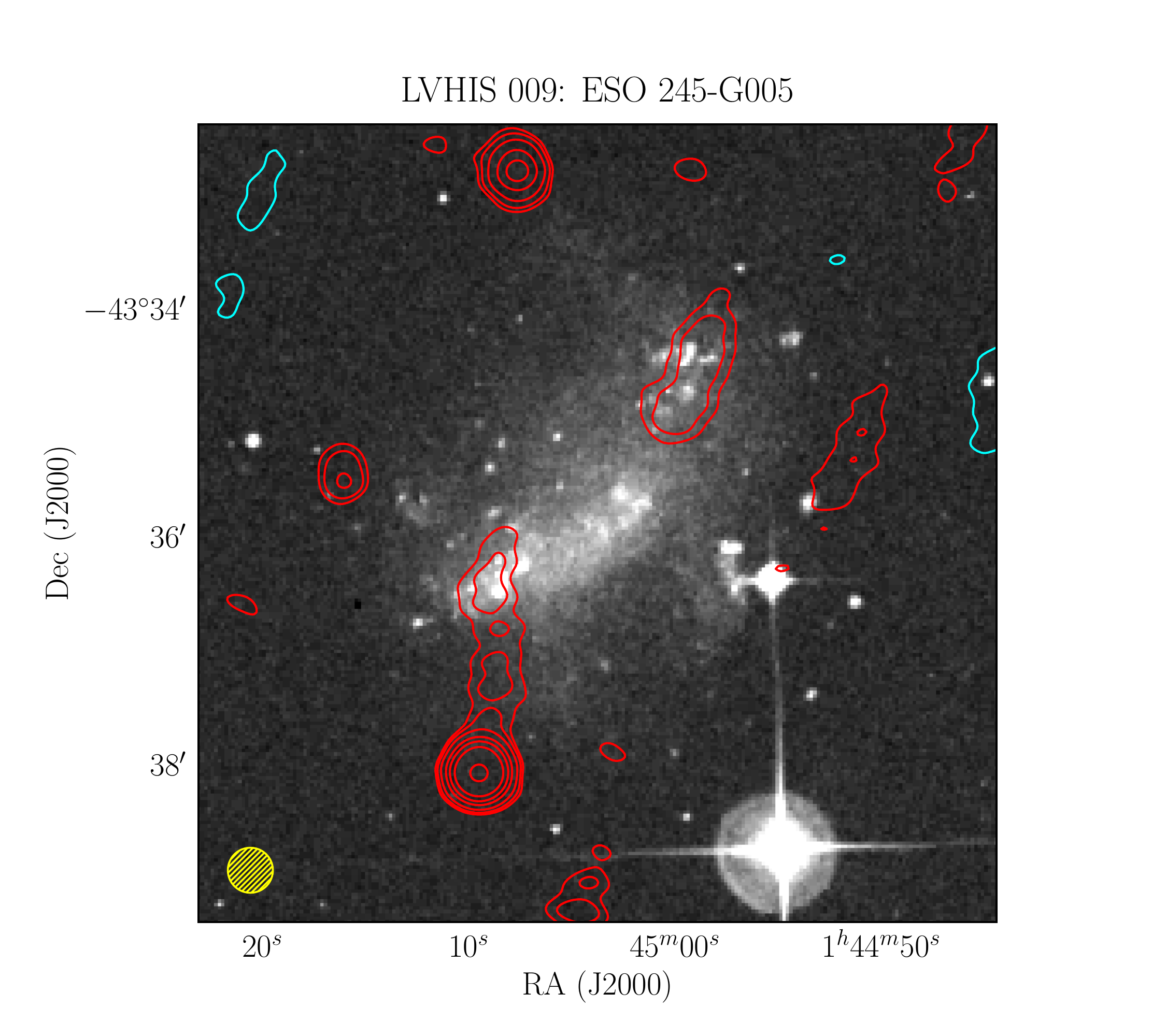} \includegraphics[width=0.47\linewidth]{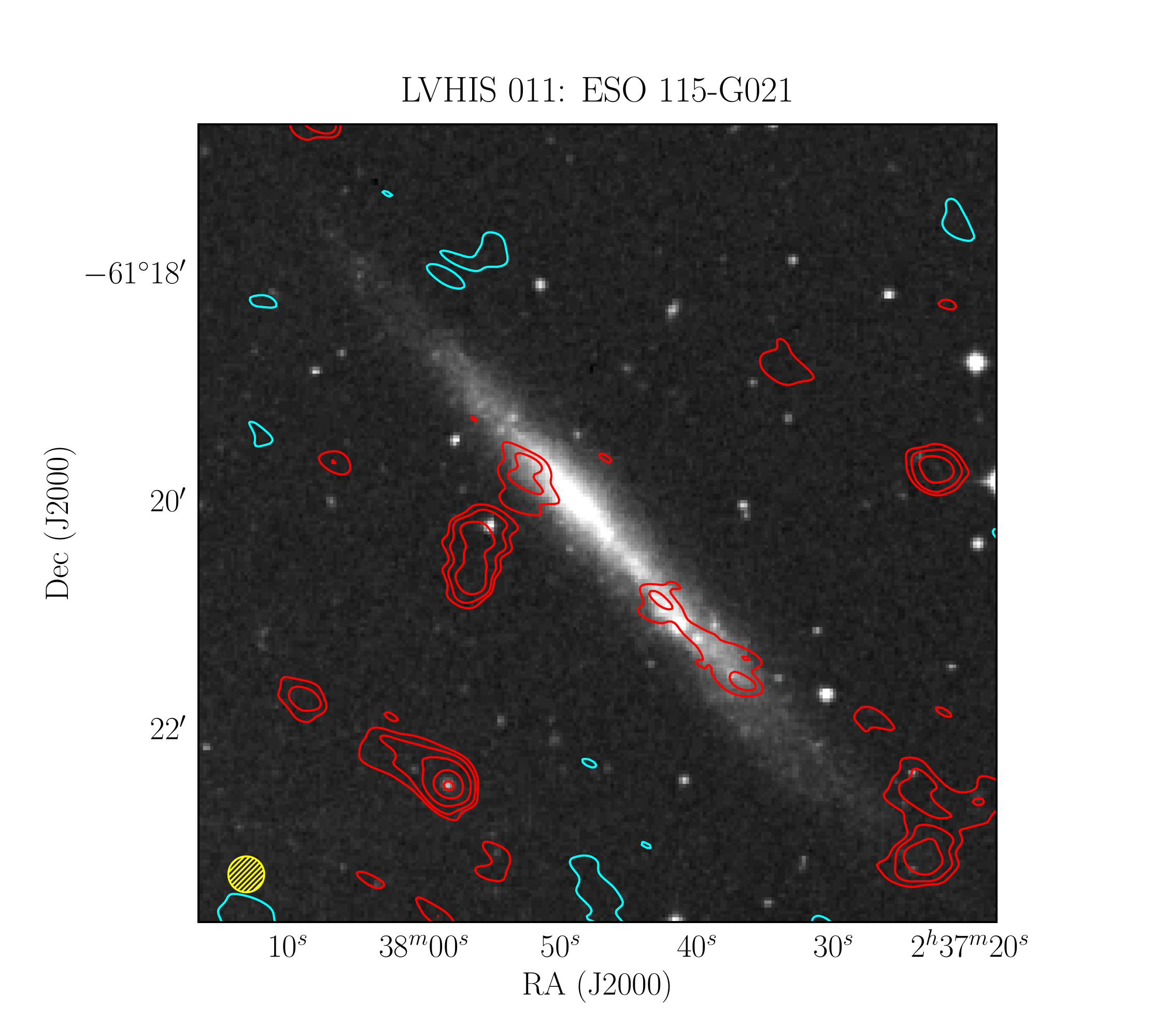} \\
\includegraphics[width=0.47\linewidth]{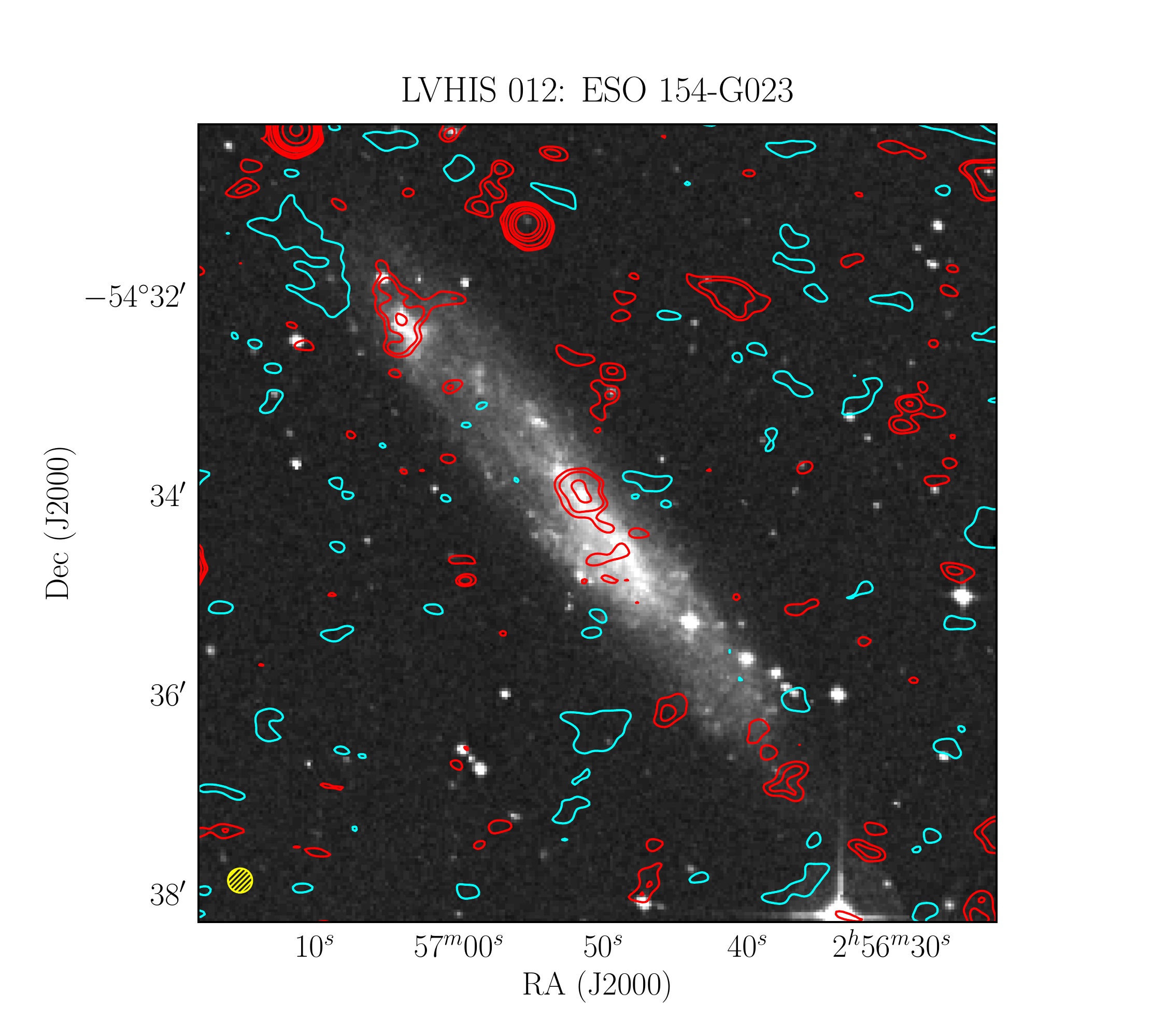} \includegraphics[width=0.47\linewidth]{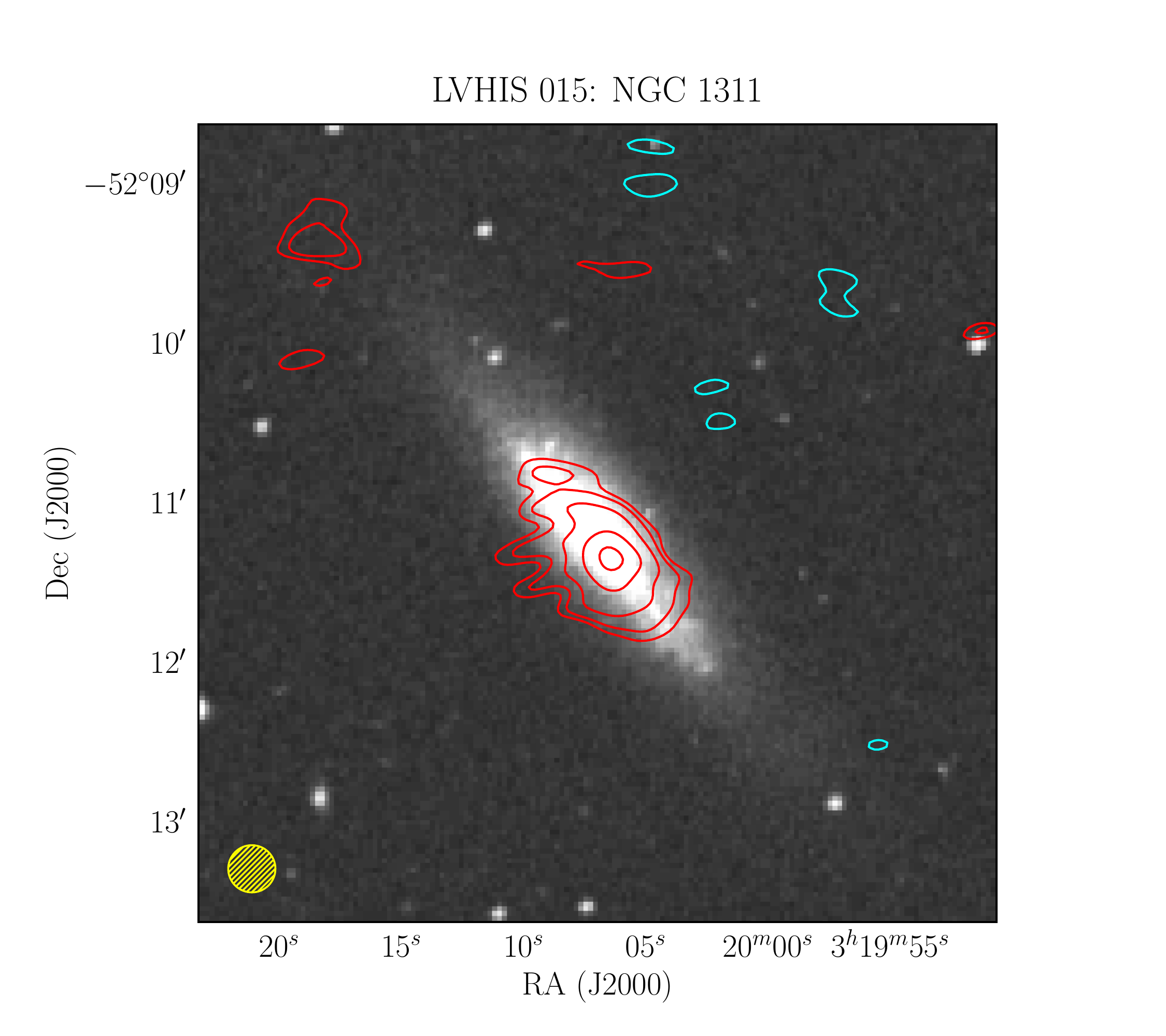} \\
\includegraphics[width=0.47\linewidth]{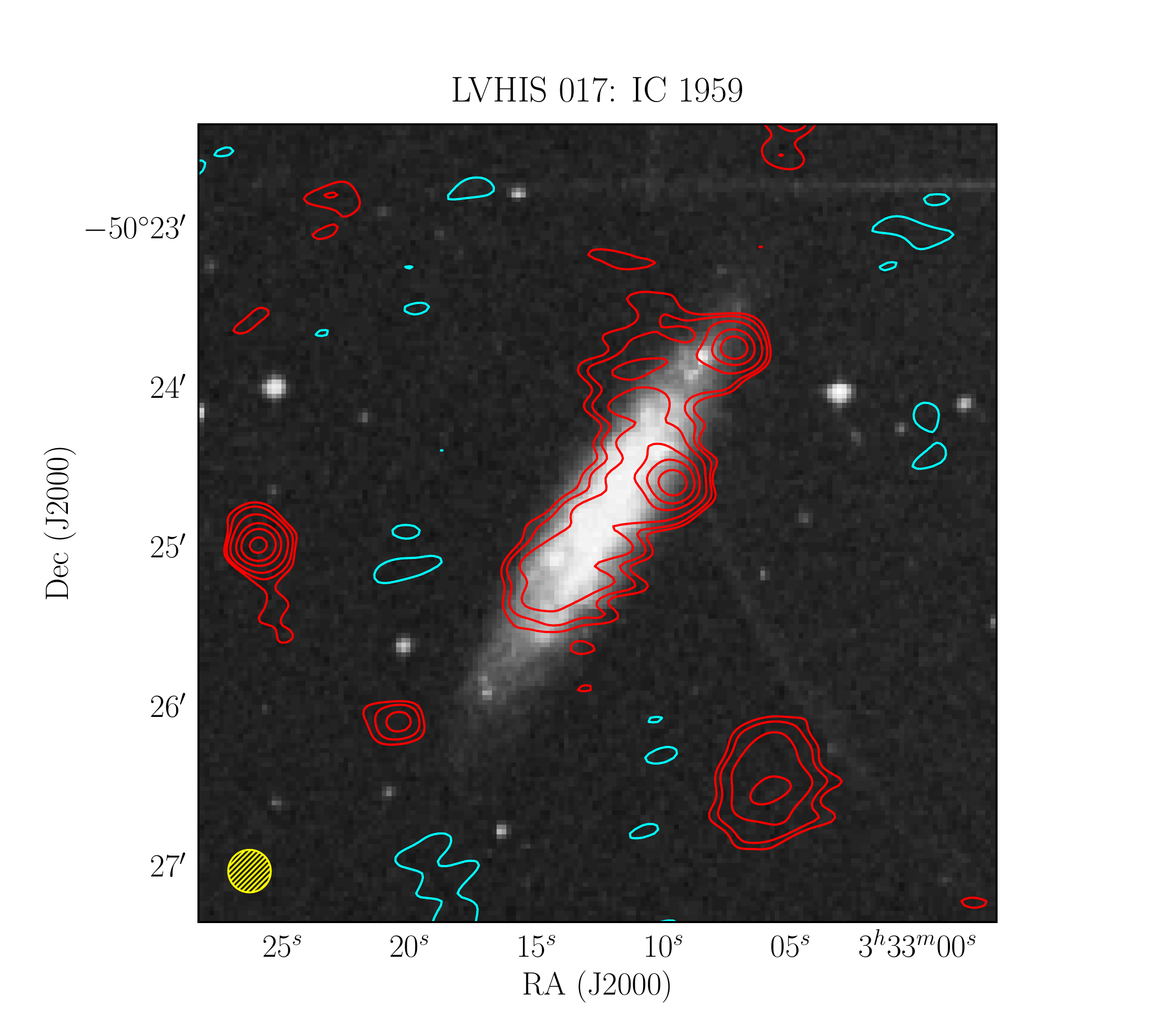} \includegraphics[width=0.47\linewidth]{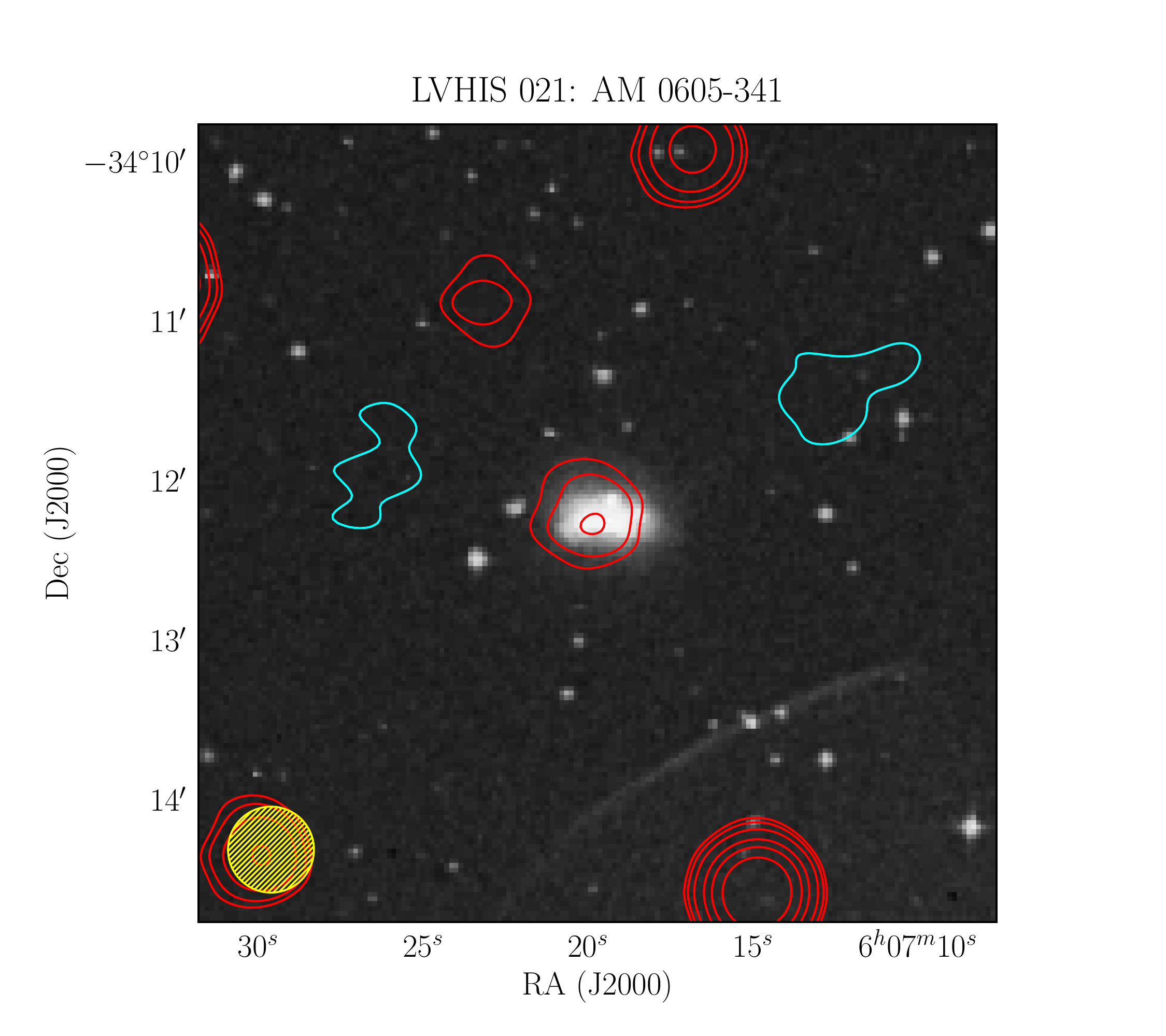} \\
\setcounter{figure}{1}
\caption{Similar to Figure \ref{fig:cont_1} but for 20 cm weak ($< 7\,\mathrm{mJy}$, but detected) LVHIS galaxies. \label{fig:cont_2}}
\end{center}
\end{figure}
\begin{figure}[h]
\begin{center}
\includegraphics[width=0.47\linewidth]{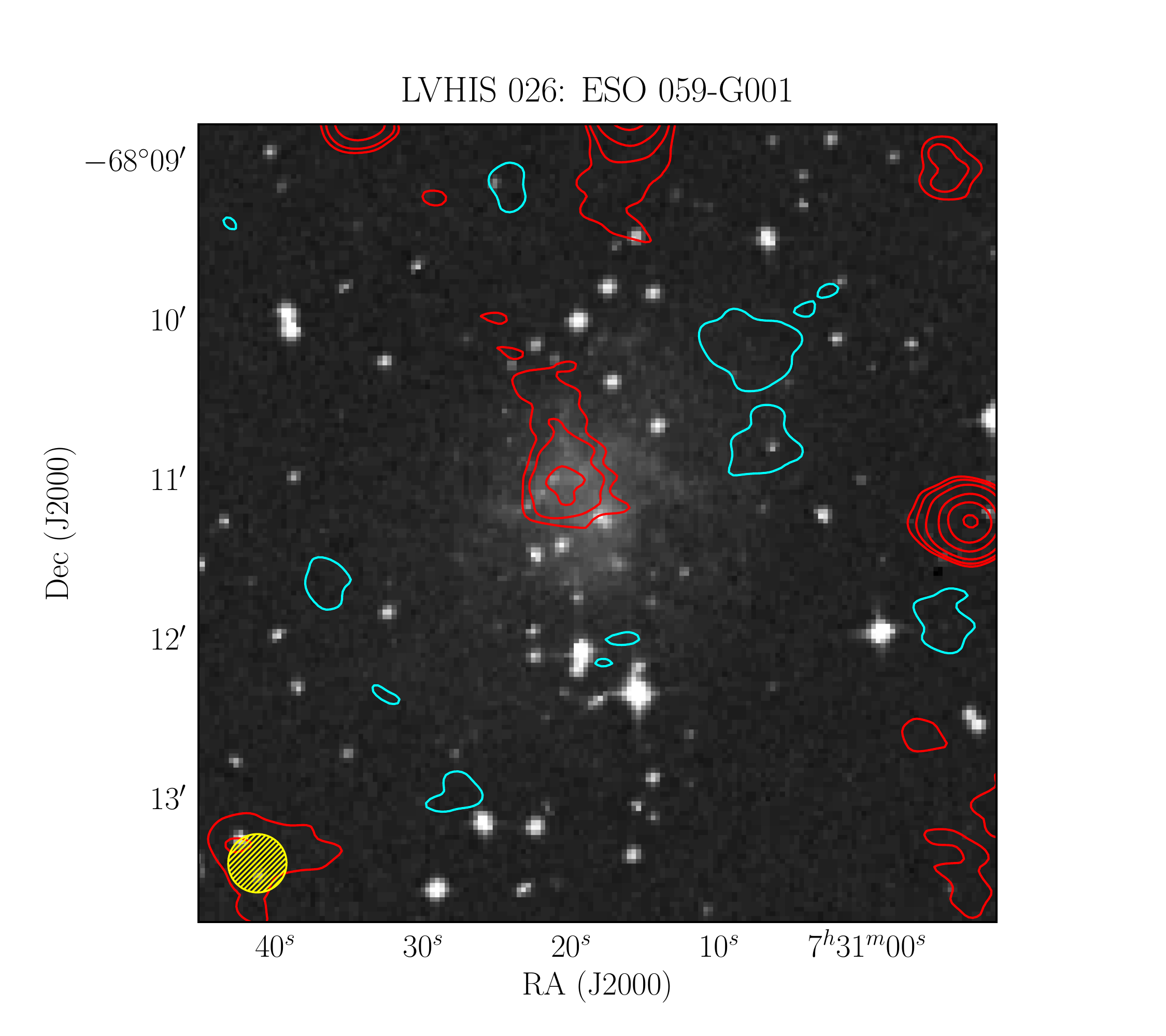} \includegraphics[width=0.47\linewidth]{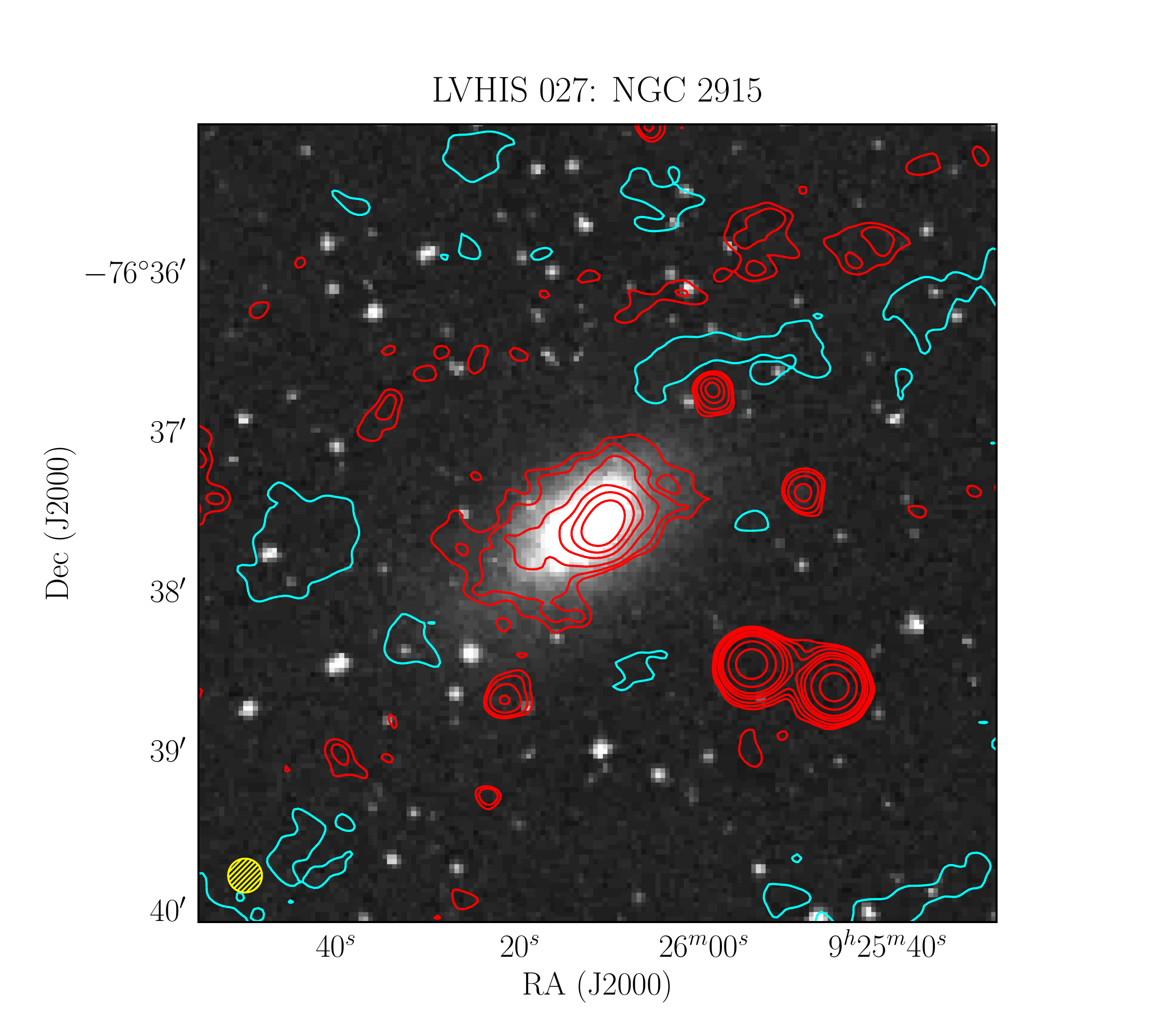} \\
\includegraphics[width=0.47\linewidth]{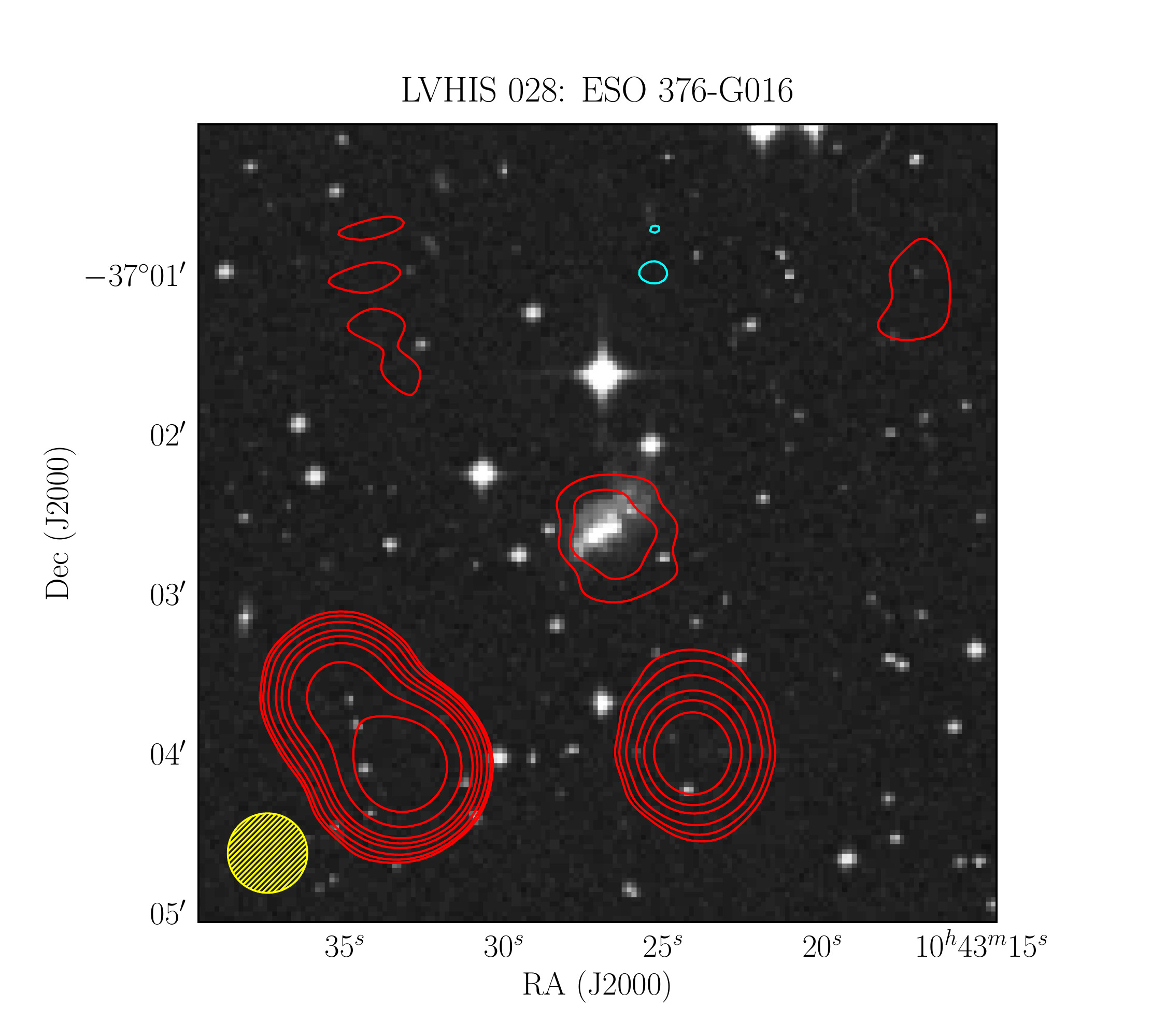} \includegraphics[width=0.47\linewidth]{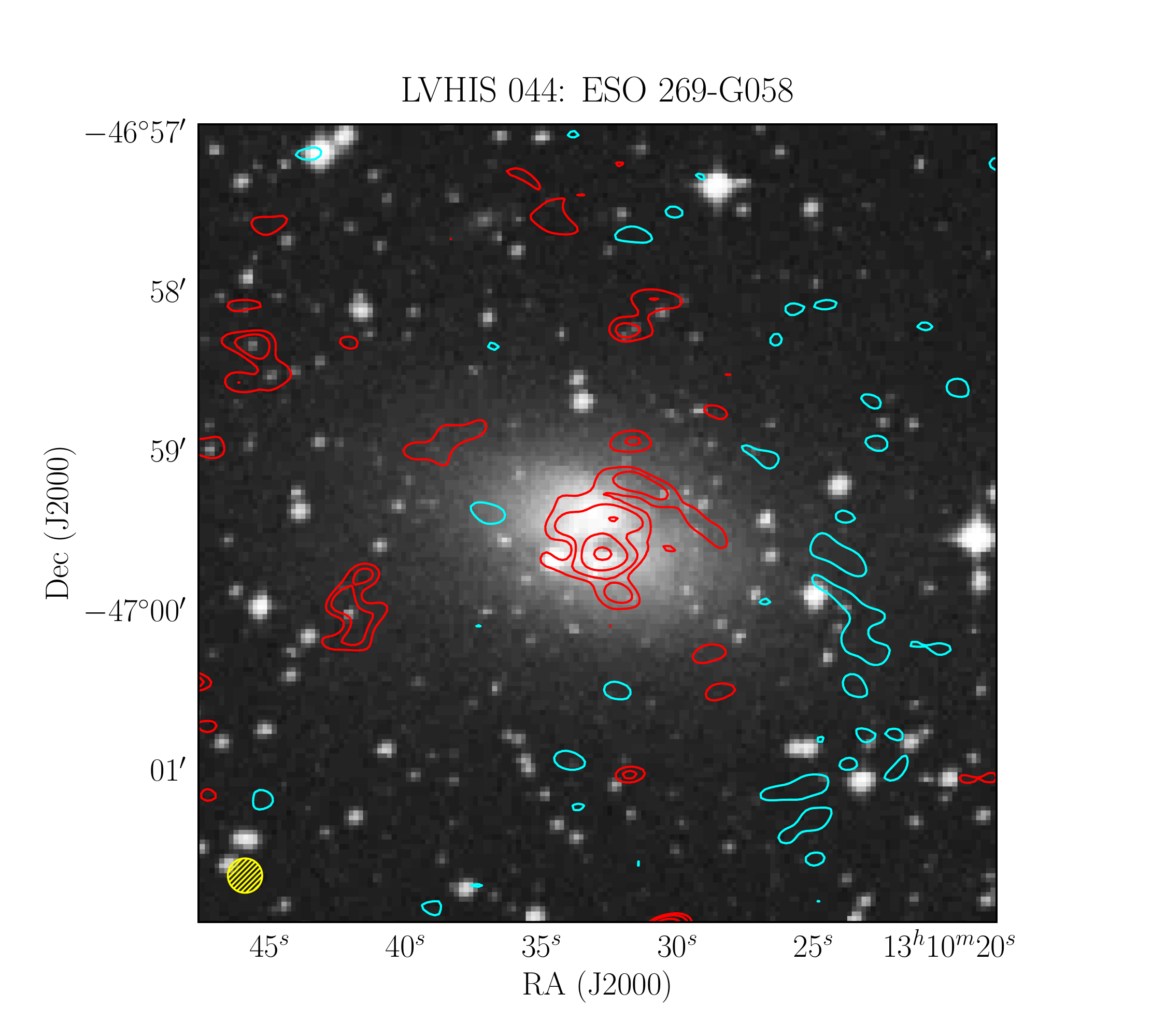} \\
\includegraphics[width=0.47\linewidth]{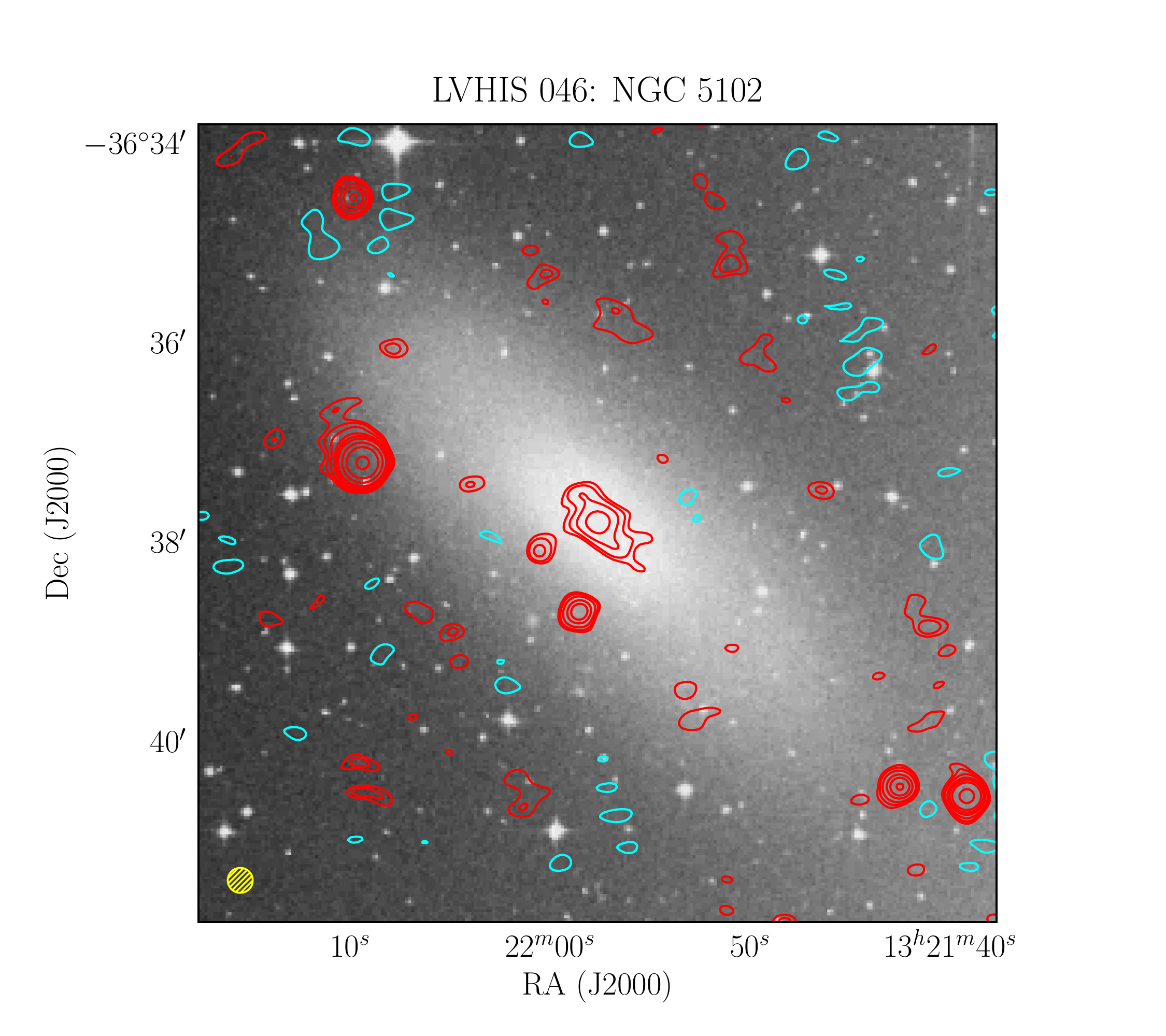} \includegraphics[width=0.47\linewidth]{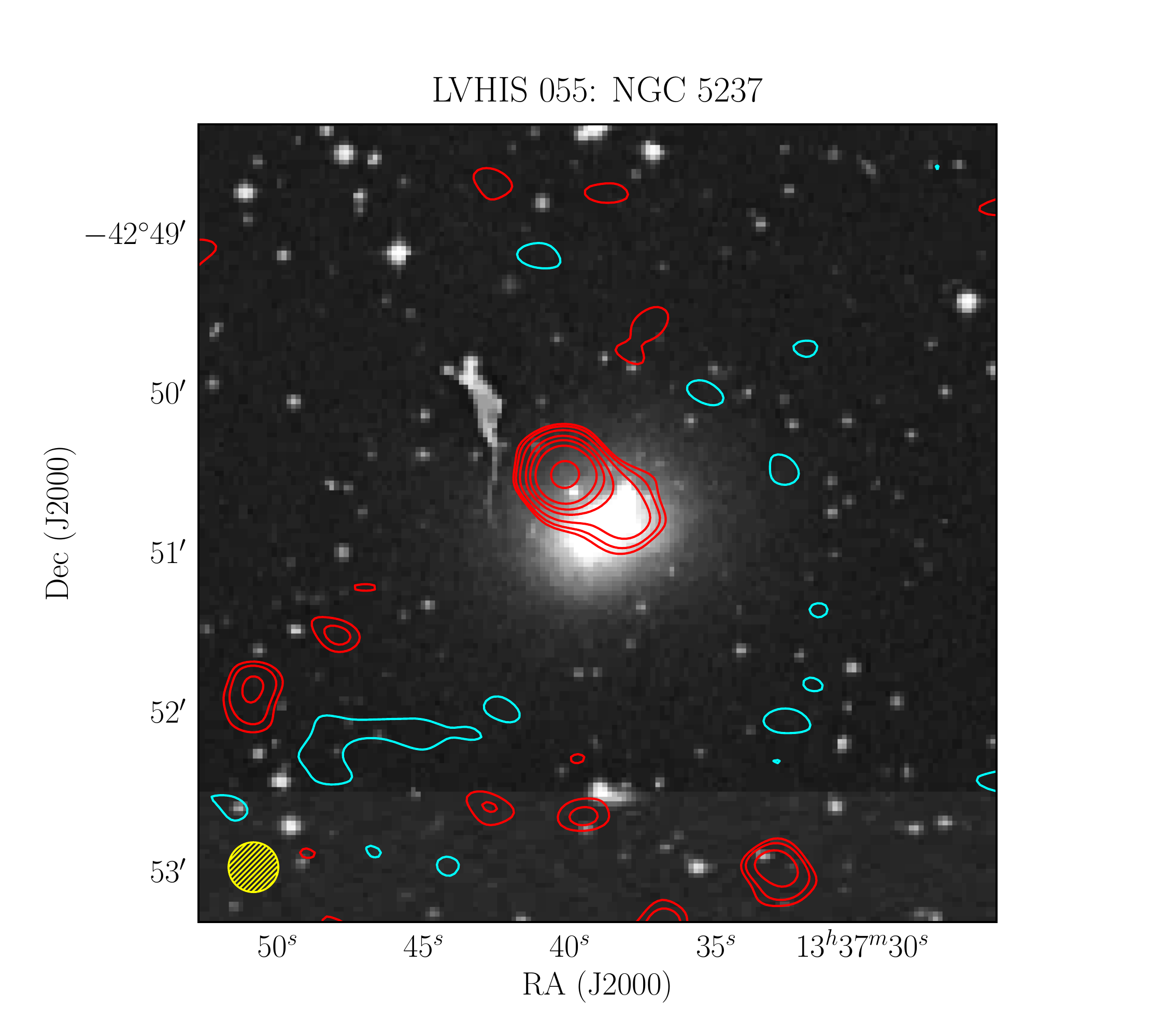} \\
\setcounter{figure}{1}
\caption{Continue: 20 cm weak LVHIS galaxies.}
\end{center}
\end{figure}
\begin{figure}[h]
\begin{center}
\includegraphics[width=0.47\linewidth]{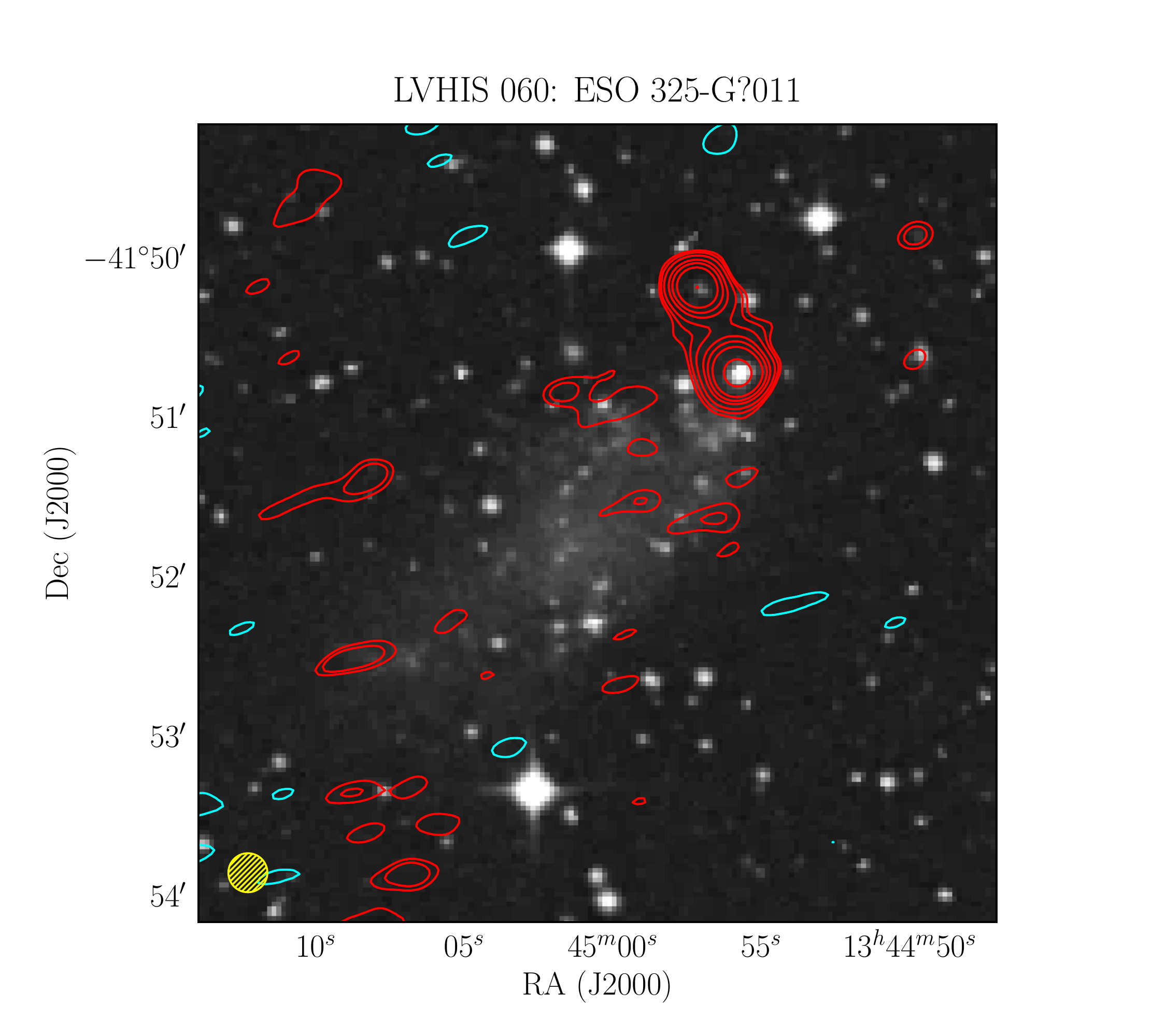} \includegraphics[width=0.47\linewidth]{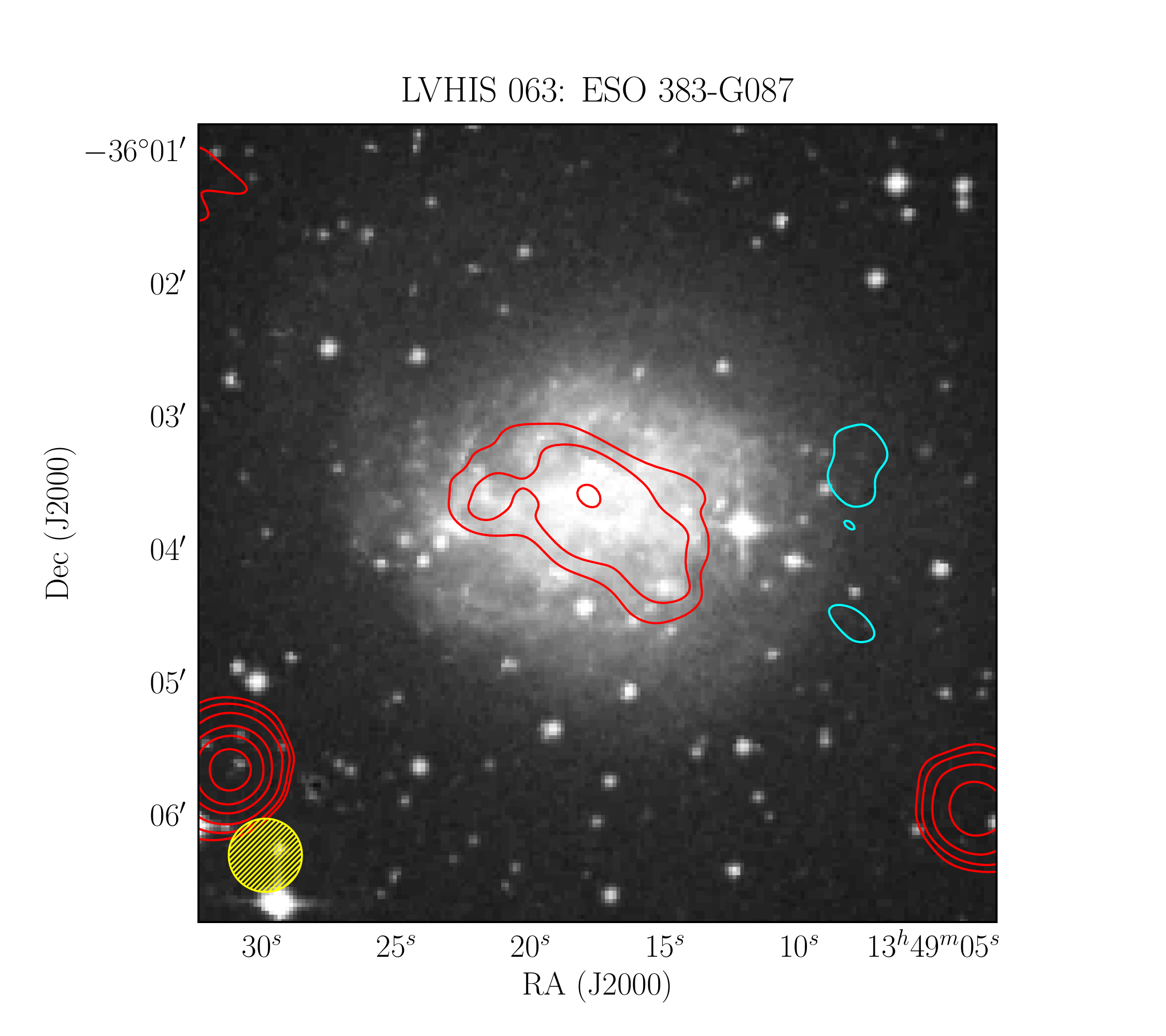} \\
\includegraphics[width=0.47\linewidth]{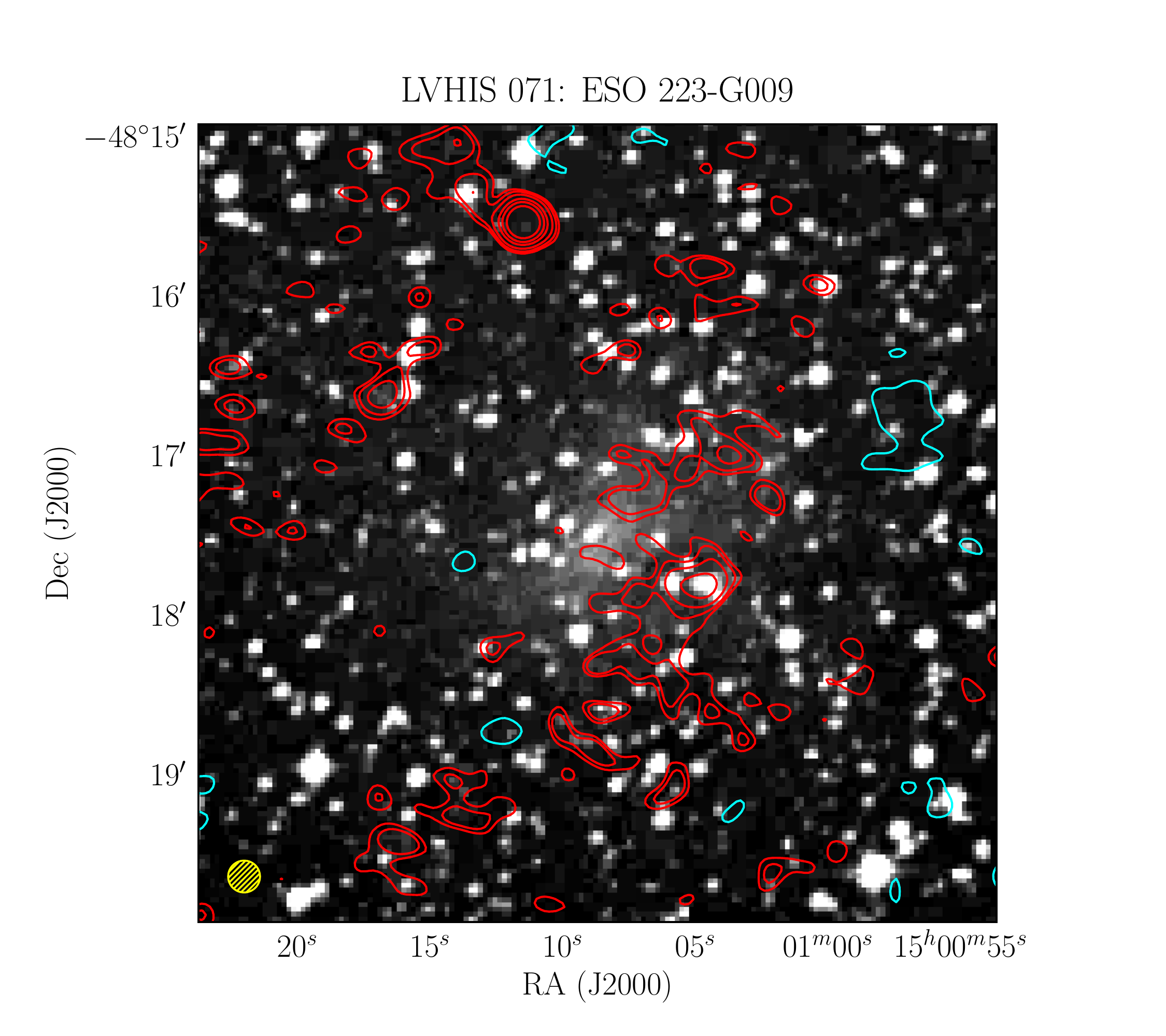} \includegraphics[width=0.47\linewidth]{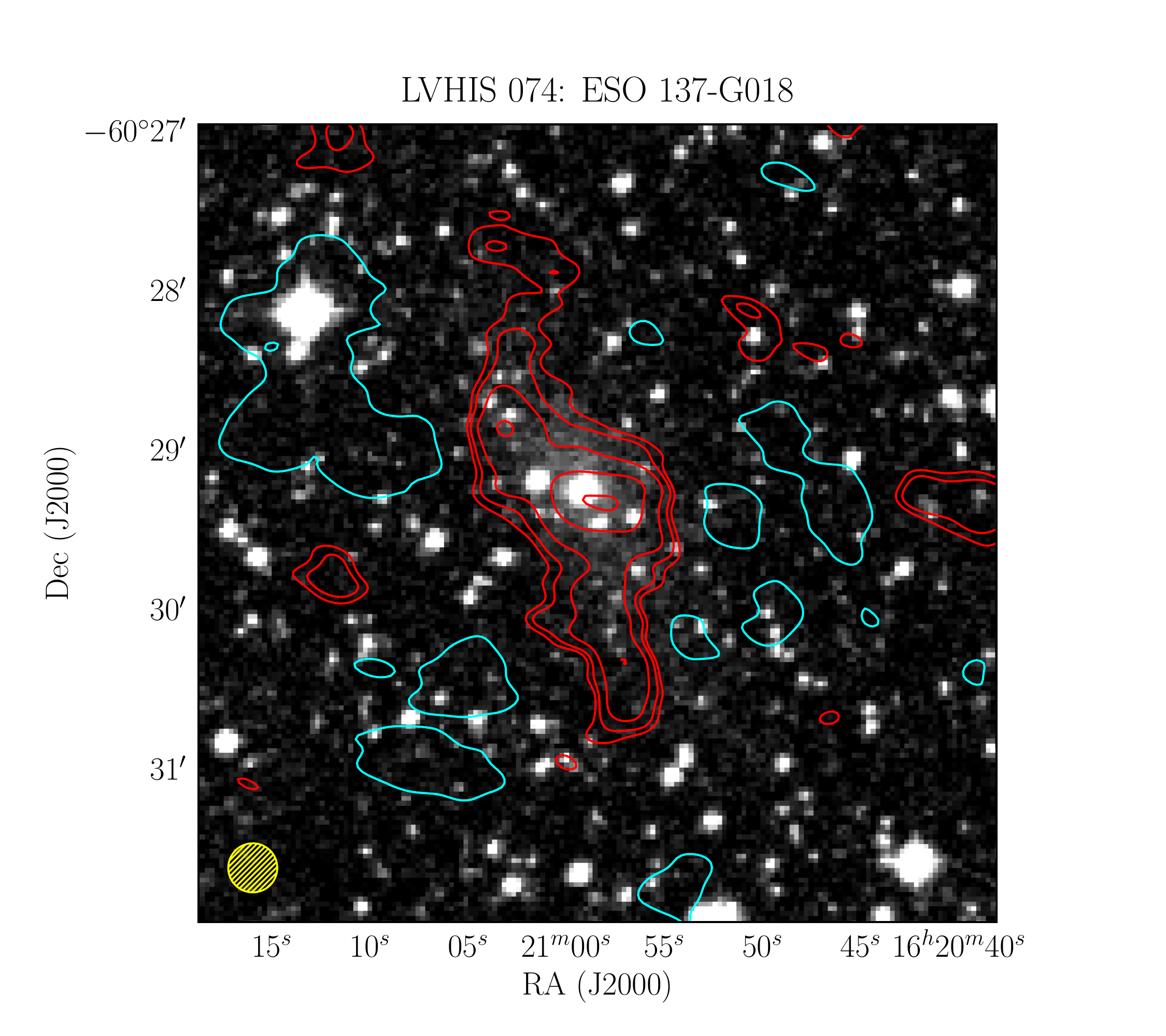} \\
\setcounter{figure}{1}
\caption{Continue: 20 cm weak LVHIS galaxies.}
\end{center}
\end{figure}

\section{Appendix: comments on individual galaxies}
\label{sec:indiv}

\subsection{LVHIS 065: NGC 5408}

In Figure \ref{fig:ngc5408} we compare the radio continuum image with Spitzer 24\mum\ image and \ha\ image. Similar to what we do for above, we plot the contours with smoothed radio data (see Section \ref{sec:radio_image}) but using the $\sigma$ estimated from the original image. The positive and negative contours are displayed in red and cyan, respectively. The negative features in \ha\ image are caused by poor broad band image subtraction, which is difficult when close to bright stars or artefacts.

The radio continuum emission is not fully consistent with the infrared or \ha\ emission. Some strong infrared/\ha\ blobs at the eastern side of the galaxy are very weak in radio. It clearly shows that the radio continuum emission does not strictly follow the star forming regions at sub-galactic scale, which makes it more complicated to explain the global FRC \citep[see also][]{2013A&A...557A.129T}.

\setcounter{figure}{0}
\begin{figure*}[hb]
  \centering \includegraphics[width=16cm]{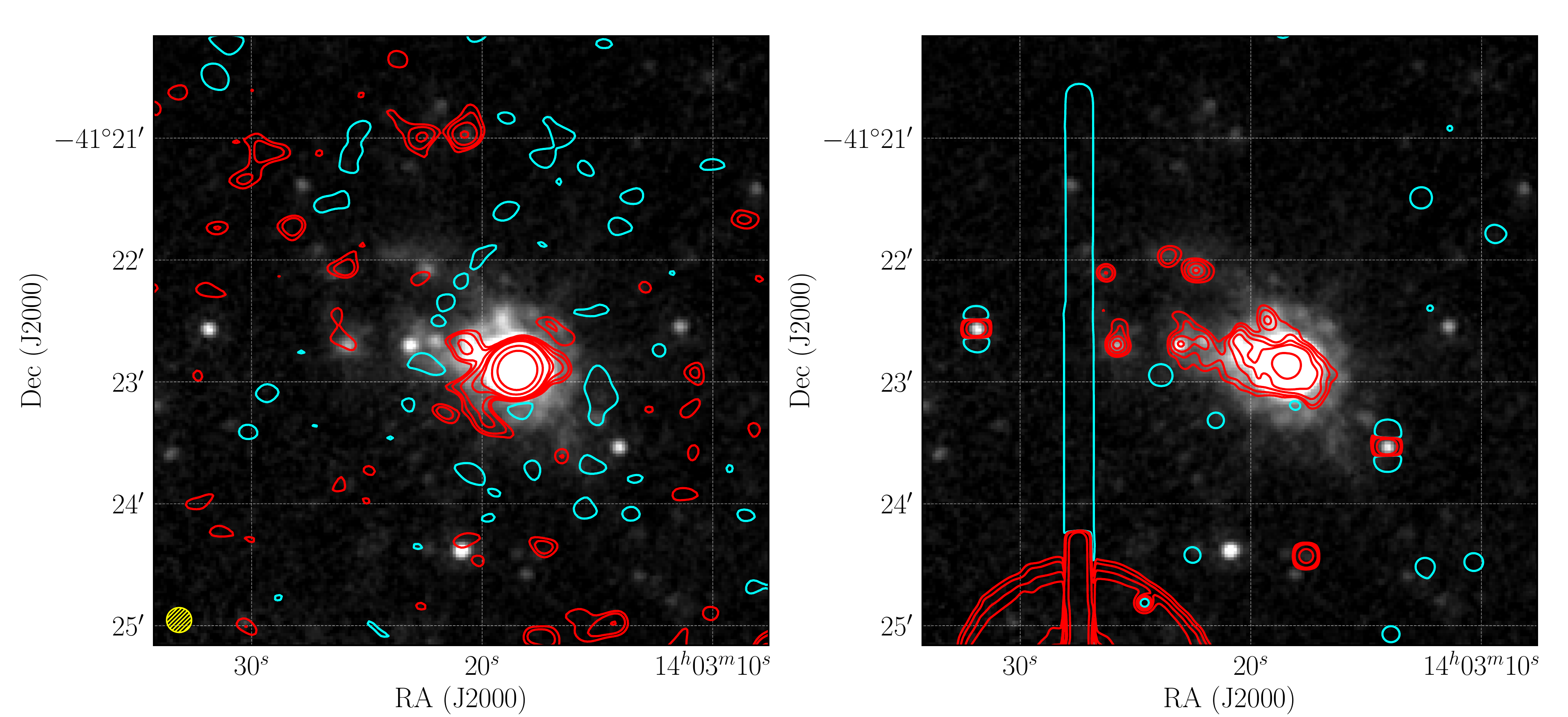}
  \caption{NGC 5408. Left: LVHIS radio continuum contours on top of Spitzer MIPS 24\mum\ image. The contour levels are -1.5, 1.5, 2, 3, 5, 7, 20 and 50-$\sigma$. Right: \ha\ contours at levels of -3, 3, 5, 9, 15 and 100-$\sigma$. See text for more details.}
  \label{fig:ngc5408}
\end{figure*}

\subsection{LVHIS 072: ESO 274--G001}

There is a strong radio source close to the optical centre of ESO 274--G001 (see Figure \ref{fig:cont_1}). However, it is unlikely associated with the galaxy itself. The \ha\ image provided by \citet{2003A&A...406..505R} shows the lack of \ha\ peak at the centre of the galaxy. The clumpy \ha\ blobs are actually distributed along the optical major axis, consistent with the extended radio continuum emission. Given the morphology, we suggest it is a strong background radio source. We subtract the central point source before calculating the radio flux density.

\subsection{LVHIS 078: IC 5152}

\begin{figure*}
  \centering \includegraphics[width=16cm]{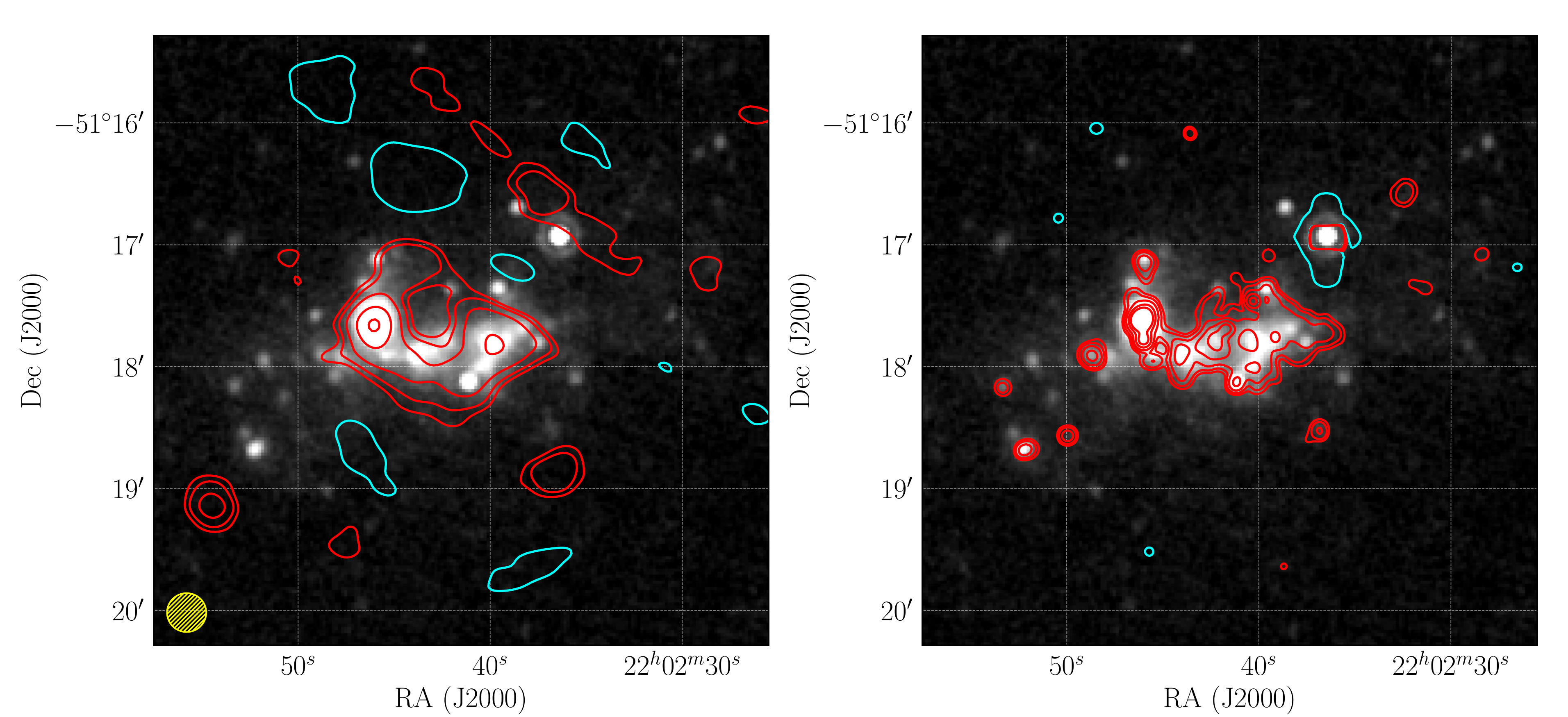}
  \caption{The same as Figure \ref{fig:ngc5408} but for IC 5152. The left panel contour levels are -1.5, 1.5, 2, 3, 5 and 7-$\sigma$. The right panel contour levels are -2, 2, 3, 5, 9 and 15-$\sigma$.}
  \label{fig:ic5152}
\end{figure*}

IC 5152 is another example of well resolved dwarf galaxy in LVHIS sample. In Figure \ref{fig:ic5152} we show the highly segmented star forming regions in this galaxy. Different from NGC 5408, the radio continuum emission is generally consistent with 24\mum\ and \ha\ emission.

\clearpage

\bibliographystyle{mnras}
\bibliography{Shao_LVHIS.bbl}

\begin{thebibliography}{}
\makeatletter
\relax
\def\mn@urlcharsother{\let\do\@makeother \do\$\do\&\do\#\do\^\do\_\do\%\do\~}
\def\mn@doi{\begingroup\mn@urlcharsother \@ifnextchar [ {\mn@doi@}
  {\mn@doi@[]}}
\def\mn@doi@[#1]#2{\def\@tempa{#1}\ifx\@tempa\@empty \href
  {http://dx.doi.org/#2} {doi:#2}\else \href {http://dx.doi.org/#2} {#1}\fi
  \endgroup}
\def\mn@eprint#1#2{\mn@eprint@#1:#2::\@nil}
\def\mn@eprint@arXiv#1{\href {http://arxiv.org/abs/#1} {{\tt arXiv:#1}}}
\def\mn@eprint@dblp#1{\href {http://dblp.uni-trier.de/rec/bibtex/#1.xml}
  {dblp:#1}}
\def\mn@eprint@#1:#2:#3:#4\@nil{\def\@tempa {#1}\def\@tempb {#2}\def\@tempc
  {#3}\ifx \@tempc \@empty \let \@tempc \@tempb \let \@tempb \@tempa \fi \ifx
  \@tempb \@empty \def\@tempb {arXiv}\fi \@ifundefined
  {mn@eprint@\@tempb}{\@tempb:\@tempc}{\expandafter \expandafter \csname
  mn@eprint@\@tempb\endcsname \expandafter{\@tempc}}}

\bibitem[\protect\citeauthoryear{{Ackermann} et~al.,}{{Ackermann}
  et~al.}{2016}]{2016A&A...586A..71A}
{Ackermann} M.,  et~al., 2016, \mn@doi [\aap] {10.1051/0004-6361/201526920},
  \href {http://adsabs.harvard.edu/abs/2016A%26A...586A..71A} {586, A71}

\bibitem[\protect\citeauthoryear{{Ascasibar}, {Gavil{\'a}n}, {Pinto}, {Casado},
  {Rosales-Ortega}  \& {D{\'{\i}}az}}{{Ascasibar}
  et~al.}{2015}]{2015MNRAS.448.2126A}
{Ascasibar} Y.,  {Gavil{\'a}n} M.,  {Pinto} N.,  {Casado} J.,  {Rosales-Ortega}
  F.,   {D{\'{\i}}az} A.~I.,  2015, \mn@doi [\mnras] {10.1093/mnras/stv098},
  \href {http://adsabs.harvard.edu/abs/2015MNRAS.448.2126A} {448, 2126}

\bibitem[\protect\citeauthoryear{{Barnes} et~al.,}{{Barnes}
  et~al.}{2001}]{2001MNRAS.322..486B}
{Barnes} D.~G.,  et~al., 2001, \mn@doi [\mnras]
  {10.1046/j.1365-8711.2001.04102.x}, \href
  {http://adsabs.harvard.edu/abs/2001MNRAS.322..486B} {322, 486}

\bibitem[\protect\citeauthoryear{{Basu}, {Roychowdhury}, {Heesen}, {Beck},
  {Brinks}, {Westcott}  \& {Hindson}}{{Basu}
  et~al.}{2017}]{2017MNRAS.471..337B}
{Basu} A.,  {Roychowdhury} S.,  {Heesen} V.,  {Beck} R.,  {Brinks} E.,
  {Westcott} J.,   {Hindson} L.,  2017, \mn@doi [\mnras]
  {10.1093/mnras/stx1567}, \href
  {http://adsabs.harvard.edu/abs/2017MNRAS.471..337B} {471, 337}

\bibitem[\protect\citeauthoryear{{Bell}}{{Bell}}{2003}]{2003ApJ...586..794B}
{Bell} E.~F.,  2003, \mn@doi [\apj] {10.1086/367829}, \href
  {http://adsabs.harvard.edu/abs/2003ApJ...586..794B} {586, 794}

\bibitem[\protect\citeauthoryear{{Boselli}, {Fossati}, {Gavazzi}, {Ciesla},
  {Buat}, {Boissier}  \& {Hughes}}{{Boselli}
  et~al.}{2015}]{2015A&A...579A.102B}
{Boselli} A.,  {Fossati} M.,  {Gavazzi} G.,  {Ciesla} L.,  {Buat} V.,
  {Boissier} S.,   {Hughes} T.~M.,  2015, \mn@doi [\aap]
  {10.1051/0004-6361/201525712}, \href
  {http://adsabs.harvard.edu/abs/2015A%26A...579A.102B} {579, A102}

\bibitem[\protect\citeauthoryear{{Bothwell}, {Maiolino}, {Kennicutt}, {Cresci},
  {Mannucci}, {Marconi}  \& {Cicone}}{{Bothwell}
  et~al.}{2013}]{2013MNRAS.433.1425B}
{Bothwell} M.~S.,  {Maiolino} R.,  {Kennicutt} R.,  {Cresci} G.,  {Mannucci}
  F.,  {Marconi} A.,   {Cicone} C.,  2013, \mn@doi [\mnras]
  {10.1093/mnras/stt817}, \href
  {http://adsabs.harvard.edu/abs/2013MNRAS.433.1425B} {433, 1425}

\bibitem[\protect\citeauthoryear{{Bothwell}, {Maiolino}, {Peng}, {Cicone},
  {Griffith}  \& {Wagg}}{{Bothwell} et~al.}{2016}]{2016MNRAS.455.1156B}
{Bothwell} M.~S.,  {Maiolino} R.,  {Peng} Y.,  {Cicone} C.,  {Griffith} H.,
  {Wagg} J.,  2016, \mn@doi [\mnras] {10.1093/mnras/stv2121}, \href
  {http://adsabs.harvard.edu/abs/2016MNRAS.455.1156B} {455, 1156}

\bibitem[\protect\citeauthoryear{{Brown} et~al.,}{{Brown}
  et~al.}{2017}]{2017ApJ...847..136B}
{Brown} M.~J.~I.,  et~al., 2017, \mn@doi [\apj] {10.3847/1538-4357/aa8ad2},
  \href {http://adsabs.harvard.edu/abs/2017ApJ...847..136B} {847, 136}

\bibitem[\protect\citeauthoryear{{Buat} et~al.,}{{Buat}
  et~al.}{2005}]{2005ApJ...619L..51B}
{Buat} V.,  et~al., 2005, \mn@doi [\apjl] {10.1086/423241}, \href
  {http://adsabs.harvard.edu/abs/2005ApJ...619L..51B} {619, L51}

\bibitem[\protect\citeauthoryear{{Calzetti}}{{Calzetti}}{2013}]{2013seg..book..419C}
{Calzetti} D.,  2013, {Star Formation Rate Indicators}.
p.~419

\bibitem[\protect\citeauthoryear{{Calzetti}, {Armus}, {Bohlin}, {Kinney},
  {Koornneef}  \& {Storchi-Bergmann}}{{Calzetti}
  et~al.}{2000}]{2000ApJ...533..682C}
{Calzetti} D.,  {Armus} L.,  {Bohlin} R.~C.,  {Kinney} A.~L.,  {Koornneef} J.,
   {Storchi-Bergmann} T.,  2000, \mn@doi [\apj] {10.1086/308692}, \href
  {http://adsabs.harvard.edu/abs/2000ApJ...533..682C} {533, 682}

\bibitem[\protect\citeauthoryear{{Calzetti} et~al.,}{{Calzetti}
  et~al.}{2010}]{2010ApJ...714.1256C}
{Calzetti} D.,  et~al., 2010, \mn@doi [\apj] {10.1088/0004-637X/714/2/1256},
  \href {http://adsabs.harvard.edu/abs/2010ApJ...714.1256C} {714, 1256}

\bibitem[\protect\citeauthoryear{{Cannon} \& {Skillman}}{{Cannon} \&
  {Skillman}}{2004}]{2004ApJ...610..772C}
{Cannon} J.~M.,  {Skillman} E.~D.,  2004, \mn@doi [\apj] {10.1086/421903},
  \href {http://adsabs.harvard.edu/abs/2004ApJ...610..772C} {610, 772}

\bibitem[\protect\citeauthoryear{{Chy{\.z}y}, {We{\.z}gowiec}, {Beck}  \&
  {Bomans}}{{Chy{\.z}y} et~al.}{2011}]{2011A&A...529A..94C}
{Chy{\.z}y} K.~T.,  {We{\.z}gowiec} M.,  {Beck} R.,   {Bomans} D.~J.,  2011,
  \mn@doi [\aap] {10.1051/0004-6361/201015393}, \href
  {http://adsabs.harvard.edu/abs/2011A%26A...529A..94C} {529, A94}

\bibitem[\protect\citeauthoryear{{Condon}}{{Condon}}{1992}]{1992ARA&A..30..575C}
{Condon} J.~J.,  1992, \mn@doi [\araa] {10.1146/annurev.aa.30.090192.003043},
  \href {http://adsabs.harvard.edu/abs/1992ARA%26A..30..575C} {30, 575}

\bibitem[\protect\citeauthoryear{{Condon}, {Helou}, {Sanders}  \&
  {Soifer}}{{Condon} et~al.}{1996}]{1996ApJS..103...81C}
{Condon} J.~J.,  {Helou} G.,  {Sanders} D.~B.,   {Soifer} B.~T.,  1996, \mn@doi
  [\apjs] {10.1086/192270}, \href
  {http://adsabs.harvard.edu/abs/1996ApJS..103...81C} {103, 81}

\bibitem[\protect\citeauthoryear{{Condon}, {Cotton}, {Greisen}, {Yin},
  {Perley}, {Taylor}  \& {Broderick}}{{Condon}
  et~al.}{1998}]{1998AJ....115.1693C}
{Condon} J.~J.,  {Cotton} W.~D.,  {Greisen} E.~W.,  {Yin} Q.~F.,  {Perley}
  R.~A.,  {Taylor} G.~B.,   {Broderick} J.~J.,  1998, \mn@doi [\aj]
  {10.1086/300337}, \href {http://adsabs.harvard.edu/abs/1998AJ....115.1693C}
  {115, 1693}

\bibitem[\protect\citeauthoryear{{Condon}, {Cotton}  \& {Broderick}}{{Condon}
  et~al.}{2002}]{2002AJ....124..675C}
{Condon} J.~J.,  {Cotton} W.~D.,   {Broderick} J.~J.,  2002, \mn@doi [\aj]
  {10.1086/341650}, \href {http://adsabs.harvard.edu/abs/2002AJ....124..675C}
  {124, 675}

\bibitem[\protect\citeauthoryear{{Dole} et~al.,}{{Dole}
  et~al.}{2006}]{2006A&A...451..417D}
{Dole} H.,  et~al., 2006, \mn@doi [\aap] {10.1051/0004-6361:20054446}, \href
  {http://adsabs.harvard.edu/abs/2006A%26A...451..417D} {451, 417}

\bibitem[\protect\citeauthoryear{{Driver} et~al.,}{{Driver}
  et~al.}{2011}]{2011MNRAS.413..971D}
{Driver} S.~P.,  et~al., 2011, \mn@doi [\mnras]
  {10.1111/j.1365-2966.2010.18188.x}, \href
  {http://adsabs.harvard.edu/abs/2011MNRAS.413..971D} {413, 971}

\bibitem[\protect\citeauthoryear{{Elmouttie}, {Haynes}, {Jones}, {Ehle},
  {Beck}, {Harnett}  \& {Wielebinski}}{{Elmouttie}
  et~al.}{1997}]{1997MNRAS.284..830E}
{Elmouttie} M.,  {Haynes} R.~F.,  {Jones} K.~L.,  {Ehle} M.,  {Beck} R.,
  {Harnett} J.~I.,   {Wielebinski} R.,  1997, \mn@doi [\mnras]
  {10.1093/mnras/284.4.830}, \href
  {http://adsabs.harvard.edu/abs/1997MNRAS.284..830E} {284, 830}

\bibitem[\protect\citeauthoryear{{Faisst} et~al.,}{{Faisst}
  et~al.}{2016}]{2016ApJ...822...29F}
{Faisst} A.~L.,  et~al., 2016, \mn@doi [\apj] {10.3847/0004-637X/822/1/29},
  \href {http://adsabs.harvard.edu/abs/2016ApJ...822...29F} {822, 29}

\bibitem[\protect\citeauthoryear{{Galametz}, {Madden}, {Galliano}, {Hony},
  {Bendo}  \& {Sauvage}}{{Galametz} et~al.}{2011}]{2011A&A...532A..56G}
{Galametz} M.,  {Madden} S.~C.,  {Galliano} F.,  {Hony} S.,  {Bendo} G.~J.,
  {Sauvage} M.,  2011, \mn@doi [\aap] {10.1051/0004-6361/201014904}, \href
  {http://adsabs.harvard.edu/abs/2011A%26A...532A..56G} {532, A56}

\bibitem[\protect\citeauthoryear{{Grootes} et~al.,}{{Grootes}
  et~al.}{2013}]{2013ApJ...766...59G}
{Grootes} M.~W.,  et~al., 2013, \mn@doi [\apj] {10.1088/0004-637X/766/1/59},
  \href {http://adsabs.harvard.edu/abs/2013ApJ...766...59G} {766, 59}

\bibitem[\protect\citeauthoryear{{Heckman}, {Robert}, {Leitherer}, {Garnett}
  \& {van der Rydt}}{{Heckman} et~al.}{1998}]{1998ApJ...503..646H}
{Heckman} T.~M.,  {Robert} C.,  {Leitherer} C.,  {Garnett} D.~R.,   {van der
  Rydt} F.,  1998, \mn@doi [\apj] {10.1086/306035}, \href
  {http://adsabs.harvard.edu/abs/1998ApJ...503..646H} {503, 646}

\bibitem[\protect\citeauthoryear{{Heesen}, {Rau}, {Rupen}, {Brinks}  \&
  {Hunter}}{{Heesen} et~al.}{2011}]{2011ApJ...739L..23H}
{Heesen} V.,  {Rau} U.,  {Rupen} M.~P.,  {Brinks} E.,   {Hunter} D.~A.,  2011,
  \mn@doi [\apjl] {10.1088/2041-8205/739/1/L23}, \href
  {http://adsabs.harvard.edu/abs/2011ApJ...739L..23H} {739, L23}

\bibitem[\protect\citeauthoryear{{Heesen}, {Brinks}, {Leroy}, {Heald}, {Braun},
  {Bigiel}  \& {Beck}}{{Heesen} et~al.}{2014}]{2014AJ....147..103H}
{Heesen} V.,  {Brinks} E.,  {Leroy} A.~K.,  {Heald} G.,  {Braun} R.,  {Bigiel}
  F.,   {Beck} R.,  2014, \mn@doi [\aj] {10.1088/0004-6256/147/5/103}, \href
  {http://adsabs.harvard.edu/abs/2014AJ....147..103H} {147, 103}

\bibitem[\protect\citeauthoryear{{Helou} \& {Bicay}}{{Helou} \&
  {Bicay}}{1993}]{1993ApJ...415...93H}
{Helou} G.,  {Bicay} M.~D.,  1993, \mn@doi [\apj] {10.1086/173146}, \href
  {http://adsabs.harvard.edu/abs/1993ApJ...415...93H} {415, 93}

\bibitem[\protect\citeauthoryear{{Helou}, {Soifer}  \&
  {Rowan-Robinson}}{{Helou} et~al.}{1985}]{1985ApJ...298L...7H}
{Helou} G.,  {Soifer} B.~T.,   {Rowan-Robinson} M.,  1985, \mn@doi [\apjl]
  {10.1086/184556}, \href {http://adsabs.harvard.edu/abs/1985ApJ...298L...7H}
  {298, L7}

\bibitem[\protect\citeauthoryear{{Helou}, {Khan}, {Malek}  \&
  {Boehmer}}{{Helou} et~al.}{1988}]{1988ApJS...68..151H}
{Helou} G.,  {Khan} I.~R.,  {Malek} L.,   {Boehmer} L.,  1988, \mn@doi [\apjs]
  {10.1086/191285}, \href {http://adsabs.harvard.edu/abs/1988ApJS...68..151H}
  {68, 151}

\bibitem[\protect\citeauthoryear{{Hopkins}, {Miller}, {Connolly}, {Genovese},
  {Nichol}  \& {Wasserman}}{{Hopkins} et~al.}{2002}]{2002AJ....123.1086H}
{Hopkins} A.~M.,  {Miller} C.~J.,  {Connolly} A.~J.,  {Genovese} C.,  {Nichol}
  R.~C.,   {Wasserman} L.,  2002, \mn@doi [\aj] {10.1086/338316}, \href
  {http://adsabs.harvard.edu/abs/2002AJ....123.1086H} {123, 1086}

\bibitem[\protect\citeauthoryear{{Hughes}, {Wong}, {Ekers}, {Staveley-Smith},
  {Filipovic}, {Maddison}, {Fukui}  \& {Mizuno}}{{Hughes}
  et~al.}{2006}]{2006MNRAS.370..363H}
{Hughes} A.,  {Wong} T.,  {Ekers} R.,  {Staveley-Smith} L.,  {Filipovic} M.,
  {Maddison} S.,  {Fukui} Y.,   {Mizuno} N.,  2006, \mn@doi [\mnras]
  {10.1111/j.1365-2966.2006.10483.x}, \href
  {http://adsabs.harvard.edu/abs/2006MNRAS.370..363H} {370, 363}

\bibitem[\protect\citeauthoryear{{Hughes}, {Cortese}, {Boselli}, {Gavazzi}  \&
  {Davies}}{{Hughes} et~al.}{2013}]{2013A&A...550A.115H}
{Hughes} T.~M.,  {Cortese} L.,  {Boselli} A.,  {Gavazzi} G.,   {Davies} J.~I.,
  2013, \mn@doi [\aap] {10.1051/0004-6361/201218822}, \href
  {http://adsabs.harvard.edu/abs/2013A%26A...550A.115H} {550, A115}

\bibitem[\protect\citeauthoryear{{Jarrett} et~al.,}{{Jarrett}
  et~al.}{2013}]{2013AJ....145....6J}
{Jarrett} T.~H.,  et~al., 2013, \mn@doi [\aj] {10.1088/0004-6256/145/1/6},
  \href {http://adsabs.harvard.edu/abs/2013AJ....145....6J} {145, 6}

\bibitem[\protect\citeauthoryear{{Kennicutt}}{{Kennicutt}}{1998a}]{1998ARA&A..36..189K}
{Kennicutt} Jr. R.~C.,  1998a, \mn@doi [\araa]
  {10.1146/annurev.astro.36.1.189}, \href
  {http://adsabs.harvard.edu/abs/1998ARA%26A..36..189K} {36, 189}

\bibitem[\protect\citeauthoryear{{Kennicutt}}{{Kennicutt}}{1998b}]{1998ApJ...498..541K}
{Kennicutt} Jr. R.~C.,  1998b, \mn@doi [\apj] {10.1086/305588}, \href
  {http://adsabs.harvard.edu/abs/1998ApJ...498..541K} {498, 541}

\bibitem[\protect\citeauthoryear{{Koribalski} et~al.,}{{Koribalski}
  et~al.}{2018}]{2018MNRAS.tmp..467K}
{Koribalski} B.~S.,  et~al., 2018, \mn@doi [\mnras] {10.1093/mnras/sty479},
  \href {http://adsabs.harvard.edu/abs/2018MNRAS.tmp..467K} {}

\bibitem[\protect\citeauthoryear{{Kroupa}}{{Kroupa}}{2001}]{2001MNRAS.322..231K}
{Kroupa} P.,  2001, \mn@doi [\mnras] {10.1046/j.1365-8711.2001.04022.x}, \href
  {http://adsabs.harvard.edu/abs/2001MNRAS.322..231K} {322, 231}

\bibitem[\protect\citeauthoryear{{Lacki} \& {Thompson}}{{Lacki} \&
  {Thompson}}{2010}]{2010ApJ...717..196L}
{Lacki} B.~C.,  {Thompson} T.~A.,  2010, \mn@doi [\apj]
  {10.1088/0004-637X/717/1/196}, \href
  {http://adsabs.harvard.edu/abs/2010ApJ...717..196L} {717, 196}

\bibitem[\protect\citeauthoryear{{Lacki}, {Thompson}  \& {Quataert}}{{Lacki}
  et~al.}{2010}]{2010ApJ...717....1L}
{Lacki} B.~C.,  {Thompson} T.~A.,   {Quataert} E.,  2010, \mn@doi [\apj]
  {10.1088/0004-637X/717/1/1}, \href
  {http://adsabs.harvard.edu/abs/2010ApJ...717....1L} {717, 1}

\bibitem[\protect\citeauthoryear{{Lara-L{\'o}pez} et~al.,}{{Lara-L{\'o}pez}
  et~al.}{2013}]{2013MNRAS.433L..35L}
{Lara-L{\'o}pez} M.~A.,  et~al., 2013, \mn@doi [\mnras]
  {10.1093/mnrasl/slt054}, \href
  {http://adsabs.harvard.edu/abs/2013MNRAS.433L..35L} {433, L35}

\bibitem[\protect\citeauthoryear{{Lauberts} \& {Valentijn}}{{Lauberts} \&
  {Valentijn}}{1989}]{1989spce.book.....L}
{Lauberts} A.,  {Valentijn} E.~A.,  1989, {The surface photometry catalogue of
  the ESO-Uppsala galaxies}

\bibitem[\protect\citeauthoryear{{Lisenfeld}, {Voelk}  \& {Xu}}{{Lisenfeld}
  et~al.}{1996}]{1996A&A...306..677L}
{Lisenfeld} U.,  {Voelk} H.~J.,   {Xu} C.,  1996, \aap, \href
  {http://adsabs.harvard.edu/abs/1996A%26A...306..677L} {306, 677}

\bibitem[\protect\citeauthoryear{{Martin} et~al.,}{{Martin}
  et~al.}{2005}]{2005ApJ...619L...1M}
{Martin} D.~C.,  et~al., 2005, \mn@doi [\apjl] {10.1086/426387}, \href
  {http://adsabs.harvard.edu/abs/2005ApJ...619L...1M} {619, L1}

\bibitem[\protect\citeauthoryear{{Murphy}}{{Murphy}}{2009}]{2009ApJ...706..482M}
{Murphy} E.~J.,  2009, \mn@doi [\apj] {10.1088/0004-637X/706/1/482}, \href
  {http://adsabs.harvard.edu/abs/2009ApJ...706..482M} {706, 482}

\bibitem[\protect\citeauthoryear{{Murphy} et~al.,}{{Murphy}
  et~al.}{2011}]{2011ApJ...737...67M}
{Murphy} E.~J.,  et~al., 2011, \mn@doi [\apj] {10.1088/0004-637X/737/2/67},
  \href {http://adsabs.harvard.edu/abs/2011ApJ...737...67M} {737, 67}

\bibitem[\protect\citeauthoryear{{Neugebauer} et~al.,}{{Neugebauer}
  et~al.}{1984}]{1984ApJ...278L...1N}
{Neugebauer} G.,  et~al., 1984, \mn@doi [\apjl] {10.1086/184209}, \href
  {http://adsabs.harvard.edu/abs/1984ApJ...278L...1N} {278, L1}

\bibitem[\protect\citeauthoryear{{Niklas} \& {Beck}}{{Niklas} \&
  {Beck}}{1997}]{1997A&A...320...54N}
{Niklas} S.,  {Beck} R.,  1997, \aap, \href
  {http://adsabs.harvard.edu/abs/1997A%26A...320...54N} {320, 54}

\bibitem[\protect\citeauthoryear{{Parker}}{{Parker}}{1966}]{1966ApJ...145..811P}
{Parker} E.~N.,  1966, \mn@doi [\apj] {10.1086/148828}, \href
  {http://adsabs.harvard.edu/abs/1966ApJ...145..811P} {145, 811}

\bibitem[\protect\citeauthoryear{{Paturel}, {Fouque}, {Lauberts}, {Valentijn},
  {Corwin}  \& {de Vaucouleurs}}{{Paturel} et~al.}{1987}]{1987A&A...184...86P}
{Paturel} G.,  {Fouque} P.,  {Lauberts} A.,  {Valentijn} E.~A.,  {Corwin}
  H.~G.,   {de Vaucouleurs} G.,  1987, \aap, \href
  {http://adsabs.harvard.edu/abs/1987A%26A...184...86P} {184, 86}

\bibitem[\protect\citeauthoryear{{Qiu}, {Shi}, {Wang}, {Zhang}  \&
  {Zhou}}{{Qiu} et~al.}{2017}]{2017ApJ...846...68Q}
{Qiu} J.,  {Shi} Y.,  {Wang} J.,  {Zhang} Z.-Y.,   {Zhou} L.,  2017, \mn@doi
  [\apj] {10.3847/1538-4357/aa832c}, \href
  {http://adsabs.harvard.edu/abs/2017ApJ...846...68Q} {846, 68}

\bibitem[\protect\citeauthoryear{{Reddy}, {Erb}, {Pettini}, {Steidel}  \&
  {Shapley}}{{Reddy} et~al.}{2010}]{2010ApJ...712.1070R}
{Reddy} N.~A.,  {Erb} D.~K.,  {Pettini} M.,  {Steidel} C.~C.,   {Shapley}
  A.~E.,  2010, \mn@doi [\apj] {10.1088/0004-637X/712/2/1070}, \href
  {http://adsabs.harvard.edu/abs/2010ApJ...712.1070R} {712, 1070}

\bibitem[\protect\citeauthoryear{{Robishaw}, {Quataert}  \&
  {Heiles}}{{Robishaw} et~al.}{2008}]{2008ApJ...680..981R}
{Robishaw} T.,  {Quataert} E.,   {Heiles} C.,  2008, \mn@doi [\apj]
  {10.1086/588031}, \href {http://adsabs.harvard.edu/abs/2008ApJ...680..981R}
  {680, 981}

\bibitem[\protect\citeauthoryear{{Roseboom} et~al.,}{{Roseboom}
  et~al.}{2012}]{2012MNRAS.426.1782R}
{Roseboom} I.~G.,  et~al., 2012, \mn@doi [\mnras]
  {10.1111/j.1365-2966.2012.21777.x}, \href
  {http://adsabs.harvard.edu/abs/2012MNRAS.426.1782R} {426, 1782}

\bibitem[\protect\citeauthoryear{{Rossa} \& {Dettmar}}{{Rossa} \&
  {Dettmar}}{2003}]{2003A&A...406..505R}
{Rossa} J.,  {Dettmar} R.-J.,  2003, \mn@doi [\aap]
  {10.1051/0004-6361:20030698}, \href
  {http://adsabs.harvard.edu/abs/2003A%26A...406..505R} {406, 505}

\bibitem[\protect\citeauthoryear{{Roussel}, {Helou}, {Beck}, {Condon}, {Bosma},
  {Matthews}  \& {Jarrett}}{{Roussel} et~al.}{2003}]{2003ApJ...593..733R}
{Roussel} H.,  {Helou} G.,  {Beck} R.,  {Condon} J.~J.,  {Bosma} A.,
  {Matthews} K.,   {Jarrett} T.~H.,  2003, \mn@doi [\apj] {10.1086/376691},
  \href {http://adsabs.harvard.edu/abs/2003ApJ...593..733R} {593, 733}

\bibitem[\protect\citeauthoryear{{Roychowdhury} \& {Chengalur}}{{Roychowdhury}
  \& {Chengalur}}{2012}]{2012MNRAS.423L.127R}
{Roychowdhury} S.,  {Chengalur} J.~N.,  2012, \mn@doi [\mnras]
  {10.1111/j.1745-3933.2012.01273.x}, \href
  {http://adsabs.harvard.edu/abs/2012MNRAS.423L.127R} {423, L127}

\bibitem[\protect\citeauthoryear{{Schleicher} \& {Beck}}{{Schleicher} \&
  {Beck}}{2016}]{2016A&A...593A..77S}
{Schleicher} D.~R.~G.,  {Beck} R.,  2016, \mn@doi [\aap]
  {10.1051/0004-6361/201628843}, \href
  {http://adsabs.harvard.edu/abs/2016A%26A...593A..77S} {593, A77}

\bibitem[\protect\citeauthoryear{{Schmidt}}{{Schmidt}}{1959}]{1959ApJ...129..243S}
{Schmidt} M.,  1959, \mn@doi [\apj] {10.1086/146614}, \href
  {http://adsabs.harvard.edu/abs/1959ApJ...129..243S} {129, 243}

\bibitem[\protect\citeauthoryear{{Schmitt}}{{Schmitt}}{1985}]{1985ApJ...293..178S}
{Schmitt} J.~H.~M.~M.,  1985, \mn@doi [\apj] {10.1086/163224}, \href
  {http://adsabs.harvard.edu/abs/1985ApJ...293..178S} {293, 178}

\bibitem[\protect\citeauthoryear{{Schmitt}, {Calzetti}, {Armus}, {Giavalisco},
  {Heckman}, {Kennicutt}, {Leitherer}  \& {Meurer}}{{Schmitt}
  et~al.}{2006}]{2006ApJ...643..173S}
{Schmitt} H.~R.,  {Calzetti} D.,  {Armus} L.,  {Giavalisco} M.,  {Heckman}
  T.~M.,  {Kennicutt} Jr. R.~C.,  {Leitherer} C.,   {Meurer} G.~R.,  2006,
  \mn@doi [\apj] {10.1086/501512}, \href
  {http://adsabs.harvard.edu/abs/2006ApJ...643..173S} {643, 173}

\bibitem[\protect\citeauthoryear{{Shao}, {Xiao}, {Shen}, {Mo}, {Xia}  \&
  {Deng}}{{Shao} et~al.}{2007}]{2007ApJ...659.1159S}
{Shao} Z.,  {Xiao} Q.,  {Shen} S.,  {Mo} H.~J.,  {Xia} X.,   {Deng} Z.,  2007,
  \mn@doi [\apj] {10.1086/511131}, \href
  {http://adsabs.harvard.edu/abs/2007ApJ...659.1159S} {659, 1159}

\bibitem[\protect\citeauthoryear{{Tabatabaei} et~al.,}{{Tabatabaei}
  et~al.}{2013a}]{2013A&A...552A..19T}
{Tabatabaei} F.~S.,  et~al., 2013a, \mn@doi [\aap]
  {10.1051/0004-6361/201220249}, \href
  {http://adsabs.harvard.edu/abs/2013A%26A...552A..19T} {552, A19}

\bibitem[\protect\citeauthoryear{{Tabatabaei}, {Berkhuijsen}, {Frick}, {Beck}
  \& {Schinnerer}}{{Tabatabaei} et~al.}{2013b}]{2013A&A...557A.129T}
{Tabatabaei} F.~S.,  {Berkhuijsen} E.~M.,  {Frick} P.,  {Beck} R.,
  {Schinnerer} E.,  2013b, \mn@doi [\aap] {10.1051/0004-6361/201218909}, \href
  {http://adsabs.harvard.edu/abs/2013A%26A...557A.129T} {557, A129}

\bibitem[\protect\citeauthoryear{{Tabatabaei} et~al.,}{{Tabatabaei}
  et~al.}{2017}]{2017ApJ...836..185T}
{Tabatabaei} F.~S.,  et~al., 2017, \mn@doi [\apj]
  {10.3847/1538-4357/836/2/185}, \href
  {http://adsabs.harvard.edu/abs/2017ApJ...836..185T} {836, 185}

\bibitem[\protect\citeauthoryear{{Tingay}, {Jauncey}, {King}, {Tzioumis},
  {Lovell}  \& {Edwards}}{{Tingay} et~al.}{2003}]{2003PASJ...55..351T}
{Tingay} S.~J.,  {Jauncey} D.~L.,  {King} E.~A.,  {Tzioumis} A.~K.,  {Lovell}
  J.~E.~J.,   {Edwards} P.~G.,  2003, \mn@doi [\pasj] {10.1093/pasj/55.2.351},
  \href {http://adsabs.harvard.edu/abs/2003PASJ...55..351T} {55, 351}

\bibitem[\protect\citeauthoryear{{Voelk}}{{Voelk}}{1989}]{1989A&A...218...67V}
{Voelk} H.~J.,  1989, \aap, \href
  {http://adsabs.harvard.edu/abs/1989A%26A...218...67V} {218, 67}

\bibitem[\protect\citeauthoryear{{V{\"o}lk} \& {Xu}}{{V{\"o}lk} \&
  {Xu}}{1994}]{1994InPhT..35..527V}
{V{\"o}lk} H.~J.,  {Xu} C.,  1994, \mn@doi [Infrared Physics and Technology]
  {10.1016/1350-4495(94)90108-2}, \href
  {http://adsabs.harvard.edu/abs/1994InPhT..35..527V} {35, 527}

\bibitem[\protect\citeauthoryear{{Wang} \& {Heckman}}{{Wang} \&
  {Heckman}}{1996}]{1996ApJ...457..645W}
{Wang} B.,  {Heckman} T.~M.,  1996, \mn@doi [\apj] {10.1086/176760}, \href
  {http://adsabs.harvard.edu/abs/1996ApJ...457..645W} {457, 645}

\bibitem[\protect\citeauthoryear{{Wang} et~al.,}{{Wang}
  et~al.}{2017}]{2017MNRAS.472.3029W}
{Wang} J.,  et~al., 2017, \mn@doi [\mnras] {10.1093/mnras/stx2073}, \href
  {http://adsabs.harvard.edu/abs/2017MNRAS.472.3029W} {472, 3029}

\bibitem[\protect\citeauthoryear{{Werner} et~al.,}{{Werner}
  et~al.}{2004}]{2004ApJS..154....1W}
{Werner} M.~W.,  et~al., 2004, \mn@doi [\apjs] {10.1086/422992}, \href
  {http://adsabs.harvard.edu/abs/2004ApJS..154....1W} {154, 1}

\bibitem[\protect\citeauthoryear{{Williams} \& {Bower}}{{Williams} \&
  {Bower}}{2010}]{2010ApJ...710.1462W}
{Williams} P.~K.~G.,  {Bower} G.~C.,  2010, \mn@doi [\apj]
  {10.1088/0004-637X/710/2/1462}, \href
  {http://adsabs.harvard.edu/abs/2010ApJ...710.1462W} {710, 1462}

\bibitem[\protect\citeauthoryear{{Wright} \& {Otrupcek}}{{Wright} \&
  {Otrupcek}}{1990}]{1990PKS...C......0W}
{Wright} A.,  {Otrupcek} R.,  1990, in PKS Catalog (1990).

\bibitem[\protect\citeauthoryear{{Wright} et~al.,}{{Wright}
  et~al.}{2010}]{2010AJ....140.1868W}
{Wright} E.~L.,  et~al., 2010, \mn@doi [\aj] {10.1088/0004-6256/140/6/1868},
  \href {http://adsabs.harvard.edu/abs/2010AJ....140.1868W} {140, 1868}

\bibitem[\protect\citeauthoryear{{Wu}, {Charmandaris}, {Houck},
  {Bernard-Salas}, {Lebouteiller}, {Brandl}  \& {Farrah}}{{Wu}
  et~al.}{2008}]{2008ApJ...676..970W}
{Wu} Y.,  {Charmandaris} V.,  {Houck} J.~R.,  {Bernard-Salas} J.,
  {Lebouteiller} V.,  {Brandl} B.~R.,   {Farrah} D.,  2008, \mn@doi [\apj]
  {10.1086/527288}, \href {http://adsabs.harvard.edu/abs/2008ApJ...676..970W}
  {676, 970}

\bibitem[\protect\citeauthoryear{{Yates} \& {Kauffmann}}{{Yates} \&
  {Kauffmann}}{2014}]{2014MNRAS.439.3817Y}
{Yates} R.~M.,  {Kauffmann} G.,  2014, \mn@doi [\mnras] {10.1093/mnras/stu233},
  \href {http://adsabs.harvard.edu/abs/2014MNRAS.439.3817Y} {439, 3817}

\bibitem[\protect\citeauthoryear{{Yun}, {Reddy}  \& {Condon}}{{Yun}
  et~al.}{2001}]{2001ApJ...554..803Y}
{Yun} M.~S.,  {Reddy} N.~A.,   {Condon} J.~J.,  2001, \mn@doi [\apj]
  {10.1086/323145}, \href {http://adsabs.harvard.edu/abs/2001ApJ...554..803Y}
  {554, 803}

\bibitem[\protect\citeauthoryear{{Zahid}, {Dima}, {Kudritzki}, {Kewley},
  {Geller}, {Hwang}, {Silverman}  \& {Kashino}}{{Zahid}
  et~al.}{2014}]{2014ApJ...791..130Z}
{Zahid} H.~J.,  {Dima} G.~I.,  {Kudritzki} R.-P.,  {Kewley} L.~J.,  {Geller}
  M.~J.,  {Hwang} H.~S.,  {Silverman} J.~D.,   {Kashino} D.,  2014, \mn@doi
  [\apj] {10.1088/0004-637X/791/2/130}, \href
  {http://adsabs.harvard.edu/abs/2014ApJ...791..130Z} {791, 130}

\bibitem[\protect\citeauthoryear{{de Jong}, {Klein}, {Wielebinski}  \&
  {Wunderlich}}{{de Jong} et~al.}{1985}]{1985A&A...147L...6D}
{de Jong} T.,  {Klein} U.,  {Wielebinski} R.,   {Wunderlich} E.,  1985, \aap,
  \href {http://adsabs.harvard.edu/abs/1985A%26A...147L...6D} {147, L6}

\makeatother
\end{thebibliography}

\label{lastpage}

\end{document}